\documentclass[varenna,seceqno]{cimento}
\usepackage{graphicx}  
\title{Disordered Electron Systems}
\author{Carlo~Di~Castro}
\institute{Dipartimento di Fisica, Universit\`a di Roma "La Sapienza", and
INFM Center for Statistical Mechanics and Complexity, Piazzale Aldo Moro 2, 00185 Roma, Italy}
\author{Roberto~Raimondi}
\institute{NEST-INFM and Dipartimento di Fisica, Universit\`a di Roma Tre,
Via della Vasca Navale 84, 00146 Roma, Italy}
\begin{document}

\maketitle

\section{Introduction}
These lectures provide an introduction to the theory of disordered interacting
electron systems. As for the case of superconductivity, the understanding
of the behavior of transport and thermodynamical properties of metals and
semiconductors has required the invention of new and fascinating concepts.

In these lectures our focus is mainly on the effects of disorder, and its interplay
with electron-electron interaction.
The resulting theory, although still cannot answer some important questions, is simple and elegant.
 It describes the combined effects of interaction and disorder in terms
of a renormalized Fermi liquid, whose Landau parameters become scale dependent
and provide, together with the conductance, the  couplings flowing under the action
of the   renormalization group. However, this  final simple description, which 
 has required several decades of intensive work from  many people, is built on several
 conceptual steps. It is our aim to lead the reader through the development of these
 various steps. Our hope is that the reading of these lectures should allow
 people not expert in the field to access the original literature.
 
 There are already several  review articles which give an account of the problem
from different view points and at different stages of the historical
development \cite{bergmann1984,lee1985,altshuler1985,castellani1985,castellani1985b,finkelstein1990,kramer1993,belitz1994}. However,
the most recent and complete are still quite hard to read for unprepared readers. 
For unprepared readers
we mean those people that are not familiar with the complex technical jargon
that the field experts have developed over the years. 

We will  concentrate
on those aspects that we believe are fundamental for the problem of the
metal-insulator transition due to disorder and interaction. This will force us to ignore
a number of extensions and developments of the theory. These latter, however, 
may be found in the existing reviews. 

We have also chosen to present the theory in the simple language of standard many-body
perturbation theory. The field-theoretic approach based on the derivation of an
effective non-linear $\sigma$-model\cite{finkelstein1990} is certainly more elegant and powerful, but
requires quite some effort to appreciate the beauty of it. We invite the reader to
approach it after reading these lectures. These lecture notes are self-contained.
A basic knowledge of many-body theory and diagrammatic technique
is the only prerequisite.

After these warnings, we outline the contents of these lectures. In the next
section we set the stage for the microscopic theory by introducing the reader to the
metal-insulator transition in disordered systems and to phenomenological scaling.
The necessary background for the microscopic theory is given in the following section.
The fourth section deals with the non-interacting problem. A number of key physical and technical
ingredients are introduced in a pedagogical way. Also, the experimental urgency to take into
account interaction effects is presented. The fifth section goes to the heart of the problem
by building the renormalized disordered Fermi liquid. Gauge invariance and Ward identities
are the shining lighthouses which help us to navigate through the messy waves of the perturbation
theory. Land is finally reached in the sixth section, where we discuss the renormalization
group equations and look back to our journey and compare the theoretical
understanding with the available experiments.

\section{Setting the stage for the metal-insulator transition}
In this  section, we begin by recalling the textbook theory of electrical transport
in metals. Then we move to a description of the actual physical systems where 
the phenomena, which we describe theoretically, are observed. We conclude
the section  with the scaling theory of the metal-insulator transition due to disorder. 

\subsection{The semiclassical approach of Drude-Boltzmann}
\label{drudeboltzmann}
The conventional theory of electrical transport is due to Drude. In its original
formulation, Drude suggested that electrons, under the action of an externally
applied electric field, are accelerated according to Newton's equation of motion
until they collide with the ions after a time $\tau$. The distance between successive
collisions determines the mean free path ${\sl l}$. Due to this sequence of independent
scattering events the electrical
conductivity is given by
\begin{equation}
\label{drudeformula}
\sigma_0 = \frac{e^2 n_0 \tau}{m}
\end{equation}
where $n_0$ is the density of electrons and $e$, $m$ are the charge and electron mass. 

\begin{figure}
\includegraphics{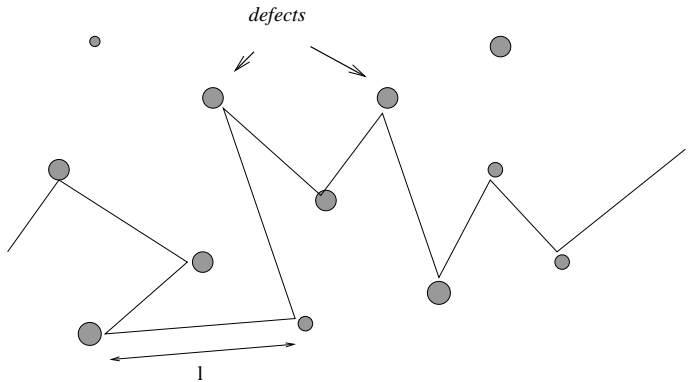}     
\caption{A pictorial representation of the semi-classical theory of transport.}
\label{drude}
\end{figure}

After the birth of quantum mechanics, Sommerfeld reformulated Drude's theory to
accomodate the Fermi statistics of electrons, providing the correct relation
between $\tau$ and ${\sl l}$ via the Fermi velocity $v_F$.  More importantly, 
with the work of Bloch, it was realized that the relaxation of electron momentum and the finite
value of the conductivity is due to imperfections of the ion lattice, {\sl i.e.} to
{\sl disorder}. Drude's law (\ref{drudeformula}) may be obtained by a {\sl semi-classical}
approach based on the Boltzmann equation for the evolution of the electron
distribution function in the presence of external fields. A pictorial description
of  Drude's model of electrical conduction is shown in fig.\ref{drude}.

As a consequence of the collisions, the electrons undergo a classical random walk of step
$l$ and a {\sl diffusive}
motion, with the diffusion coefficient $D$ related to the conductivity by 
Einstein's relation
\begin{equation}
\label{einstein1}
\sigma_0 = e^2\frac{\partial n_0}{\partial \mu} D
\end{equation}
where ${\partial n_0}/{\partial \mu}$ is the {\sl compressibility}. In the case of the
Fermi gas, ${\partial n_0}/{\partial \mu}=2 N_0$ is simply related to the density
of states per unit volume  per spin
\begin{equation}
\label{freedensityofstates}
N_0=\frac{\Omega_d}{(2\pi\hbar)^d} 2^{\frac{d-2}{2}} m^{d/2} E_F^{\frac{d-2}{2}}
\end{equation}
where $\Omega_d$ is the solid angle in $d$ dimensions and $E_F$ the
Fermi energy. From the Drude formula (\ref{drudeformula}) and  Einstein's relation
(\ref{einstein1}),  with $n_0= 2N_0 2 E_F /d$, 
one gets the diffusion coefficient $D=(2E_F \tau)/(d~m)=v_F^2\tau/d$.
Within the independent electron approximation, only one diffusion  constant $D$ controls
 the charge, spin and heat transport, leading to relations similar to eq.(\ref{einstein1})
 for the charge. In particular, the thermal conductivity, $\kappa_E$,
 \begin{equation}
 \label{thermalconductivity}
 \kappa_E=C_{V, 0} D
 \end{equation}
 where $C_{V,0}=(2\pi^2/3)k_B^2 N_0 T$ is the specific heat for the electron gas.

In concluding this subsection, we introduce the conductance related to the conductivity
by geometrical factors
\begin{equation}
\label{conductance}
G=\frac{\sigma_0 {\cal S}}{{\cal L}}=\sigma_0 { L}^{d-2}
\end{equation}
where ${\cal S}$ and ${\cal L}$ are the cross section and length of the conductor
to which we assign the typical size $L$.
By using the explicit expression of the density of states, one may rewrite $G$ 
as
\begin{equation}
\label{conductance2}
G=\frac{2e^2}{h}\frac{\Omega_d}{(2\pi)^{d-1} ~d}\frac{2E_F\tau}{\hbar}\left(\frac{p_F L}{\hbar}\right)^{d-2}
=\frac{2e^2}{h}\frac{\Omega_d}{(2\pi)^{d-1} ~d}
\frac{p_F~{\sl l}}{\hbar}\left(\frac{p_F L}{\hbar}\right)^{d-2},
\end{equation}
which shows that, in the natural conductance units ($G_0=2e^2/h=12.9 {\rm k}\Omega^{-1}$), the value of $G$ is controlled
by the dimensionless parameters $p_F {\sl l}/\hbar=2\pi {\sl l}/\lambda_F$ and 
$L/\lambda_F$. In two dimensions, in particular, conductivity and conductance have the same
physical dimensions and the ratio between the Fermi wavelength and the mean free path
is the only parameter that controls the value of the conductivity. In the semi-classical
limit, $\lambda_F\ll {\sl l}$, the Drude formula predicts a high conductivity.
The rate of collisions $\tau^{-1}$ is proportional to the impurity concentration.
By increasing the disorder in the semiclassical approach, one has that $\sigma_0$
diminishes, but remains finite. 
Ioffe and Regel\cite{ioffe1960} stated the criterion that in the metallic phase the mean free 
path $l$ cannot be smaller than the average interelectron distance proportional to $\hbar /p_F$, i.e.,
$p_F l/\hbar \geq 1$. Mott\cite{mott1972} applied this criterion to the Drude conductivity arguing that
for $d\geq 2$ there is a minimum metallic conductivity, $\sigma_{0 {\rm min}}$ when $l\approx \hbar /p_F$,
\begin{equation}
\label{sigmaminimum}
\sigma_{0 {\rm min}}=\frac{2e^2}{h}\frac{\Omega_d}{(2\pi)^{d-1}~d}\left(\frac{p_FL}{\hbar}\right)^{d-2}.
\end{equation}
As a result, there should be a discontinuity of the conductivity 
(which is universal in $d=2$) at the transition from the metal
to the insulator.  
However, when  $l\approx \hbar /p_F$ we are deeply in the quantum limit and
the Ioffe-Regel criterion cannot be naively applied to the semiclassical Drude formula.
Indeed one  expects that corrections beyond the semi-classical approximations 
will strongly modify eq.(\ref{drudeformula}) opening the way to a new perspective in the
metal-insulator transition.
Most of these lectures concern precisely this type of corrections.

\subsection{The metal-insulator transition}
\label{subsectionmit}
\begin{figure}
\includegraphics{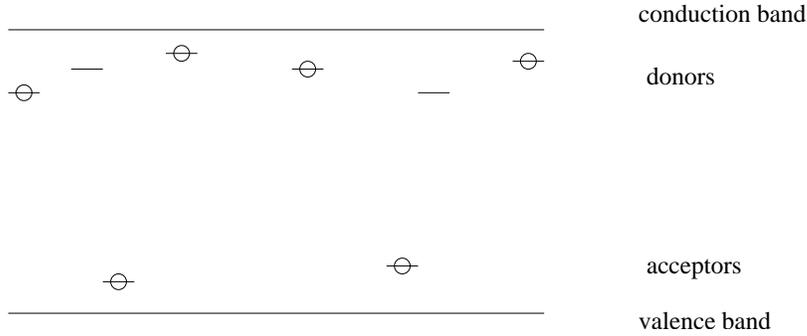}     
\caption{Energy diagram of an n-type semiconductor containing donors and acceptors.
The horizontal lines represent centres, the circles electrons in them.}
\label{impurityband}
\end{figure}

There are, of course, finite-temperature corrections to  Drude's formula.
In general, temperature-dependent corrections arise from inelastic scattering of electrons
beween them and with the phonons, resulting in the characteristic $\approx T^2$ and $\approx T^3$
behavior of the conductivity. However, typical disordered systems show, at low temperature,
strong anomalies. 
In metallic films, for instance, there are  temperature dependent   logarithmic
corrections\cite{dolan1979}
\begin{equation}
\label{anomalies2d}
\sigma (T)=\sigma_0 +m \ln T,
\end{equation} 
where $m$ is positive.
Besides metals, 
experimental realizations of disordered systems are obtained
in doped semiconductors like $Si:P$, $Ge:Sb$ and amorphous alloys as $Nb:Si$, $Al:Ge$, $Au:Ge$.
In a doped semiconductor, there are two types of conduction mechanisms. The first is due
to the thermal activated carriers and dominates at high temperature. To understand the second mechanism,
let us consider, for instance, an n-type semicondutor, as shown in fig.\ref{impurityband}.
An electron sitting at its donor atom location, has a wave function exponentially localized
around the impurity (the energy of such an electron is indicated by a short horizontal line
in the figure). Due to the small, but finite overlap of wave functions centered at different
impurity locations, the donor electrons can move around by tunneling from one impurity to another.
This gives rise to what is called impurity conduction. A doped semiconductor is compensated
when, besides the majority donor atoms, it contains also some minority acceptor atoms. In this
way, some of the donor electrons are captured by the acceptor levels, by allowing the tunneling
of a donor electron from an occupied level to an unoccupied one.
By increasing the impurity concentration, the overlap of the wave functions sitting at different
impurity sites becomes larger. 
One point to notice is that by increasing the impurity concentration, there are two
competing effects. On the one hand, disorder increases due to the larger number of
scattering centers. On the other,
at a high enough concentration of impurities, the overlap is such that the
impurity levels form a band, which behaves as an intrinsically disordered
degenerate electron gas and yields a metallic
conductivity.  Hence disorder effects are stronger at lower concentration.
The transition to metallic behavior of the impurity conduction occurs at a 
critical impurity concentration, $n_c$. 
$Si:P$, where $P$ donors sit substitutionally and randomly in a dislocation-free $Si$ lattice,
is an ideal system to study the effect of disorder on transport properties. For instance,
at enough impurity concentration to be in the metallic state,
one measures
\begin{equation}
\label{anomalies3d}
\sigma (T)=\sigma_0 +m T^n,
\end{equation}
where the coefficient $m$ can be both positive and negative and $n=1/2$
\cite{rosenbaum1983}.
By decreasing the $P$ concentration below a critical value,
the system undergoes  a metal-insulator transition at $T=0$ in the sense that
\begin{equation}
\label{mit}
\sigma_0 \approx (n -n_c)^{\mu},
\end{equation}
with the critical exponent $\mu =1/2$\cite{rosenbaum1980,rosenbaum1983}
The value of $\mu$ for uncompensated $Si:P$ is still under debate and 
depends strongly on the identification of the critical region 
in the experimental data\cite{stupp1993,rosenbaum1994,stupp1994,castner1994,stupp1994b}.
 Compensated samples\cite{thomas1982} and  the alloys
\cite{hertel1983,yamaguchi1983,rhode1987}  have $\mu =1$.

Besides the transport properties, anomalies are also seen in the 
tunneling density of states for  $Au : Ge$\cite{mcmillan1981}, $NbSi$\cite{hertel1983}, in specific
heat\cite{kobayashi1979,thomas1981,paalanen1988,lakner1989}, 
and in spin susceptibility
\cite{ikehata1985,paalanen1986,alloul1987,hirsh1992,schlager1997} in $SiP$.
As we will discuss in a more detailed way in the next sections, taking into
account corrections both in transport and thermodynamic properties will be
crucial  in developing an effective Fermi-liquid theory for these
systems.

\begin{figure}
\includegraphics{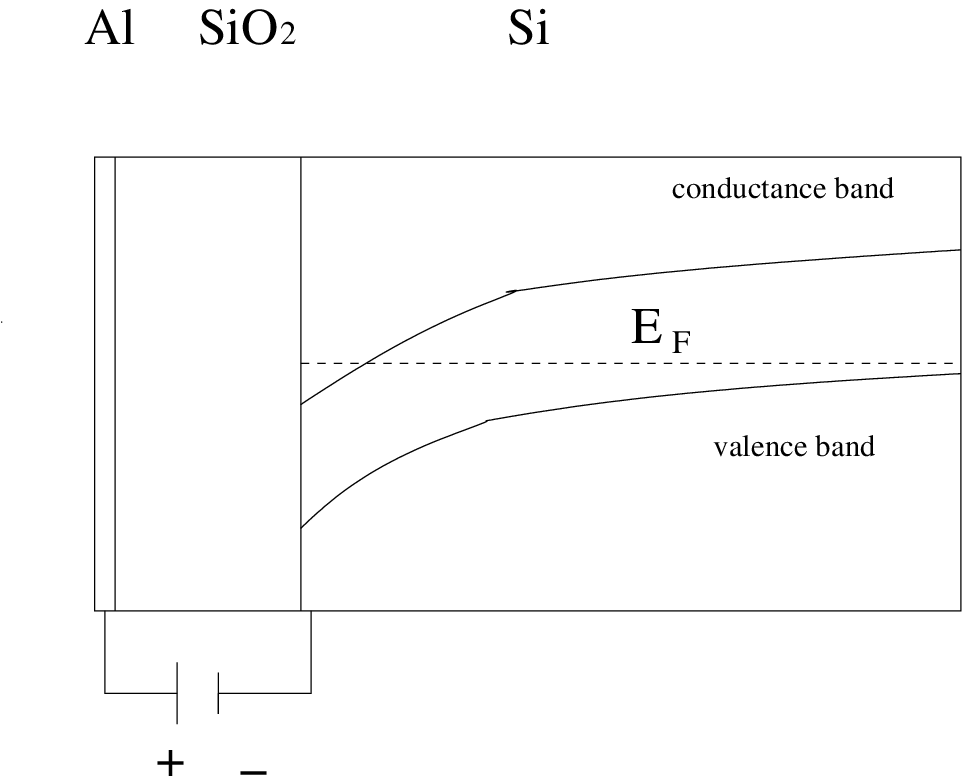}     
\caption{Scheme of a SiMOSFET device. The metallic gate ($Al$) is positevely biased so that
it attracts the electrons, which on the other hand cannot enter the insulating layer of
$SiO_2$. The electrons are then confined at the interface $Si-SiO_2$ and form a two-dimensional
electron gas. In the proximity of the interface the slope of the energy bands determines the
effective tickness of the two-dimensional electron gas.}
\label{mosfet}
\end{figure}

In more recent years, the discovery of a metal-insulator transition
in the two-dimensional electron gas\cite{kravchenko1995} has stimulated a
renovated effort to understand the interplay of disorder and interaction effects
\footnote{At  present there is not yet a general consensus on whether we have
a real zero-temperature transition or rather a crossover effect.}.
This phenomenon has been first observed in $Si$-MOSFET devices and later also
in other two-dimensional electron gas realizations as in 
semiconductor hetero-structures.
In $Si$-MOSFETs devices, the two-dimensional electron gas is formed at the interface
between the bulk silicon and an insulating layer of silicon oxide, 
as shown in fig.\ref{mosfet}.
By applying a positive bias on the metallic gate deposited on the insulating silicon
oxide layer one forces the electrons to move in the $SiO_2$-$Si$ interface, in an almost
two-dimensional environment. 
As compared with the doped semiconductor systems, these systems have the advantage
that the density of the electron gas is almost continuously controlled by the  degree of the
band bending at the interface, i.e., by the applied bias. This allows a very
fine scanning of the properties of the sample as function of  density.
Furthermore, the disorder is mainly due to scattering centers in the insulating
layer, so that in principle one varies the density  at fixed amount of disorder.
By varying the electron density, $n$, one can change the effect of the interaction since $E_F\propto n$,
and in 2D the Coulomb electron-electron interaction $E_{C}\propto n^{1/2}$.
The ratio $r_s$ between the Coulomb interaction evaluated at the average 
interparticle distance and the Fermi energy is given by
\begin{equation}
\label{coulombtokinetic}
r_s=\frac{E_C}{E_{kin}}=\frac{e^2/(\varepsilon r_{av})}{E_F},
\end{equation} 
where $\varepsilon$ is the dielectric constant.
By using $E_F=n/(2N_0)$ and $r_{av}=1/\sqrt{n}$ and recalling the expression of the
Bohr radius $a_B =\varepsilon \hbar^2/(me^2)$, one obtains
\begin{equation}
\label{rs}
r_s=\frac{1}{\pi \varepsilon a_B\sqrt{n}}.
\end{equation}
At the present, the origin of the metal-insulator transition in two-dimensional
systems is still unclear and represents a very hot issue of debate in
the literature. (A recent review may be found in refs.\cite{abrahams2001,kravchenko2004}).
For this reason, we prefer not to enter now in a detailed discussion of the
experimental features of this phenomenon, which we will point out later on when
relevant results of the theory will require it.

\subsection{The Anderson transition and quantum interference}
\label{anderson}
All of this suggests that disorder cannot be treated only within a semi-classical approach.
In 1958, Anderson invented the field of {\sl localization}, proposing that under certain
circumstances diffusion may be completely suppressed\cite{anderson1958}. 
He proposed a lattice model where
the site energies are randomly distributed. When disorder is absent, a small hopping
amplitude is enough to delocalize the electron states and form Bloch waves. However,
by increasing the disorder, the hopping processes may only spread an initially localized
state over a finite distance, which defines the localization length $\xi_0$. 
Since in the process of the impurity band formation, 
localized states are more likely to form in the band tails, Mott argued\cite{mott1967}
that
there must exist a critical energy $E_c$, called the {\sl mobility edge},
which separates localized from extendend states. When the Fermi energy $E_F$ is below
the mobility edge, the system is an insulator. When $E_F$ passes through
$E_c$, the system becomes metallic.  From this point of view, for a given model and
given amount of disorder, the problem is to compute $E_c$. For non-interacting
systems, this can be tackled numerically, by exactly solving the
Schr\"odinger equation in a disordered lattice. A review of the status of the numerical
simulations may be found in ref.\cite{kramer1993}.

Even though these concepts played an important role in shaping our
modern view of the  metal-insulator transition, a great impulse
to the development of the field came, however,
 by the discovery of the phenomenon of {\sl weak
localization}.

As remarked at the end of the previous subsection, the standard theory of transport
is based on a semi-classical approach, where in evaluating the probability for electron
diffusion one neglects the interference between the amplitudes corresponding to different
trajectories and essentially treats a classical random walk with step
$l$ and diffusion constant $D$.
 This is indeed justified in many cases, where the semi-classical theory
works. 
In fact,  in a disorderd system the phase difference for any two different trajectories
will  vary randomly. The situation changes, however,
for  {\sl self-intersecting} trajectories, which come from closed loops. In this case,
trajectories naturally come in pairs, depending on whether  one goes around the loop
clockwise or counter-clockwise.
One expects that interference between these pairs of trajectories  modifies the semi-classical
result.  There is a simple argument to estimate the probability of having a self-intersecting
trajectory. On the one hand, one has that the electron motion is described by a classical
diffusion process such that the average distance after a time $t$ is
\begin{equation}
\label{diffusionlaw}
\overline{{\bf r}^2}=Dt.
\end{equation}
On the other hand, the quantum nature of the electron may be thought of in terms of a {\sl tube}
of size $\lambda_F$ generated by the electron motion. In  a time ${\rm d}t$, the volume spanned
by the tube increases by 
\begin{equation}
\label{tubelaw}
{\rm d} V_{tube}=\lambda_F^{d-1} ~v_F {\rm d}t.
\end{equation}
Let us consider the ratio between the increase of the tube volume in time ${\rm d}t$ and 
the total volume generated by the diffusion
process. The total probability for self-intersection may be estimated by integrating this ratio
over time
\begin{equation}
\label{estimateofselfintersection}
P \sim \int_{\tau}^{\tau_{\phi}}\frac{\lambda_F^{d-1} v_F {\rm d}t}{(Dt)^{d/2}},
\end{equation}
where the lower limit $\tau$ is the time above which the diffusive regime, after a few
collisions, starts to set in. The upper limit, $\tau_{\phi}$, is the time until which phase
coherence of the wave function persists. In general, {\sl inelastic} processes at finite
temperature make $\tau_{\phi}$ a decreasing function of  temperature.
In two dimensions the probability grows logarithmically as  temperature decreases.
 At zero temperature, when
$\tau_{\phi}\rightarrow \infty$, the upper limit is provided by the system size via
the diffusion relation $L^2/D\equiv \tau_L$. In this way, the probability of self-intersection
acquires  a scale dependence
\begin{equation}
P\sim \frac{v_F \lambda_F^{d-1}}{D^{d/2}\tau^{ (d-2)/2}}\frac{2}{d-2}
\left[ \left( \frac{\tau}{\tau_L} \right)^{\frac{d-2}{2}}-1 \right]\propto
\frac{G_0}{\sigma_0 l^{d-2}}
\frac{2}{d-2}
\left[ \left(\frac{\tau}{\tau_L} \right)^{\frac{d-2}{2}}-1 \right].
\end{equation}
By assuming that the  conductivity corrections are proportional to this probability, one obtains
\begin{equation}
\label{conductivityestimate}
\frac{\delta\sigma}{\sigma_0}\propto \frac{1}{g_0 \epsilon}
\left[\left( \frac{l}{L}\right)^{\epsilon}-1 \right]
\end{equation} 
where $\epsilon =d-2$ and $g_0\equiv G(l)/G_0$ is the conductance at the scale $l$ in units of $G_0$.
Equation (\ref{conductivityestimate}) is valid at $T=0$ and the inverse scattering time or the inverse mean free
path play the role of the  ultraviolet cutoff, whereas $\tau_L^{-1}$ is the infrared cutoff.
At $d=2$, the conductivity correction is log-singular. 

At finite temperature, when $\tau_{\phi}<\tau_L$, the infrared cutoff becomes temperature dependent
and in 2D the correction becomes logarithmic in temperature.
This opens the way to the  scaling theory discussed in the next subsection. 

As a final remark to this subsection, we point out that the weak-localization
phenomenon is sensitive to any perturbation that breaks the time reversal
invariance. This is clear from the above argument of the interference between time
reversed trajectories. For instance, in the presence of a magnetic field,
the two trajectories acquire a phase difference 
$\phi_1 -\phi_2 =(2e /\hbar c)\Phi_B$, proportional to  the magnetic flux $\Phi_B$
threading the surface delimited by the closed loop. Since at finite temperature
the logarithmic singularity is cut off at time $\tau_{\phi} $,  in order to cut off the singularity 
typical magnetic 
fields must be of the order of a flux quantum over a region whose size is of the order of
the  dephasing length $L_{\phi}=\sqrt{D\tau_{\phi}}$. This gives the condition
 $B \geq (h/ec)/L_{\phi}^2$.
It is also clear that further dephasing mechanisms, as for instance, spin-flip scattering
with typical time $\tau_s$ become important when $\tau_s <\tau_{\phi}$.
\subsection{The scaling theory of the metal-insulator transition}
\label{subsectionscaling}
\begin{figure}
\includegraphics{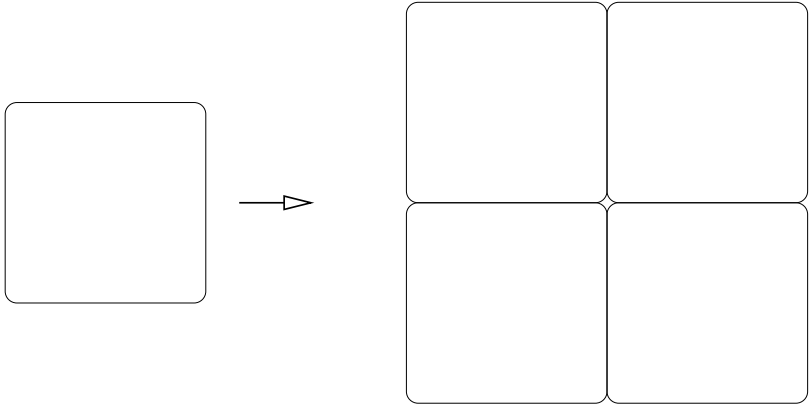}
\caption{Block of size $2L$ obtained by blocks of size $L$.}
\label{thouless}
\end{figure}

\begin{figure}
\includegraphics{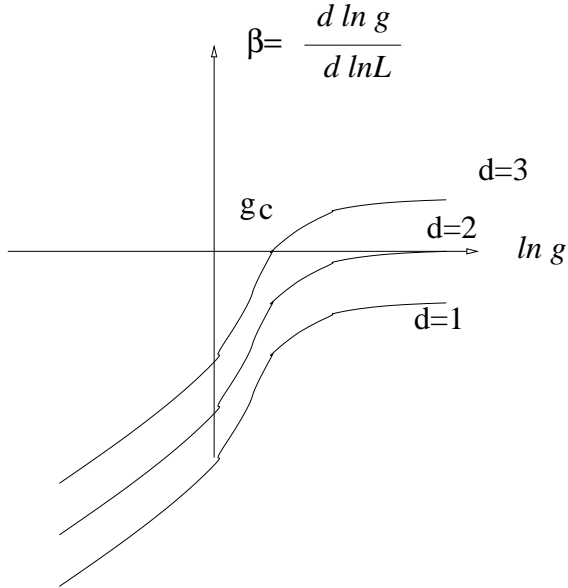}
\caption{Schematic $\beta$-function.}
\label{scaling}
\end{figure}

The starting point is the argument of Thouless concerning the evolution
of the wave function as the system size is 
increased\cite{edwards1972,licciardello1975}. To fix the ideas, let us imagine
that the system of system size $2L$ is made up by combining blocks of size $L$,
as shown in fig.\ref{thouless}.
Suppose we know the eigenstates for a block of size $L$. The level spacing for these
is $\Delta E$. We ask how are the states when we combine blocks together. Let $\delta E$
be the energy brought about by the perturbation of joining the blocks together. Clearly,
we expect that for $\delta E \ll \Delta E$ the new eigenstates for a system of
size $2L$ will differ very little from  those at scale $L$. The energy 
$\delta E$ measures the sensitivity to a change in the boundary conditions.
In a diffusive sytem this may be related to the time necessary to reach the
boundary, $\delta E =\hbar D/L^2$. The level spacing, on the other hand, is related to
the inverse density of states, $\Delta E =1/(N_0 L^d)$. By using 
Einstein's relation, the ratio of the two energies  gives
\begin{equation}
\label{thoulesscriterion}
\frac{\delta E}{\Delta E}=
N_0 L^d \frac{\hbar D}{L^2}=\frac{1}{2\pi}\frac{\sigma_0 L^{d-2}}{2e^2/h}=\frac{g(L)}{2\pi}
\end{equation}
where $g(L)$ is the conductance  at the scale $L$ in units of $2e^2/h$.
If $g\ll 1$, the new eigenstates are not modified much by the assembling of the blocks.
On the other hand, when $g\gg 1$, the new states are delocalized on all the blocks.
The scaling theory\cite{abrahams1979} assumes the $g(L)$ is the only parameter that controls
the evolution of the eigenstates when we rescale the system size.
Mathematically this is expressed by requiring that 
the conductance of a block of size $L'=bL$ is expressed in terms of the conductance of a block of size
$L$ by a function of $L'/L$ and $g(L)$ only, i.e., $g(L')=f(L'/L, g(L))$.
Its logarithmic derivative for $L'=L$, which defines 
the  $\beta$-function of the corresponding renormalization group equations, depends on the scale
$L$ only through $g(L)$ itself
\begin{equation}
\label{betafunction}
\frac{{\rm d}\ln g (L)}{{\rm d}\ln L}=\beta (g(L)).
\end{equation}
The vanishing of the $\beta$-function controls the scale-invariant limit, i.e., provides the fixed point of the
trasformation $g^*$. In the case of one-parameter equation the fixed $g^*$ point coincides with the critical
value $g_c$. Linearization of the transformation, starting from the fixed point, provides the scaling behavior 
of the physical quantities.
The $\beta$-function is relatively well known in the two limits of a good metal,
where Ohm's law is valid and $\sigma$ is a constant,
 and in the strongly localized insulating regime, where
the scale-dependent conductance  falls off exponentially over a localization length $\xi_0$ as
\begin{equation}
\label{insulatingregime}
g(L)=g_0 e^{-L/\xi_0}.
\end{equation}
One  then immediately gets
\begin{eqnarray}
\beta (g)&=&d-2,~~~g\gg  1\label{betaasymptoticsmetal}\\
\beta (g)&=&\ln\frac{g}{g_0},~~~g \ll 1\label{betaasymptoticsinsulator}
\end{eqnarray}
where $g_0$ is the conductance at some initial microscopic scale $l$. 
Under reasonable assumptions, the $\beta$-function has the qualitative behavior
shown in fig.\ref{scaling}.
A positive (negative)
 value for $\beta$ means that upon increasing the system size, $g$ increases
 (decreases) corresponding to a metallic (insulating)
 behavior. 
 
 A zero $g_c$ such that $\beta (g_c)=0$,  signals an unstable fixed
 point for the flow of $g$. This represents a metal-insulator transition.
 One consequence of eqs.(\ref{betaasymptoticsmetal}),(\ref{betaasymptoticsinsulator})
  is that, for $d \leq 2$, the system
 is always an insulator at zero temperature and all states are localized.
Close to a critical point at $d>2$ we may linearize the $\beta$-function to get
\begin{equation}
\label{linearizebeta}
\frac{{\rm d}(g-g_c)}{{\rm d}\ln L}=g_c \beta'(g_c)(g-g_c), \ \ 
\beta'(g_c)=\left(\frac{{\rm d}\beta (g)}{{\rm d} g} \right)_{g=g_c}
\end{equation}
from which
\begin{equation}
\label{linearizedg}
g(L)-g_c =(g_0 -g_c)\left(\frac{L}{\sl l}\right)^{x_g}
\end{equation}
where $g_0$ is the conductance at scale ${\sl l}$ and 
\begin{equation}
\label{criticalexponent}
x_g =g_c\beta'(g_c).
\end{equation}
By defining a correlation length, which coincides with the localization length in the insulating side,
it diverges at criticality as
\begin{equation}
\label{correlationlength}
\xi \sim (g_0 -g_c)^{-\nu}.
\end{equation}
By assuming that $\xi$ is the only relevant length, it scales
as $\xi'=\xi /(L/{\sl l})$ and one deduces $\nu =1/x_g$. Furthermore,
from the critical behavior of the conductivity
\begin{equation}
\label{criticalsigma}
\sigma \sim (g_0 -g_c)^{\mu}\sim \frac{g(\xi)}{\xi^{d-2}}\rightarrow \frac{g_c}{\xi^{d-2}}
\end{equation}
one derives the scaling law\cite{wegner1976}
\begin{equation}
\label{scalinglaw}
\mu =(d-2)\nu \equiv \epsilon \nu .
\end{equation}
In the metallic regime, where $g$ is large, we can assume that the $\beta$-function
can be expandend in a power series  in
$1/g$\cite{langer1966,abrahams1979}
\begin{equation}
\label{betametallic}
\beta (g)=d-2 -\frac{a}{g}-\frac{b}{g^2}+...~.
\end{equation}
Above two dimensions, if $a>0$, one has a fixed point $g_c =a/\epsilon$ and $\nu =1/\epsilon$
leading to $\mu =1$. On the other hand, if $a=0$, the fixed point is determined by
the second order term, $g_c^2=b/\epsilon$, which implies $\nu =1/(2\epsilon)$ and $\mu =1/2$.
Finally, when $a<0$, there is no fixed point at this order. 

At $d=2$ the scaling equation reduces to
\begin{equation}
\label{scaling2d}
\frac{{\rm d}g}{{\rm d}\ln L}=-\frac{a}{g}.
\end{equation}
For $a>0$ the system scales to an insulator, whereas for $a<0$ it scales to a perfect
conductor. This phenomenological theory does not allow for a metallic phase in $d=2$.

\section{The microscopic approach}
In this section we introduce a few general tools that will be used in building a
microscopic theory. First we discuss how physical observables may be evaluated
in terms of response functions. Secondly, we show how conservation laws impose
constraints, {\sl e.g.} Ward Identities,
 on these correlation functions, which are  useful when
performing perturbative expansions. We mainly follow ref.\cite{castellani1983}.
\subsection{Linear response theory, Kubo formula and all that}
\label{kubo}
The coupling with an external electromagnetic field is given by
the Hamiltonian\footnote{We adopt the relativistic notation: upper and lower indices
indicate contravariant and covariant vectors, respectively, {\sl i.e.}
$A^{\mu}=(\phi , {\bf A})$ and $A_{\mu}=(\phi , -{\bf A})$. }
\begin{equation}
\label{eminteraction}
H_{emi}=\frac{1}{c}\int~{\rm d}{\bf r}A^{\mu} ({\bf r})J_{\mu}({\bf r})
\end{equation}
where the Greek index runs over time ($\mu =0$) and space indices  ($\mu =1,...,d$).
The latter will be later on indicated by Latin letters .
As it is standard, lower indices have space components with a minus sign,
{\sl e.g.}, $J_{\mu}=(c\rho , - {\bf J})$.
 With ${\bf r}$ we indicate the position vector
in any dimension $d$. The external scalar, $\phi ({\bf r})$, and vector
potential, ${\bf A}({\bf r})$, are coupled with the charge and current density,
defined by
\begin{eqnletter}
\label{densities}
\rho ({\bf r})&=& e\psi^{\dagger}_{\sigma}({\bf r})\psi_{\sigma}({\bf r})\label{density}\\
{\bf J}({\bf r})&=& -{\rm i}\frac{e\hbar}{2m}\left[
\psi^{\dagger}_{\sigma}({\bf r})\nabla\psi_{\sigma}({\bf r})-
(\nabla\psi^{\dagger}_{\sigma}({\bf r}))\psi_{\sigma}({\bf r})\right]
-\frac{e^2}{mc}{\bf A}({\bf r})
\psi^{\dagger}_{\sigma}({\bf r})\psi_{\sigma}({\bf r})\label{current}\\
&\equiv &{\bf j}({\bf r})-\frac{e^2}{mc}{\bf A}({\bf r})
\psi^{\dagger}_{\sigma}({\bf r})\psi_{\sigma}({\bf r}).\label{twocurrents}
\end{eqnletter}
In the following, for the sake of simplicity we shall set $\hbar$, $c$, and
$k_B$ equal to one. In eqs.(\ref{densities}), $\psi_{\sigma} ({\bf r})$ 
($\psi^{\dagger}_{\sigma} ({\bf r})$) is the annihilation (creation) Fermion
field operator. Our goal is to study the system response to an external
electromagnetic field within linear response.
The second term in eq.(\ref{twocurrents}), the  diamagnetic contribution,
being  already linear in the  field, may be evaluated as
\begin{equation}
\label{diamagnetic}
{\bf j}^{dia}=-\frac{e^2}{m}n_0 {\bf A}
\end{equation} 
where $n_0$ is the equilibrium (number) density. 
By the compact notation $x=(t,{\bf r})$,
the linear response is given by
\begin{equation}
\label{linearresponse}
J^{\mu}(x)=\int {\rm d} x'~K^{\mu\nu}(x,x')A_{\nu}(x')
\end{equation}
where
the response kernel $K^{\mu\nu}(x,x')$ is the four-current correlation function, which includes
both the density-density and current-current correlation functions,
\begin{equation}
\label{separation}
K^{\mu\nu}(x,x')=R^{\mu\nu}(x,x')+
\frac{e^2}{m}n_0\delta^{(4)}(x-x')\delta_{\mu\nu}(1-\delta_{\nu 0})
\end{equation}
and
\begin{equation}
\label{responsefunction}
R^{\mu\nu}(x,x')=-{\rm i}\theta (t-t')<[j^{\mu}(x),j^{\nu}(x')]>,
\end{equation}
with the average taken over the appropriate statistical
 ensemble\footnote{The plus sign in front of the diamagnetic term is due to the fact that
by using a lower index for $A_{\mu}$ the space part has a minus sign.} . If the unperturbed system
is traslationally invariant and has a   time-independent Hamiltonian,
we can use Fourier transforms with respect to both ${\bf r}-{\bf r}'$ and $t-t'$,
\begin{equation}
\label{fourier}
K^{\mu\nu}({\bf r}-{\bf r}', t-t')=
\int^{\infty}_{-\infty}\frac{{\rm d}\omega}{2\pi}\sum_{\bf q}
e^{{\rm i}{\bf q}\cdot ({\bf r}-{\bf r}')-{\rm i}\omega (t-t')}
K^{\mu\nu}({\bf q}, \omega).
\end{equation}
The sum over the momenta is left unspecified for the time being. It depends on the choice
of boundary conditions. In the limit of an infinite system, the sum gets replaced
by an integral over all space in the standard way.
In Fourier space, eq.(\ref{linearresponse}) becomes local 
\begin{equation}
\label{linearresponsefourier}
J^{\mu}(q)=K^{\mu\nu}(q)A_{\nu}(q),
\end{equation}
where $q=(\omega ,{\bf q})$. Physical observables are now readily obtained. For instance,
the DC electrical conductivity, by making the  choice of a time-dependent vector gauge,
${\bf E}=-\partial_t {\bf A}(t)$, reads
\begin{equation}
\label{conductivitykubo}
\sigma_{ij}=-\lim_{\omega\rightarrow 0}\frac{K^{ij}({\bf 0},\omega )}{{\rm i}\omega}.
\end{equation}
\subsection{Conservation laws and gauge invariance}
\label{gauge}
Charge conservation is expressed by the continuity equation 
\begin{equation}
\label{continuity}
\partial_t \rho +\nabla\cdot {\bf J}=0,
\end{equation}
while gauge invariance requires that the physics is unchanged by the
replacement
\begin{equation}
\label{gaugeinvariance}
A^{\mu}(x)\rightarrow A^{\mu}(x)-\partial^{\mu}f(x),
\end{equation}
with $f$ an arbitrary function and $\partial^{\mu}=(\partial_t ,-\nabla )$.
Equations (\ref{continuity},\ref{gaugeinvariance}) imply
\begin{eqnletter}
\label{constraints}
q_{\mu}K^{\mu\nu}&=&0,\label{constraint1}\\
K^{\mu\nu}q_{\nu}&=&0\label{constraint2}.
\end{eqnletter}
More explicitly, one has the following relations connecting the various correlation
functions:
\begin{eqnletter}
\omega K^{00}&=&q^jK^{j0},\\
\omega K^{0i}&=&q^jK^{ji},\\
\omega K^{00}&=&q^jK^{0j},\\
\omega K^{i0}&=&q^jK^{ij},
\end{eqnletter}
from which $K^{0i}=K^{i0}$, $K^{ij}=K^{ji}$, and 
\begin{equation}
\label{connection}
\omega^2 K^{00}=q^i K^{ij}q^j.
\end{equation}
 The conductivity tensor may be decomposed
into  {\sl longitudinal} and {\sl transverse} components as
\begin{equation}
\label{decomposition}
\sigma_{ij}=\frac{q^iq^j}{{\bf q}^2}\sigma_L+\left(\delta_{ij}-\frac{q^iq^j}{{\bf q}^2}\right)\sigma_T
\end{equation}
so that eq.(\ref{conductivitykubo}) reads
\begin{equation}
\label{kubo2}
\sigma_L ={\rm i}\lim_{\omega\rightarrow 0}\lim_{|{\bf q}|\rightarrow 0}\frac{\omega}{{\bf q}^2}
K^{00}({\bf q},\omega).
\end{equation}
The charge response to a static and homogeneous external potential, {\sl e.g.} the compressibility,
is given as
\begin{equation}
\label{compressibility}
\frac{\partial n}{\partial \mu} =- \frac{1}{e^2}\lim_{|{\bf q}|\rightarrow 0} K^{00}({\bf q},0).
\end{equation}
To appreciate the physical meaning of  eqs.(\ref{kubo2},\ref{compressibility}),
let us consider the {\sl phenomenological} expression of the current for a good metal
\begin{equation}
\label{phenomenologicalcurrent}
{\bf J}=\sigma_L {\bf E}-D{\bf \nabla}\rho
\end{equation}
where $D$ is the diffusion coefficient, which, under general statistical considerations,
is related to $\sigma_L$ by  Einstein's relation
\begin{equation}
\label{einstein}
\sigma_L =e^2\frac{\partial n}{\partial \mu} D.
\end{equation}
Equation (\ref{phenomenologicalcurrent}) may be used together with the continuity equation (\ref{continuity})
to find an expression for the {\sl density-density}, $K^{00}$, response function.
By taking the divergence of eq.(\ref{phenomenologicalcurrent}) and replacing it into eq.(\ref{continuity}),
one gets
\begin{equation}
\label{densityfluctuations}
(\partial_t-D\nabla^2 )\rho =\sigma_L\nabla^2 \phi,
\end{equation}
from which, after Fourier transforming, the density-density response function reads
\begin{equation}
\label{densitydensity}
K^{00}=-\sigma_L\frac{{\bf q}^2}{-{\rm i}\omega+D{\bf q}^2}=-e^2\frac{\partial n}{\partial \mu}
\frac{D{\bf q}^2}{-{\rm i}\omega+D{\bf q}^2}.
\end{equation}
The above equation, of course, agrees with eq.(\ref{kubo2}) and gives the compressibility
(\ref{compressibility}) as required by  Einstein's relation. Notice that the latter, within
the linear response, is derived from the eq.(\ref{connection}), connecting the
density-density and current-current response functions. The task of a {\sl microscopic}
theory, as it will be shown in the following sections, is to derive the expression for the
current instead of {\sl phenomenologically} assuming eq.(\ref{phenomenologicalcurrent}).
\subsection{Response functions and Ward identities}
\label{ward}
We begin by introducing a  vertex function 
\begin{equation}
\label{reduciblevertex}
\Lambda^{\mu}(x,x',x'')=<T_t J^{\mu}(x)\psi (x')\psi^{\dagger} (x'')>,
\end{equation}
where $T_t$ is the time-ordering operator and the average is over a statistical ensemble.
In this subsection, for the sake of simplicity, we  confine ourselves to the zero-temperature
limit with the average taken over the ground state. We also neglect spin indices for a little
while to keep the notation as simple as possible. When the derivative $\partial/\partial_{x_{\mu}}$ acts
on the right-hand side of eq.(\ref{reduciblevertex}), one obtains two contributions. One is due to 
the derivative acting on $J^{\mu}$ and gives zero due to the continuity equation (\ref{continuity}).
A second contribution comes from the time-derivative of the $T_t$-product. As a result one gets
the following Ward Identity:
\begin{equation}
\label{wardidentity}
\frac{\partial}{\partial x_{\mu}}\Lambda^{\mu}(x,x',x'')=
{\rm i}e~\delta (x-x'')G(x',x)-{\rm i}e~\delta (x-x')G(x,x''),
\end{equation}
where we have introduced the single-particle Green function
\begin{equation}
\label{greenfunction}
G(x,x')=-{\rm i}<T_t \psi (x)\psi^{\dagger}(x')>.
\end{equation}
One may also consider Fourier transforms with respect to the relative coordinates
$x-x''$ and $x'-x$ (The arguments of the two Green's functions in the right-hand side
of eq.(\ref{wardidentity})). We define the Fourier transform of eq.(\ref{reduciblevertex}) as
\begin{equation}
\label{vertexfourier}
\Lambda^{\mu}(x,x',x'')=\int {\rm d}q~{\rm d}p~e^{{\rm i}(p-q/2)(x'-x)} 
~e^{{\rm i}(p+q/2)(x-x'')}~\Lambda^{\mu}(p,q)
\end{equation}
in terms of which the Ward Identity becomes
\begin{equation}
\label{wardidentityfourier}
q_{\mu}\Lambda^{\mu}(p,q)=e ~G(p-q/2)-e ~G(p+q/2).
\end{equation}
In the above $p=(\epsilon , {\bf p})$ and in eq.(\ref{vertexfourier})
$${\rm d}p~=\frac{{\rm d}\epsilon}{2\pi}\sum_{\bf p},
$$
and similarly for $q$.
The connection between the vertex function and response functions is  obtained by
 introducing the truncated
vertex $\Gamma^{\mu}$ defined as
\begin{equation}
\label{irreduciblevertex}
\Lambda^{\mu}(p,q)=G(p+q/2)\Gamma^{\mu}(p,q)G(p-q/2).
\end{equation}
In terms of $\Gamma^{\mu}$ the response functions read
\begin{equation}
\label{responsefunctiongamma}
K^{\mu\nu}(q)=-{\rm i}  \int {\rm d}p~\gamma^{\mu}(p,q)G(p+q/2)\Gamma^{\nu}(p,q)G(p-q/2),
\end{equation}
where $\gamma^{\mu}=(e, e~{\bf p}/m)\equiv (e, \gamma^i_{\bf p})$ is the {\sl bare}  vertex. 
As a check, the bare vertex is found by writing the Ward Identity in terms of $\Gamma^{\mu}$
\begin{equation}
\label{wardidentitygamma}
q_{\mu}\Gamma^{\mu}(p,q)=e ~G^{-1}(p+q/2)-e ~G^{-1}(p-q/2),
\end{equation}
and using the {\sl bare} Green's function expression
\begin{equation}
\label{greenfunctionfree}
G(p)=\frac{1}{\epsilon -\xi_{\bf p}+{\rm i}~sign~(|{\bf p}|-p_F)},
\end{equation}
$p_F$ being the Fermi momentum and $\xi_{\bf p}={\bf p}^2/2m -\mu$.

We conclude this section by giving a few more consequences of the Ward identity 
(\ref{wardidentityfourier}). First, we notice that, while in the {\sl static }
limit, the density-density response function gives the compressibility (compare
eq.(\ref{compressibility})), in the {\sl dynamic} limit we have
\begin{equation}
\label{dynamic}
\lim_{\omega \rightarrow 0} K^{00}({\bf 0},\omega )=0.
\end{equation}
The above result, which is a mathematical formulation
of the particle number conservation, 
follows from eq.(\ref{wardidentityfourier}) after taking the ${\bf q}$-zero limit
and upon integration over momentum ${\bf p}$ and the entire energy $\epsilon$ range.

Finally, by restricting the frequencies to the region $(\epsilon +\omega /2)(\epsilon-\omega /2) <0$,
and taking advantage of the Ward Identity in the zero-momentum limit,
one gets for  the density response function
\begin{eqnarray}
\label{dynamicresponse}
K^{00}_{+-}({\bf 0},\omega )&=&-{\rm i} e 
\int^{\omega /2}_{-\omega /2} \frac{{\rm d}\epsilon}{2\pi}\sum_{\bf p}~
\Lambda^{0}(\epsilon, {\bf p};\omega ,{\bf 0})\nonumber\\
&=&-{\rm i} e^2 
\int^{\omega /2}_{-\omega /2} \frac{{\rm d}\epsilon}{2\pi}\sum_{\bf p}
\frac{1}{\omega}\left(  G^R({\bf p}, \epsilon -\omega /2 ) -
G^A({\bf p}, \epsilon +\omega /2 )\right)\nonumber \\
&\approx_{\omega \rightarrow 0}&{\rm i} e^2\frac{1}{2\pi}\sum_{\bf p}
\left(  G^R({\bf p}, 0 ) -
G^A({\bf p}, 0)\right),
\end{eqnarray}
where we made use of the fact that the sign of the frequency determines  whether
the Green's function is analytical in the upper (retarded R) or in the lower (advanced A)
half of the complex plane as a function of  frequency.
By recalling the expression for the single-particle density of states
$$
N(\epsilon )=-\frac{1}{\pi}\sum_{\bf p}{\cal I}m~ G^R({\bf p}, \epsilon ),
$$
one obtains an expression for the single-particle density of states at the Fermi energy
\begin{equation}
\label{densityofstates}
N(0)\equiv N=\lim_{\omega\rightarrow 0}\frac{1}{2 e^2}K^{00}_{+-}({\bf 0},\omega ),
\end{equation}
where the factor of $2$ in the denominator is due to the spin degeneracy.

\section{Non-interacting Disordered Electrons}
\subsection{Self-consistent Born approximation}
\label{subsectionborn}
Quite generally, non-interacting electrons in the presence
of disorder are described by the following Hamiltonian:
\begin{equation}
\label{noninteracting}
H=\int~{\rm d}{\bf r}~\psi_{\sigma}^{\dagger}({\bf r})
\left[-\frac{\nabla^2}{2m}+u({\bf r}) \right]\psi_{\sigma}({\bf r})
\end{equation}
where $u({\bf r})$ is taken as a {\sl Gaussian} random variable defined by
\begin{equation}
{\overline {u({\bf r})}}=0,~~{\overline {u({\bf r})u({\bf r}')}}={\overline {u^2}}
\delta ({\bf r}-{\bf r}')\equiv
\frac{1}{2\pi N_0\tau}\delta ({\bf r}-{\bf r}').
\end{equation}
In the above, $N_0$ is the {\sl free} single-particle density of states per spin and
$\tau$ is a parameter inversely proportional to the impurity concentration and
 whose physical meaning will be evident in a few moments.
\begin{figure}
\includegraphics{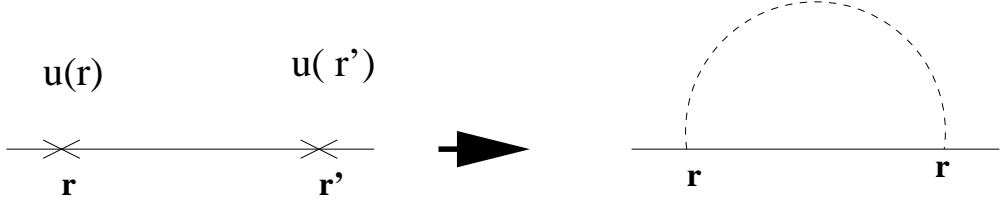}     
\caption{Self-energy in the Born approximation before and after averaging over
the impurity distribution. The dashed line represents the average
of two impurity insertions. When the internal Green's function line (solid line) is
replaced with the dressed Green's function one obtains the self-consistent
Born approximation.}
\label{born}
\end{figure}
In the Born approximation\cite{abrikosov1975}
 one has for the self-energy the expression (see fig.\ref{born})
\begin{equation}
\label{firstorderselfenergy}
\Sigma^{1} ({\bf r},t;{\bf r}',t')=\frac{\delta ({\bf r}-{\bf r}')}{2\pi N_0\tau}
G^{0}({\bf r},t;{\bf r},t').
\end{equation}
The superscript $0$ for $G$ indicates that we are considering the unperturbed expression
(\ref{greenfunctionfree}). In Fourier space, eq.(\ref{firstorderselfenergy}) reads
\begin{equation}
\label{firstorderselfenergyfourier}
\Sigma^{1}({\bf p},\epsilon )=\frac{1}{2\pi N_0\tau}
\sum_{{\bf p}'}\frac{1}{\epsilon -\xi_{\bf p'}+{\rm i}~sign~(|{\bf p}'|-p_F)}.
\end{equation}
For large values of ${\bf p}'$, the real part of the sum over ${\bf p}'$ diverges, but
its value {\sl does} not depend on the energy $\epsilon$. This divergency is a consequence
of the simple model taken for the scattering potential. A more realistic momentum-dependent scattering potential
will generally cure the divergence and give rise to a finite contribution that may be
absorbed into a redefinition of the chemical potential. The main contribution to the energy
dependence of  the
sum comes from the values of ${\bf p}'$ close to the Fermi surface. By following
the standard procedure we pass from momentum to energy integration
\begin{equation}
\label{standard}
\sum_{{\bf p}}~...=N_0~\int_{-\mu}^{\infty}~{\rm d}\xi_{\bf p}...\approx N_0~\int^{\infty}_{-\infty}~
{\rm d}\xi_{\bf p}...
\end{equation}
where we have sent to minus infinity the lower limit of integration, since $\mu\approx E_F$ is
the biggest energy scale in the problem. Then, by residue integration, we obtain
\begin{equation}
\label{residue}
 \Sigma^{1}({\bf p},\epsilon )=-\frac{{\rm i}}{2\tau}sign (\epsilon ).
 \end{equation}
To proceed in the perturbative expansion one replaces the above result into the
Green's function and computes $\Sigma^2$. At second order one has exactly the same
expression as before except that the pole of the Green's function is now moved away 
 from the real axis by the quantity $1/2\tau$. However, the residue integration
does not depend on the distance of the pole from the real axis and one realizes
that the right-hand side of eq.(\ref{residue}) is indeed a self-consistent solution
which yields
\begin{equation}
\label{selfconsistentborn}
G({\bf p},\epsilon )=
\frac{1}{\epsilon -\xi_{\bf p}+\frac{{\rm i}}{2\tau}~sign~(\epsilon)},
\end{equation}
from which emerges the meaning of $\tau$ as the {\sl elastic} quasiparticle relaxation
time. We stress that the consistency of the above approximation for evaluating integrals
over momentum is based on the fact that the distance from the real axis of the pole in
the Green's function remains small compared to the Fermi energy, which corresponds
to the condition
\begin{equation}
\label{expansioncondition}
E_F\tau \gg 1
\end{equation}
and one finds the {\sl effective} expansion parameter discussed previously. 
The {\sl self-consistent} Born approximation effectively selects a subset
of diagrams characterized by the absence of {\sl crossing} of impurity average
lines. This sequence of independent scattering events 
leads to a ladder resummation of diagrams for the vertex part,
as it will be shown in the next subsection.
\subsection{Vertex part and diffuson ladder}
\label{subsectionvertex}
\begin{figure}
\includegraphics{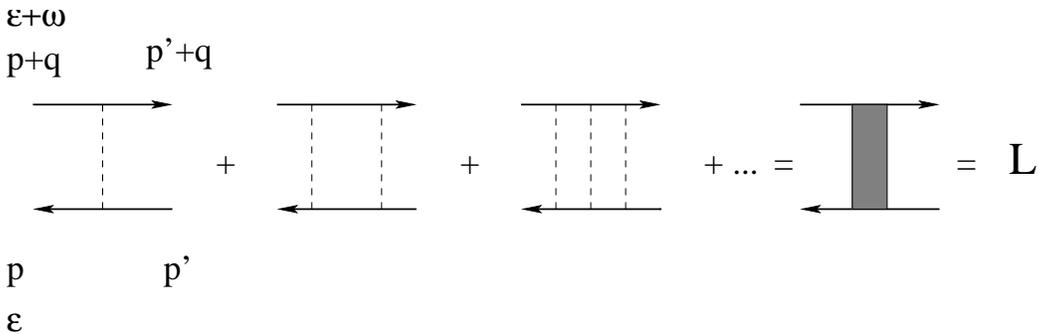}     
\caption{Ladder resummation.}
\label{ladder}
\end{figure}

This subsection has a twofold aim. On the one hand we show that the microscopic
approach, at leading order in the parameter $1/E_F\tau $, reproduces the semiclassical
theory of Drude-Boltzmann. On the other hand, we introduce a number of technical ingredients
that will be used extensively  in these lectures.
In particular, we will perform the evaluation of both the density-density and current-current
correlation function. 
\begin{figure}
\includegraphics{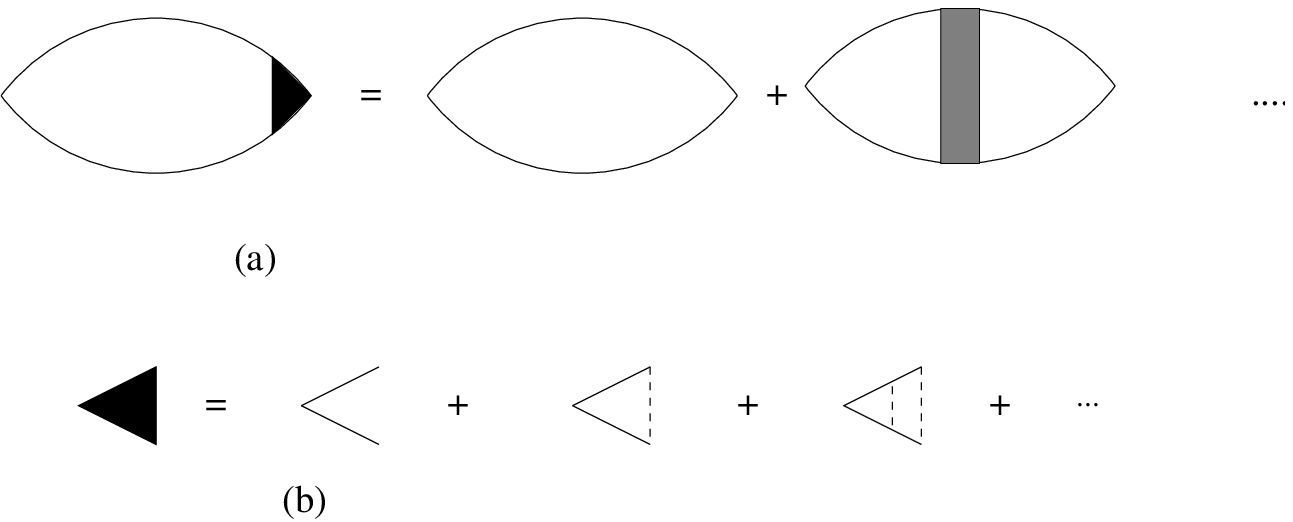}     
\caption{(a) Diagram for the correlation function. The black triangle indicates
the vertex part. (b) The vertex part is obtained as an infinite resummation of
non crossing impurity lines.}
\label{vertex}
\end{figure}
We begin with the density-density response function. The expression to evaluate reads
\begin{equation}
\label{k00}
K^{00}({\bf q},\omega )
=-2{\rm i} e^2 \int^{\infty}_{-\infty}~\frac{{\rm d}\epsilon}{2\pi}~\sum_{{\bf p}}~
G\left({\bf p}_+,\epsilon_+  \right) \Gamma^0 ({\bf p},\epsilon ;{\bf q},\omega )
G\left({\bf p}_-,\epsilon_-  \right)
\end{equation}
where $ {\bf p}_{\pm}\equiv {\bf p}\pm {\bf q}/2$ and $ \epsilon_{\pm}\equiv \epsilon\pm \omega/2$
and we have introduced a factor of two due to the spin.
The truncated vertex $\Gamma^0$ is indicated by a black triangle in fig.\ref{vertex}.
The Green's functions appearing in the diagrams are those evaluated within the self-consistent
Born approximation,  as explained in the previous subsection. The evaluation of the vertex
requires the evaluation of the series of ladder diagrams shown in fig.\ref{ladder}.
The series of ladder diagrams, to be called ladder from now on, may be evaluated
by solving the integral equation
\begin{equation}
\label{ladderintegral}
L_{{\bf p},{\bf p}',\epsilon}({\bf q},\omega )=L^{(0)}+
L^{(0)}\sum_{{\bf p}''}
G\left({\bf p''}_+,\epsilon_+  \right)
G\left({\bf p''}_-,\epsilon_-  \right)
L_{{\bf p}'',{\bf p}',\epsilon}({\bf q},\omega ),
\end{equation}
where $L^{(0)}=\frac{1}{2\pi N_0\tau}$.
The ladder generally depends on three momenta and two energies. However, as it will be
shown in a few moments, the actual dependence is only on ${\bf q}$ and $\omega$.
This is the reason of the notation adopted. Later on we will drop the subscripts
${\bf p}$,${\bf p'}$ and $\epsilon$. 
The kernel of the integral equation (\ref{ladderintegral}) gives a non-vanishing contribution
when the poles of the two Green's functions lie on opposite sides of the real axis.
To see this, let us consider the term with two impurity lines in the ladder series.
It reads
\begin{equation}
\label{fundamentalintegral}
\Pi=
\sum_{{\bf p}}
G\left({\bf p}_+,\epsilon_+  \right)
G\left({\bf p}_-,\epsilon_-  \right).
\end{equation}
To evaluate the integral we go to the energy variable introduced in eq.(\ref{standard}).
We have $\xi_{{\bf p}\pm {\bf q}/2}=\xi_{\bf p}\pm {\bf p}\cdot {\bf q}/2m+ q^2/8m$.
Since the integral is dominated by the contribution coming from the poles we set to
$p_F$ the absolute value of ${\bf p}$ in the scalar product with ${\bf q}$. The integral
over the energy $\xi_{\bf p}$ may then be carried out by residue method as explained
in the previous subsection. Also we notice that the condition of having poles on opposite
sides of the real axis implies a restriction on the frequencies, i.e.,
$\epsilon+\omega /2 >0$ and $\epsilon -\omega /2 <0$. As a result we get
\begin{equation}
\label{fundamental2}
\Pi={\theta (\omega^2/4-\epsilon^2)}{2\pi N_0\tau}\int~\frac{{\rm d}\Omega_{\hat {\bf p}}}{\Omega_{\hat {\bf p}}}~
\frac{1}{1-{\rm i}\omega \tau +{\rm i}l q\cos (\theta )},
\end{equation}
where $\Omega_{\hat {\bf p}}$ is the solid angle and $\theta$ is the angle between ${\bf p}$ and 
${\bf q}$. $l=v_F\tau$ represents the mean free path due to {\sl elastic} scattering.
Although the angular integral may be evaluated exactly, in the following we will be
interested in the limits $\omega \tau \ll 1$ and $l q \ll 1$, which define the
{\sl diffusive}
\footnote{This expansion is sufficient in the low temperature regime
when $T\tau \ll 1$. At higher temperature with $T\tau \ge 1$ one must retain the
full frequency and momentum dependence of $\Pi$. This defines the
quasi-ballistic regime. A detailed discussion can be found in ref.\cite{zala2001}.}
 transport regime. It is then convenient to expand for small frequency and
momentum the right-hand side of eq.(\ref{fundamental2}) and perform the angular integral.
We get finally
\begin{equation}
\label{fundamental3}
\Pi={\theta (\omega^2/4-\epsilon^2)}{2\pi N_0\tau}
\left(1+{\rm i}\omega\tau -D q^2\tau \right),
\end{equation}
where the diffusion coefficient is given by $D=v_Fl/d$, $d$ being the dimensionality.
As anticipated, the ladder does not depend on the momenta ${\bf p}$ and ${\bf p}'$.
It depends on the difference of the incoming and outgoing momenta only and
the integral equation becomes an algebraic one. The final solution reads
\begin{eqnarray}
\label{ladderfinal}
L( q,\omega )&=&\frac{L^{(0)}}{1- L^{(0)}\Pi}\nonumber\\
&=&\frac{1}{2\pi N_0 \tau^2}
\frac{\theta (\omega^2/4-\epsilon^2)}{-{\rm i}\omega +D q^2}.
\end{eqnarray}
This is the most important equation of this subsection. It shows how the diffusive pole
one expects emerges from the repeated elastic scattering. 
In terms of the ladder the vertex reads
\begin{eqnarray}
\label{vertexladder}
\Gamma^0 ({\bf p},\epsilon ,{\bf q},\omega )&\equiv&\gamma_0+\delta\Gamma^0\nonumber\\
&=&e\left(1+\sum_{\bf p'}
G\left({\bf p'}_+,\epsilon_+  \right)
G\left({\bf p'}_-,\epsilon_-  \right)
L_{{\bf p}',{\bf p},\epsilon}({\bf q},\omega )\right)\nonumber\\
&=&e( 1+2\pi N_0 \tau L( q,\omega ))
=e\left(1+\frac{\theta (\omega^2/4-\epsilon^2)}{-{\rm i}\omega\tau +D\tau q^2} \right).
\end{eqnarray}
One notices that in the zero-momentum limit the expression for $\Gamma^0$ may
be obtained directly from the Ward Identity (\ref{wardidentitygamma}) 
and agrees with the above equation.  With the vertex we may now complete the
evaluation of the density-density response function. It is natural to split it in two
parts corresponding to the two contributions in the vertex. 
 The contribution, which does not contain the diffusive pole and does not have
 restrictions on the frequency range, reads
\begin{eqnarray}
\label{staticlimit}
K^{00}_{++}+K^{00}_{--}&=&
-2{\rm i} e^2 \int^{\infty}_{-\infty}~\frac{{\rm d}\epsilon}{2\pi}~\sum_{{\bf p}}~
G\left({\bf p}_+,\epsilon_+  \right) ~
G\left({\bf p}_-,\epsilon_-  \right) \nonumber\\
&\approx&-2{\rm i} e^2 \int^{\infty}_{-\infty}~\frac{{\rm d}\epsilon}{2\pi}~\sum_{{\bf p}}~
G\left({\bf p},\epsilon \right)^2\nonumber\\
&=&-2{\rm i} e^2\frac{1}{2\pi} ~\sum_{{\bf p}}~\left[G^R\left({\bf p},0 \right)-
G^A\left({\bf p},0 \right) \right]\nonumber\\
&=&-2e^2N_0,
\end{eqnarray}
where the superscripts $R(A)$ indicate the retarded and advanced Green's functions
\begin{equation}
\label{retardedadvanced}
G^{R(A)}({\bf p},\epsilon )=
\frac{1}{\epsilon -\xi_{\bf p}\pm\frac{{\rm i}}{2\tau}}.
\end{equation}
In obtaining the above result, we have taken the {\sl static} limit (frequency goes to zero first,
and then momentum). 
Given the restriction in the frequencies for the second part in the vertex,
one could naively think that in the small frequency and momentum limit, 
this contribution  would be vanishingly small. This  however depends
on the order the two limits are performed. Due to the presence of the diffusive pole
in the ladder, one obtains for the second {\sl dynamic} contribution
\begin{eqnarray}
\label{dynamiclimit}
K^{00}_{+-}&=&  -2{\rm i} e^2 \int^{\omega /2}_{-\omega /2}~\frac{{\rm d}\epsilon}{2\pi}~\sum_{{\bf p}}~
G\left({\bf p}_+,\epsilon_+  \right) ~
G\left({\bf p}_-,\epsilon_-  \right)
\frac{1}{-{\rm i}\omega\tau +D\tau q^2}\nonumber\\
&\approx&-2 e^2 N_0 \frac{{\rm i}\omega}{-{\rm i}\omega +D q^2},
\end{eqnarray}
where due to the diffusive pole one may set  ${\bf q=0}$ and $\omega =0$ in the Green's functions
and perform the integral with residue methods. By combining together
eqs.(\ref{staticlimit}),(\ref{dynamiclimit}) one obtains the total density-density response
function at leading order in $E_F\tau$, in complete agreement with the result of 
eq.(\ref{densitydensity}) based on the phenomenological expression of the current.
We also note that, as expected, both the compressibility and the single-particle density
of states are given by the Fermi gas expression $N_0$.

\begin{figure}
\includegraphics{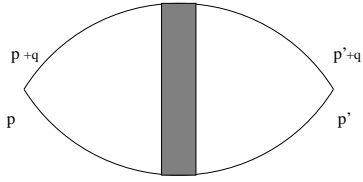}     
\caption{Diagram contributing to the leading correction
to the current-current response function in the dynamical region.
In the ladder flows the momentum ${\bf q}$.}
\label{conductivity_ladder}
\end{figure}

We now turn to the evaluation of the current-current response function. For the purpose
of computing the electrical conductivity, this is not strictly necessary since
the Ward identity (\ref{kubo2}) allows us to  get $\sigma$ from the density-density correlation 
function(\ref{densitydensity}).
We believe however that it is instructive to show how the calculation goes. 
The expression reads
\begin{equation}
\label{kij}
R^{ij}({\bf q},\omega )
=-2{\rm i}  \int^{\infty}_{-\infty}~\frac{{\rm d}\epsilon}{2\pi}~\sum_{{\bf p}}~
\gamma^i_{\bf p} ~G\left({\bf p}_+,\epsilon_+  \right)~ \Gamma^j ({\bf p},\epsilon ;{\bf q},\omega )
G\left({\bf p}_-,\epsilon_-  \right).
\end{equation}
The first observation concerns the vectorial nature of the vertex. By going through
the same steps as for the density-density response function, the integral 
 contains a vertex $\gamma^i$ which makes the integral  vanish upon
angular integration. As a result the current vertex $\Gamma^i$ remains unrenormalized.
This can be illustrated with the help of fig.\ref{conductivity_ladder}.
Since the region of small ${\bf q}$ gives the dominant contribution to the integral,
one can set ${\bf q}=0$ in the Green's functions.
 As a consequence the two ${\bf p}$ and ${\bf p'}$ integrations in fig.\ref{conductivity_ladder}
 decouple from one another and vanish for the presence of the vectorial vertex. 
 The diagram of fig.\ref{conductivity_ladder}, which is the diffusive polar contribution
 to the density-density response function, does not contribute to the current-current response
 functions.
However, the evaluation of the remaining part of the response function is more delicate
as compared to the case of the density-density response. The reason is due to the fact 
 that in performing
the small-frequency limit we need to divide by $\omega$ according to eq.(\ref{conductivitykubo}).
One then cannot simply take the zero-frequency limit by setting $\omega =0$ before performing
the integral, but it is necessary to make an expansion in powers of $\omega$.
After a few manipulations we get
\begin{eqnarray}
\label{currentstatic}
R^{ij}({\bf 0},\omega )&=&-2{\rm i}~\sum_{{\bf p}}~\gamma^i_{\bf p}\gamma^j_{\bf p}\left[ 
\int^{\infty}_{-\infty}~\frac{{\rm d}\epsilon}{2\pi}~G^2\left({\bf p},\epsilon 
\right)\right.\nonumber\\
&-&\left. \frac{\omega}{2}\frac{1}{2\pi}\left(
G^R\left({\bf p},0 \right)-G^A\left({\bf p},0  \right) \right)^2\right]\nonumber\\
&=&-\frac{e^2n_0}{m}-{\rm i}\omega \frac{e^2 n_0 \tau}{m}.
\end{eqnarray}
The first term,  by recalling eq.(\ref{separation}),  cancels exactly with the
diamagnetic contribution. The remaining term, by using eq.(\ref{conductivitykubo}), gives
the Drude formula for the electrical conductivity.

To summarize, in this subsection we have evaluated the linear response of a disordered 
Fermi gas to an electromagnetic external field to the leading order in the parameter
$1/(E_F\tau )$.  This parameter is also the natural dimensionless coupling  of the
present microscopic problem which deals with the Fermi gas (characterized by the unique
energy scale $E_F$) in the presence of disorder which introduces the (other) energy scale
$N_0 {\overline { u^2}}\sim \tau^{-1}$.
At this order, one recovers the results of the semiclassical approach
of Drude-Boltzmann. However, we have developed a formalism within which next-to-leading
corrections may be investigated systematically. This will be the subject of the
next subsections as far as non-interacting electrons are concerned. Interaction
effects will be considered in the next section. 
\subsection{Weak localization}
\label{wl}
\begin{figure}
\includegraphics{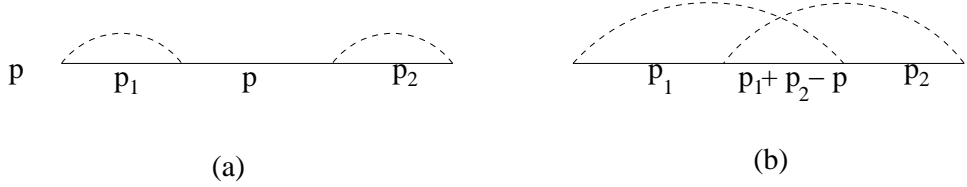}     
\caption{(a) A diagram without crossing. (b) A diagram with crossing.}
\label{crossing}
\end{figure}

\begin{figure}
\includegraphics{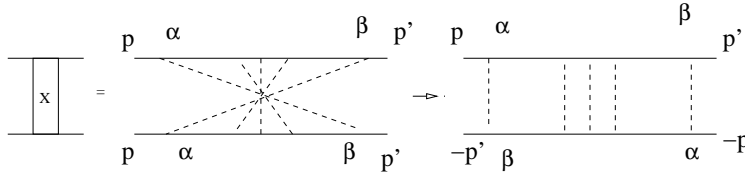}     
\caption{Crossed ladder expressed in term of the direct ladder. Notice that, upon the reversal of
the bottom Green's function line, the in and out combination of spin indices are $\alpha \beta$
and $\beta \alpha$, respectively. }
\label{cooperonladder}
\end{figure}

We have stated that the leading approximation in an expansion in the parameter
$1/(E_F\tau)$ is obtained by considering diagrams without crossing. We begin our
discussion of the next-to-leading corrections, by showing how crossing of
impurity lines increases the order of a diagram. A simple example is shown in
fig.(\ref{crossing}), where both diagrams are of the same order in the
impurity lines. Let us estimate these diagrams. By recalling the self-energy
expression (\ref{residue}) and the Green's function (\ref{selfconsistentborn}),
the diagram (a) reads
\begin{eqnarray}
\label{nocrossing}
(a)&=&\left( \frac{{\rm i}sign \epsilon}{2\tau}\right)^2 G({\bf p},\epsilon) \nonumber\\
&\approx_{\epsilon, \xi_{\bf p}\rightarrow 0}&\frac{1}{\tau^2}\tau =\frac{1}{\tau}\nonumber
\end{eqnarray}
while the diagram (b) yields
\begin{eqnarray}
\label{withcrossing}
(b)&=&\left(\frac{1}{2\pi N_0 \tau }\right)^2 
\sum_{{\bf p}_1,{\bf p}_2} G({\bf p}_1,\epsilon) G({\bf p}_2,\epsilon)
 G({\bf p}_1+{\bf p}_2-{\bf p},\epsilon) \nonumber\\
 &=&\frac{1}{(2\pi N_0 \tau)^2}\sum_{{\bf p}_1,{\bf p}_2}
 \frac{1}{(\epsilon -\xi_1+{\rm i}\delta_{\epsilon})(\epsilon -\xi_2+{\rm i}\delta_{\epsilon})
 (\epsilon -\xi_1-\xi_2-\xi+2\mu -\frac{{\bf p}_1\cdot {\bf p}_2}{m}+\frac{{\bf p}\cdot
 ({\bf p}_1+ {\bf p}_2)}{m}
 +{\rm i}\delta_{\epsilon})}\nonumber\\
 &\approx &\frac{1}{\tau^2 \mu},\nonumber
\end{eqnarray}
where for brevity $\delta_{\epsilon}=sign(\epsilon )/2\tau$ and $\xi_1\equiv \xi_{{\bf p}_1}$
and similarly for ${\bf p }_2$ and ${\bf p}$.  The above result is obtained  by
noting that, due to the poles of the first two Green functions, we can set $\xi_1=\xi_2=0$
in the third one. Clearly (b) is smaller of (a) by a factor $1/\mu\tau$.
\begin{figure}
\includegraphics{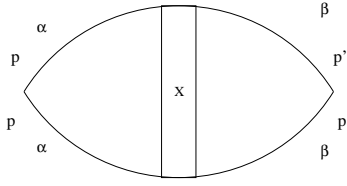}     
\caption{Diagram contributing to the next-to-leading correction
to the current-current response function. The Greek indices label the spin of the Green's
function line. Notice that at the density vertex the spin is conserved.}
\label{weaklocalization}
\end{figure}
In general to take diagrams with crossing of impurity lines becomes a very complicated
problem. There is, however, a subset of diagrams, the so-called maximally crossed diagrams,
which can be evaluated. The contribution of the series of these diagrams to the current-current
response function is shown in fig.\ref{weaklocalization}. The corresponding  expression
reads
\begin{equation}
\label{currentweaklocalization}
\delta K^{ij}({\bf 0},\omega )=-2{\rm i}~\sum_{{\bf p},{\bf p}'}~
\gamma_{\bf p}^i\gamma_{\bf p'}^j 
\int^{\infty}_{-\infty}~\frac{{\rm d}\epsilon}{2\pi}~
G\left({\bf p},\epsilon_+  \right)~ 
G\left({\bf p},\epsilon_-  \right)~L_c({\bf p},{\bf p'},\omega)
G\left({\bf p'},\epsilon_+  \right)~ 
G\left({\bf p'},\epsilon_-  \right),
\end{equation}
where with  $L_c({\bf p},{\bf p'},\omega)$ we have indicated
the series of the maximally crossed diagrams. As for the 
 leading order calculation, the first step requires the evaluation
of the series giving $L_c({\bf p},{\bf p'},\omega)$.
This can be done by observing that the series of the maximally crossed diagrams,
from now on called crossed ladder, $L_c$,  may be
expressed in terms of the direct ladder by reversing one of the electron
Green's function lines, as shown in fig.\ref{cooperonladder}.
This corresponds to
\begin{equation}
\label{cooperon}
L_c({\bf p},{\bf p'},\omega) =L({\bf p}+{\bf p'},\omega)-\frac{1}{2\pi N_0\tau}.
\end{equation}
Hence, the crossed ladder has a diffusive form with respect to the combination
 ${\bf p}+{\bf p'}$  since one of the two Green's function lines has been reversed.
   This means that the dominant contribution comes
 from the region ${\bf p}\sim -{\bf p'}$ and the integration over the two vector vertices
 is no longer decoupled as in the leading order case with the direct ladder.
 The direct ladder studied in the previous subsection
describes the propagation of a particle-hole pair. The crossed ladder, when
one of the lines is reversed, describes the propagation of a particle-particle pair
and the diffusive pole is with respect to the total incoming momentum of the
pair. This is reminiscent of the interaction scattering channel relevant
for  superconductivity and for this reason the crossed ladder
is called the cooperon ladder. The direct ladder, on the other hand,
is, in the technical jargon, called the diffuson. A key point to keep in mind
about the difference between the diffuson and the cooperon is the total
electric charge of the pair. While for the diffuson this is zero, it is
twice the electron charge for the cooperon. As a consequence, in the diffusive regime,
the cooperon is affected by the presence of a magnetic field, while the diffuson is not
touched. We will see how a magnetic field affects the cooperon in the next subsection.
We are now ready to complete our derivation of the correction to the current response.
By performing the integrals (see Appendix A) in the standard way, eq.(\ref{currentweaklocalization})
 becomes
\begin{eqnarray}
\label{weakloc2}
\delta K^{ij}&=&
-2{\rm i} 
\int^{\omega /2}_{-\omega /2}~\frac{{\rm d}\epsilon}{2\pi}~
~\sum_{{\bf p},{\bf p}'}~ \gamma_{\bf p}^i\gamma_{\bf p'}^j 
G^R\left({\bf p},\epsilon  \right)~ 
G^A\left({\bf p},\epsilon  \right)
G^R\left({\bf p'},\epsilon  \right) 
G^A\left({\bf p'},\epsilon  \right)\nonumber\\
&&
\frac{1}{2\pi N_0 \tau^2}\frac{1}{-{\rm i}\omega +D ({\bf p}+{\bf p'})^2}\nonumber\\
&=&-2{\rm i}~\int^{\omega /2}_{-\omega /2}~\frac{{\rm d}\epsilon}{2\pi}~\sum_{{\bf p}}\sum_{\bf Q}
~\gamma_{\bf p}^i\gamma_{\bf Q -p}^j 
G^R\left({\bf p},\epsilon  \right)~ 
G^A\left({\bf p},\epsilon  \right)
G^R\left({\bf Q-p},\epsilon  \right) 
G^A\left({\bf Q-p},\epsilon  \right)\nonumber\\
&&
\frac{1}{2\pi N_0 \tau^2}\frac{1}{-{\rm i}\omega +D {\bf Q}^2}\nonumber\\
&=&d{\rm i}~\omega~\frac{e^2}{\pi}\sum_{\bf Q}
\frac{1}{-{\rm i}\omega /D + {\bf Q}^2}\nonumber
\end{eqnarray}
from which and from eq.(\ref{conductivitykubo}), at $d=2$, we get the correction to the
conductivity
\begin{equation}
\label{weakloc3}
\delta \sigma =-\frac{e^2}{\pi^2\hbar}\ln \left(\frac{L}{l} \right)
=-\sigma_0 2t \ln \left(\frac{L}{l}\right),
\end{equation}
where we have resumed the physical units.
The above equation represents the {\sl weak-localization}
correction\cite{gorkov1979} and $t=(4\pi^2N_0D\hbar )^{-1}=(2\pi E_F \tau / \hbar )^{-1}$ 
is the effective
small expansion parameter in the metallic regime. To make connection with the phenomenological
scaling theory, we note that the parameter $t$ coincides with the inverse conductance in units of $G_0$ divided by
$2\pi$, $t=1/(2\pi g)$.
The logarithmic divergence has been cutoff at large ${\bf Q}$ by the inverse
mean free path, {\sl i.e.} the distance beyond which diffusive motion starts
to set in. The small ${\bf Q}$ cutoff is instead provided by the inverse of
some length scale, $L$. In the zero frequency and zero temperature limit, this 
scale is given by the system size. At finite temperature, inelastic processes
provide a so called dephasing length $L_{\phi}$. 

As discussed in the first section, the physical origin of the weak-localization
correction is due to quantum interference. We are now in a position to appreciate
the exact meaning of this statement. In a Feynman diagram a Green's function line
describes the amplitude for going from one point to another. In the response
functions, the two Green's function lines represent the operation of taking the
product of one amplitude with its complex conjugate. Each amplitude  is 
a sum over all possible paths so that in the product there will appear interference
terms. The leading approximation, by restricting to ladder diagrams, makes
an {\sl effective} selection of paths. 
When we average over the impurity configurations, by connecting two impurity insertions by
a dashed line, this corresponds to the fact that both the upper
and lower Green's function lines are going through the same scattering center,
{\sl i.e.},  one is considering the product of the amplitude of a given path with 
its complex conjugate. Hence, in the leading approximation there is no
interference.  When considering, on the other hand, the maximally crossed diagrams, one observes
that the upper Green's function line goes through a sequence of scattering events
which is exactly the opposite of the one followed by the lower line.
This represents the interference between trajectories that are one the time reversed of the
other. One also notices that these trajectories are made by closed loops and  always
come in pairs, due to the fact the loop may be gone around clock- or anticlockwise.

\begin{figure}
\includegraphics{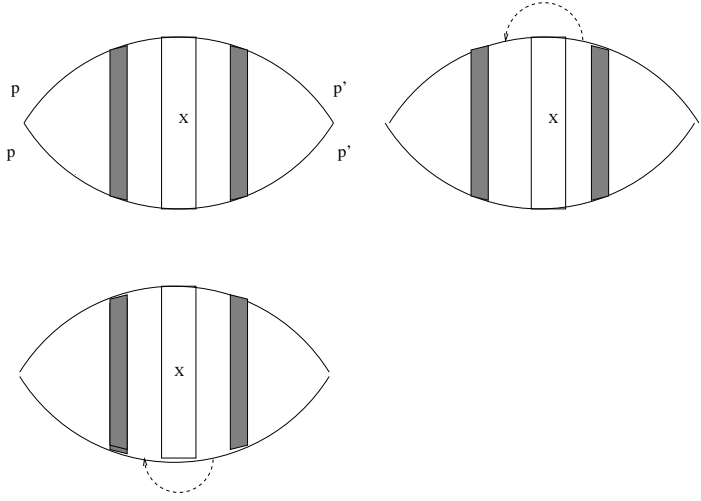}     
\caption{Diagram contributing to the next-to-leading correction
to the density-density response function.}
\label{densitydiagram}
\end{figure}

So far we have established that the electrical conductivity acquires a logarithmic
correction at order $1/E_F\tau$. The correction has a negative sign and signals
a slowing-down of the electron diffusion. It is then legitimate to ask what this
implies for the full momentum and frequency dependence of the density response function.
To this end, we now consider the corresponding corrections with the bare density vertex $\gamma_0 =e$.
In the presence of  the scalar vertex, as we have seen in subsec. 4.2 already for the Drude 
approximation, the direct ladder contributes to the density response function
to obtain the diffusive form (\ref{densitydensity}). At the order $t$ we are now considering,
many more diagrams have to be taken into account. Most of them cancel each other
\cite{castellani1983} and we are left with those shown 
 in fig.\ref{densitydiagram}. The  expression for the first diagram reads
\begin{equation}
\label{densitydensitycorrection}
\delta K^{00}_{+-}=-\frac{2{\rm i}~\omega }{2\pi}
~\sum_{{\bf p},{\bf p}'}~\delta\Gamma^0 ({\bf q},\omega )
G^R\left({\bf p}_+ \right)~ 
G^A\left({\bf p}_-  \right)~
L_c({\bf p},{\bf p'},\omega)
G^R\left({\bf p'}_+  \right) 
G^A\left({\bf p'}_-  \right)
\delta\Gamma^0 ({\bf q},\omega ).
\end{equation}
In eq.(\ref{densitydensitycorrection}) we have already performed the sum over the frequency
$\epsilon$, which gives rise to the factor $\omega$ in front. In the Green's functions,
consistently with the approximations used up to now, we have set the frequency to zero and
only the momentum argument is explicitly shown. The two vertex corrections $\delta\Gamma^0$ at the extreme
left and extreme right describes the direct ladders appearing in the diagram and are given by eq.(\ref{vertexladder}). 
 To evaluate the integrals over the momenta ${\bf p}$ and ${\bf p'}$,
we note that the cooperon ladder gives a big contribution to the integral when 
${\bf p}+{\bf p'}$ is small. It is then convenient to introduce the variable
${\bf Q}={\bf p}+{\bf p'}$ and set ${\bf Q}=0$ everywhere except that in the cooperon
ladder. We are then left with an integral  over the cooperon ladder and
a second over a product of four Green's functions. However, this is not yet the
full story. It turns out that there exist further diagrams (the second and the third)
that may be obtained by simply
{\sl decorating} the first diagram of fig.\ref{densitydiagram} with an extra impurity line.
Such a decoration only adds two Green's functions and an extra summation over a fast momentum.
To proceed we have then to integrate the Green's functions in the diagrams of
fig.\ref{densitydiagram} according to the expression
\begin{eqnarray}
\label{hikamibox}
I_H&=&\sum_{\bf p}G^R({\bf p}_+)G^A(-{\bf p}_+)G^R(-{\bf p}_-)G^A({\bf p}_-)\nonumber\\
&+&\frac{1}{2\pi N_0\tau}\sum_{\bf p}G^R({\bf p}_+)G^R(-{\bf p}_-) G^A({\bf p}_-)  \sum_{\bf p'}
G^R({\bf p'}_+) G^A(-{\bf p'}_+) G^R(-{\bf p'}_-)\nonumber\\
&+&\frac{1}{2\pi N_0\tau}\sum_{\bf p}G^R({\bf p}_+)G^A(-{\bf p}_+)G^A({\bf p}_-)
\sum_{\bf p'}G^A(-{\bf p'}_+)G^A({\bf p'}_-)G^R(-{\bf p'}_-).
\end{eqnarray}
Notice the factor $\frac{1}{2\pi N_0\tau}$ for the extra impurity line.
Since we are interested in the small-momentum limit, we make an expansion in powers
of ${\bf q}$. After a straightforward, but lengthy calculation, one gets
\begin{equation}
\label{hikamiresult}
I_H=4\pi \tau^4N_0 D {\bf q}^2.
\end{equation}
Finally, from eq.(\ref{vertexladder}) and the expression (\ref{cooperon}) for the crossed ladder, we get
as a correction to first order in $t$ to eq.(\ref{dynamiclimit}) for $K^{00}_{+-}$
\begin{equation}
\label{densitydensitycorrection2}
\delta K^{00}_{+-}=-\frac{2{\rm i}~\omega  D {\bf q}^2}{(D{\bf q}^2-{\rm i}\omega )^2}
\frac{e^2}{\pi}\sum_{\bf Q}\frac{1}{D{\bf Q}^2-{\rm i}\omega}\equiv
+\frac{2e^2{\rm i}~\omega N_0  {\bf q}^2}{(D{\bf q}^2-{\rm i}\omega )^2}\delta D,
\end{equation}
where 
$$\delta D=\frac{\delta \sigma}{2e^2N_0}=-\frac{D}{\pi N_0}\sum_{\bf Q}\frac{1}{D{\bf Q}^2-{\rm i}\omega}
$$
agrees with the expression for $\delta\sigma$ found in eq.(\ref{weakloc3}) derived from the current-current
response function.
The above equation shows that the correction to $K^{00}$, made of many different contributions,
 can at the end be absorbed into
a renormalization of the diffusion coefficient $D_R= D+\delta D$. Hence we conclude
that at this order eq.(\ref{dynamiclimit}) changes into
\begin{equation}
\label{renormalizedk00}
K^{00}_{+-}=-2 e^2 N_0 \frac{{\rm i}\omega}{-{\rm i}\omega +D_R {\bf q}^2}
\end{equation}
and by virtue of eq.(\ref{compressibility}) and eq.(\ref{densityofstates}) 
the compressibility and the density of states are not renormalized.
Only one renormalization ($D\rightarrow D_R$) is required in this case.
 The one-parameter scaling theory follows. Given the expression (\ref{weakloc3}) for
 $\delta\sigma$, the group equation for the conductance $g$ has an expansion in $t$ of its $\beta$-function
 (eq.(\ref{betametallic})) with the coefficient $a=1/\pi$. The critical index for the conductivity, at order
 $\epsilon$ in $d=3$, is $\mu =1$. The frequency cuts off  the singularity in the diffusion ladder and acts
 in this transition as an external field in ordinary transitions. Its scaling index $x_{\omega}$ has the same value
 as the dimension of the $D_Rq^2$, i.e., $x_{\omega}=\epsilon +2 =d$. 


\subsection{Effect of a magnetic field}
\label{bfield}
\begin{figure}
\includegraphics{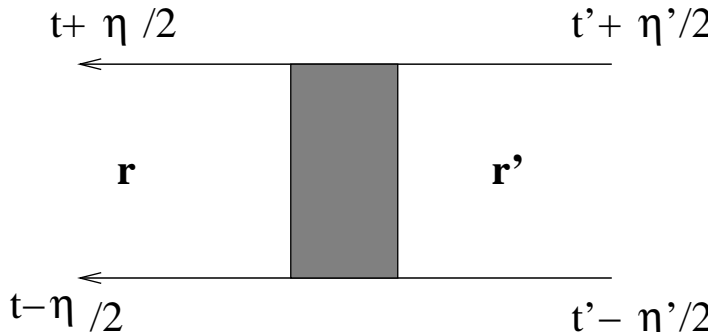}     
\caption{The cooperon ladder in time and space representation. Notice that both particles 
in the incoming (outgoing) pair have the same (center-of-mass) coordinate ${\bf r'}$ (${\bf r}$).
The lack of explicit dependence on the relative coordinate corresponds to the fact the ladder
resummation does not depend on the fast momentum.}
\label{cooperon_times}
\end{figure}

As we have pointed out in sec.2.3, the magnetic field will cut off the weak-localization corrections.
To see this explictly, it is useful to switch to a space and time representation
of the cooperon ladder, as shown in fig.\ref{cooperon_times}.
The cooperon describes the propagation of a pair of electrons that
have coiciding space coordinates (within the spatial resolution given by
the mean free path $l$). $t$, $t'$ and $\eta$, $\eta '$ are center-of-mass
and relative times of the electron pair. For instance, an incoming pair
has a temporal evolution factor
\begin{equation}
e^{-{\rm i}(\epsilon +\omega /2)(t'+\eta ' /2)}e^{-{\rm i}(\epsilon -\omega /2)
(t'-\eta ' /2)}=
e^{-2{\rm i} \epsilon t'}e^{-{\rm i}\omega \eta ' /2}.
\end{equation} 
In this representation the cooperon reads
\begin{equation}
L^{\eta ,\eta'}_c({\bf r},{\bf r}')=\frac{\theta (\eta -\eta ')}{2\pi N_0\tau^2}
\frac{e^{-|{\bf r}- {\bf r}'|^2/D(\eta -\eta ')}}{(2\pi D (\eta -\eta '))^{d/2}},
\end{equation}
and obeys the diffusion equation
\begin{equation}
\left( 2 \frac{\partial}{\partial\eta} -D \nabla^2_{\bf r} \right)
L^{ \eta ,\eta'}_c({\bf r},{\bf r}')=\frac{1}{2\pi N_0\tau^2}\delta (\eta -\eta ')
\delta ({\bf r}- {\bf r}').
\end{equation}
We note that the formula for the weak-localization correction of eq.(\ref{weakloc3}) involves the Fourier transform with respect to time of the cooperon at coinciding space points. 
In terms of $L^{\eta ,\eta'}_c({\bf r},{\bf r}')$, the weak-localization
correction eq.(\ref{weakloc3}), in $d=2$,  is recovered as
\begin{equation}
\label{wl_realspace}
\delta \sigma =-e^2 4 N_0 D\tau^2 \int_{\tau}^{\tau_{\phi}}
{\rm d}\eta L^{\eta ,-\eta}_c ({\bf r},{\bf r})=
-\frac{e^2}{\pi^2\hbar }\ln \frac{L_{\phi}}{l}.
\end{equation}
The space and time representation makes more transparent the physical origin
of the weak localization correction. The cooperon propagator
$L^{\eta ,-\eta}_c({\bf r},{\bf r})$ represents the propagation
of a pair of electrons going around the same closed trajectory in opposite
directions, or one electron going through the time reversed trajectory of the other
electron. The time nedeed to go around the loop is $2\eta$ and one has to
integrate over all possible values of $\eta$ between $\tau$ and $\tau_{\phi}$. The lower limit, $\tau$,
sets the time over which diffusive behavior develops. The upper limit, $\tau_{\phi}$,  sets the
time over which phase coherence between the two electrons going around the loop
is maintained.  For both times, we switch to the corresponding lengths via the
diffusion constant.

We are now ready to consider the effect of an external magnetic field.
It enters the diffusion equation via the minimal substitution as
\begin{equation}
\nabla \rightarrow \nabla -2{\rm i}e{\bf A}({\bf r}),
\end{equation}
which is to be expected for the minimal substitution of the two electrons
described by the cooperon. In fact, as the established name cooperon may
suggest, this is completely analogous to the minimal substitution
that it is usually made when considering the Landau-Ginzburg equations
for superconductivity.  In the absence of the external magnetic field,
the diffusion equation is solved via the knowledge of the eigenvalues
of the laplacian operator exactly as for the Schrodinger equation at imaginary
time. The analogy is made precise by saying that one may define particle-like
parameters as a mass $m^* =1/(2D)$ and a charge $e^* =2e$. 
A magnetic field will modify this mass and act as a cutoff for the singularity in $\sigma$.
In the presence
of a uniform external magnetic field, the diffusion equation may be solved
in terms of Landau levels. More precisely, in two dimensions for
a magnetic field perpendicular to the plane, the cooperon at coinciding space points reads
\begin{equation}
\label{cooperon_magneticfield}
L^{ \eta,\eta '}_c ({\bf r},{\bf r})=\frac{\theta (\eta -\eta ')}{2\pi N_0 \tau^2}
g_s\sum_{n=0}^{\infty}e^{-E_n (\eta -\eta ')/2}=
\frac{\theta (\eta -\eta ')}{2\pi N_0\tau^2}
g_s\frac{e^{-\omega_c (\eta -\eta ')/4}}{1-e^{-\omega_c (\eta -\eta ')/2}},
\end{equation}
where $g_s =e^*B/(2\pi)$, $E_n=\omega_c (n+1/2)$, and $\omega_c =2e^*DB$
are the degeneracy and energy of the effective Landau level, and the ciclotron
frequency, respectively. By inserting the eq.(\ref{cooperon_magneticfield})
into the expression for the weak-localization correction of
eq.(\ref{wl_realspace}) one gets
\begin{equation}
\label{magnetoconductance}
\delta \sigma (B)=-e^2D \frac{g_s}{\pi} \int_{\tau}^{\tau_{\phi}}
{\rm d}\eta ~\frac{1}{\sinh \left({\omega_c \eta/2 }\right)}=
-\frac{e^2}{2\pi^2\hbar}
\ln\left[ 
\frac{\tanh\left(\omega_c \tau_{\phi}/4\right)}{\tanh\left(\omega_c \tau /4\right)}
\right].
\end{equation}
In the metallic regime, where  $\tau \ll \tau_{\phi}$, at 
low magnetic field ($\omega_c \tau_{\phi} \ll 1 $), 
the above correction reads 
\begin{equation}
\delta \sigma (B)=-\frac{e^2}{2\pi^2\hbar }\left[\ln\left(\frac{\tau_{\phi}}{\tau }\right)-
 \frac{4}{3}\left(\frac{\Phi (B)}{\Phi_0}\right)^2\right],
\end{equation}
where $\Phi_0 =h c/2e$ and $\Phi (B)=\pi L_{\phi}^2B$ is the magnetic flux through
a circle with radius the dephasing length. 
The effect of the magnetic field is felt for fields
such that $\Phi (B)$ is of the order of the  flux quantum $\Phi_0$.
Then it suppresses the
weak localization correction and gives rise to a negative
magnetoresistance. 
Experimentally, by measuring the magnetoconductance one may obtain the value of the
dephasing time at a given temperature, $\tau_{\phi}(T)$. 

In the opposite limit,  such that $\omega_c\tau_{\phi}\gg 1$ (but $\omega_c\tau \ll 1$),
eq.(\ref{magnetoconductance}) reads
\begin{equation}
\label{strongfields}
\delta \sigma (B)=\frac{e^2}{2\pi^2\hbar}\ln\left(\omega_c\tau\right),
\end{equation}which is no longer singular.
The crossover between the two regimes occurs when the infrared cut off $\tau_{\phi}^{-1}$ is replaced
by $\omega_c$. This condition amounts to $\Phi (B)\sim \Phi_0$, as  was already stressed in
subsec.2.3.

As far as the metal-insulator transition
is concerned, the suppression of the weak localization correction corresponds to the vanishing
of the coefficient $a$ of the perturbative expansion of the $\beta$-function 
(cf. eq.(\ref{betametallic})). Then, corrections arise when a crossing of two  direct ladders
is considered in current-current response function and   appear at
the second order in the expansion in power of $t\sim 1/g$. This, as mentioned at the end
of subsection \ref{anderson},   implies a change of the critical conductivity exponent
from $\mu =1$ to $\mu =1/2$. This is contrast with several experiments, as discussed at the end
of this section.

\subsection{Spin effects}
\label{spins}

The weak-localization correction is also affected by the presence of magnetic
impurities. These latter make  the cooperon propagator massive. 
In the presence of spin-dependent scattering, it is useful to decompose the ladder
into singlet and triplet components with respect to the total spin for the incoming
pair. The latter is a  particle-hole and particle-particle pair for the diffuson and
 cooperon, respectively.  In Appendix \ref{magneticimpurities}, we give the details
 on how to derive the expression for the ladder in the presence of spin-flip scattering
 by magnetic impurities. 
The final result, in the limit when the spin-flip scattering time  is much
larger than the elastic scattering time, $\tau_s\gg\tau$, reads
\begin{eqnarray}
L^{S} &=&\frac{1}{2\pi N_0 \tau^2}\frac{1}{Dq^2-{\rm i}\omega},\label{spinsingletd}\\
L^{T}&=&\frac{1}{2\pi N_0 \tau^2}\frac{1}{Dq^2-{\rm i}\omega +\frac{4}{3\tau_s}},\label{spinstripletd}\\
L^{S}_{c}&=&\frac{1}{2\pi N_0 \tau^2}\frac{1}{Dq^2-{\rm i}\omega +\frac{2}{\tau_s}},\label{spinsingletc}\\
L^{T}_{c}&=&\frac{1}{2\pi N_0 \tau^2}\frac{1}{Dq^2-{\rm i}\omega +\frac{2}{3\tau_s}}\label{spintripletc}.
\end{eqnarray}
One sees that only the singlet channel of the diffuson remains massless.
While it is obvious that both triplet channels become massive, magnetic impurities as well as magnetic
field break the time reversal symmetry and introduce a mass in the cooperon singlet as well.
The  $a$-coefficient of the $\beta$-function expansion is again vanishing
and  the singlet channel diffuson gives rise to corrections to second order in
$t\sim 1/g$, yielding the value $\mu =1/2$. Comparison with available experiments is
postponed at the end of the section.

Spin-orbit scattering has the global effect of reversing the sign of the quantum
interference contribution to the conductivity. In Appendix \ref{spinorbitscattering} we derive for
both the diffuson and cooperon ladders the expression
\begin{eqnarray}
L^{S} &=&\frac{1}{2\pi N_0 \tau^2}\frac{1}{Dq^2-{\rm i}\omega},\label{orbitsingletd}\\
L^{T}&=&\frac{1}{2\pi N_0 \tau^2}\frac{1}{Dq^2-{\rm i}\omega +\frac{d+1}{d\tau_{so}}},\label{orbittripletd}\\
L^{S}_{c}&=&\frac{1}{2\pi N_0 \tau^2}\frac{1}{Dq^2-{\rm i}\omega }\label{orbitsingletc},\\
L^{T}_{c}&=&\frac{1}{2\pi N_0 \tau^2}\frac{1}{Dq^2-{\rm i}\omega
+\frac{d+1}{d\tau_{so}}}\label{orbittripletc}.
\end{eqnarray}
We see that in this case the cooperon singlet remains massless in contrast to the magnetic impurities case.
In the non magnetic impurity case the weak localization correction to the conductivity comes
from both the cooperon singlet and triplet channels. As  is shown in Appendix C, the singlet contribution
is antilocalizing and amounts to $1/3$ of the localizing triplet contribution. The latter being now suppressed,
we are left with the singlet antilocalizing interference contribution. 
  In terms of the $\beta$-function expansion, this means that the coefficient
$a$ changes sign. 
Also in this case, the comparison with the experiments is postponed to the next subsection.

In all the cases discussed above, no corrections arise to the thermodynamic quantities like the single-particle
density of states, the specific heat, and the spin susceptibility.
\subsection{A review of the experimental situation}
\label{experiments}
It is now time to compare the theoretical predictions obtained from the microscopic approach
with the experiments. As we will see, there are a number of facts that suggest that
a proper description of the metal-insulator transition cannot be obtained without
taking into account the effects of the electron-electron interaction.
1) In semiconductor-metal alloys, as we have already  pointed out,
the metal-insulator transition is observed with a conductivity critical exponent $\mu =1$.
This is in agreement with the non-interacting scaling theory. On the other hand, tunneling
measurements reveal that  the density of states has strong anomalies. 
For instance, in \cite{mcmillan1981}, the single particle density of states of $Ge_{1-x}Au_x$
is measured for different values of the $Au$ concentration above and below the critical
concentration for the metal-insulator transition. On the metallic side, the density of states
shows a dip at the Fermi energy, which in the insulating regime, at lower $Au$ concentrations,
develops in a gap. In the case of the $Nb_xSi_{1-x}$ alloy, the value of the density of states,
as obtained from tunneling measurements, is plotted as a function of the $Nb$ concentration, and it
is seen scaling to zero linearly by approaching the critical concentration
for the metal-insulator transition\cite{hertel1983}.
2) The presence of magnetic impurities, in the non-interacting theory,
should lead to a value $\mu =1/2$. However, in metal films of $Cu:Mn$, one finds
experimentally $\mu =1$\cite{okuma1988}.
3) In the amorphous alloy $Si Au$, it has been observed\cite{nishida1984}  $\mu =1$ both in the
absence and presence of a magnetic field of $5$ Tesla. 
This systems has a strong spin-orbit coupling which, within the single-particle scheme,
should switch from an antilocalizing term in the absence of the magnetic field, to a transition
with $\mu =1/2$ in the presence of the field.
4) Similarly,  a value $\mu =1$, both with and without a magnetic field, has also been  observed
in $Al_{0.3}Ga_{0.7}As:Si$ with a fine tuning of the electron concentration close
to the metal-insulator transition, by using the photoconductivity effect\cite{katsumoto1987,katsumoto1988}.
Also in this system, it is estimated that spin-orbit scattering is relevant. 
5) The experiments in uncompensated doped semiconductors are even more puzzling. 
First, as we mentioned in the introduction, there is the problem of the experimental
determination of the value of the critical conductivity exponent in 
$Si:P$, whose value $\mu =1/2$\cite{rosenbaum1980,paalanen1982,rosenbaum1983} has been
questioned\cite{stupp1993}.  In another $n-type$ system, e.g., $Si:As$ 
\cite{shafarman1989}, also a value $\mu =1/2$  has been observed, whereas 
a  close value of  $\mu =0.65$ has been reported for a  $p$-type
system as $Si:B$\cite{dai1991}, where the spin-orbit scattering is expected
to be strong. The situation is further complicated by the experimental observation
that the introduction of a magnetic field
changes the value of the conductivity exponent to $\mu =1$\cite{dai1993}
as for the alloys. Besides the interpretation of the issues raised by the transport
measurements, the experiments also show that there
is a strong enhancement at low temperature of the electronic
specific heat\cite{thomas1981} and the spin susceptibility\cite{paalanen1986}.
6) Finally, we want to comment on the problem of the metal-insulator transition
in the two-dimensional electron gas.
Although there is quite a rich experimental literature on this phenomenon, 
very little is really understood. First, the real occurrence of a zero-temperature
metal-insulator transition has not gathered the general consensus. In any case,
from the point of view of the present discussion about the relevance of
the electron-electron interaction in disordered systems, $Si$-MOSFET devices
and semiconducting heterostructures are even a stronger case. In fact, if there is
a metal-insulator transition, this is clearly beyond the conventional non-interacting
scaling theory for which all states are localized in two dimensions for any value
of the disorder. Secondly,
 the Coulomb interaction in these systems is expected
to be very strong. 

By considering  the effective mass of $Si$, $m=0.19m_0$ ($m_0$ being the bare electron mass),
and taking a  value for $\varepsilon \sim 11.9$, at typical electron densities $n\sim 10^{11} cm^{-2}$
from eq.(\ref{rs}) one gets $r_s\sim 2-3$, which has also led to suggest that  Wigner cristallization may play a role
\cite{attaccalite2002}.  
A last point to make about two dimensional systems is their strong parallel magnetic field
magnetoresistance. When the applied field gives a Zeeman energy $g\mu_B B\sim E_F$, the
resistivity increases by more than an order of magnitude\cite{simonian1997}. 
   
The interplay between disorder and electron-electron interaction will account for most of the
questions arisen above.
   
\section{Interacting Disordered Electrons}
In this section, we consider the interplay of disorder and interaction\cite{altshuler1979,altshuler1980,altshuler1980b}. 
In addition to the Hamiltonian (\ref{noninteracting}) we have now to consider the interaction term
\begin{eqnarray}
\label{interactinghamiltonian}
H_I&=&\frac{1}{2}\int {\rm d}{\bf r} \ {\rm d} {\bf r'}
\psi_{\alpha}^{\dagger}({\bf r}  )\psi_{\beta}^{\dagger}( {\bf r'} )
V( {\bf r}-{\bf r'})\psi_{\beta}( {\bf r'} ) \psi_{\alpha}({\bf r}  )\nonumber\\
&=&\sum_{{\bf p},{\bf p'},{\bf q}}V ({\bf q})
a^{\dagger}_{\alpha{\bf p}}a^{\dagger}_{\beta{\bf p'}+{\bf q}}
a_{\beta{\bf p'}}a_{\alpha{\bf p}+{\bf q}}.
\end{eqnarray}
In the above summation over repeated spin indices is understood.
We show that adding the interaction
 leads, in perturbation theory,
 to additional logarithmic corrections in two dimensions to both thermodynamic and transport
quantities. 
We discuss how these corrections may be interpreted in terms of a disorder-renormalized
Fermi liquid\cite{altshuler1983b,finkelstein1983,castellani1984,
castellani1986b,castellani1987}.
To achieve such a goal, a key step is the identification of which additional parameters
besides $t$
are required to take into account the interaction in the scaling description of disordered systems\cite{castellani1983b}.
In the next subsection, we will see how the above mentioned logarithmic
corrections arise in the  single-particle
density of states, the electrical conductivity (which was the only quantity
affected by disorder in the non-interacting case), the specific heat, and the spin susceptibility.
The last two will be evaluated via the correction to the thermodynamic potential.
We will recognize how disorder selects particular regions of transferred momentum
in electron-electron interaction giving rise to the effective couplings of the theory.
In the following subsection,
these will turn out to be  related to the Landau scattering amplitudes.
Once  the skeleton structure of the theory is developed,
via the Ward identities we will identify the singular scale-dependent
terms with the Landau parameters themselves and derive the corresponding group equations
in $d=2+\epsilon $. 
\subsection{Perturbation theory and the search for the effective couplings}
\subsubsection{Density of states}
\label{subsectiondos}
\begin{figure}
\includegraphics{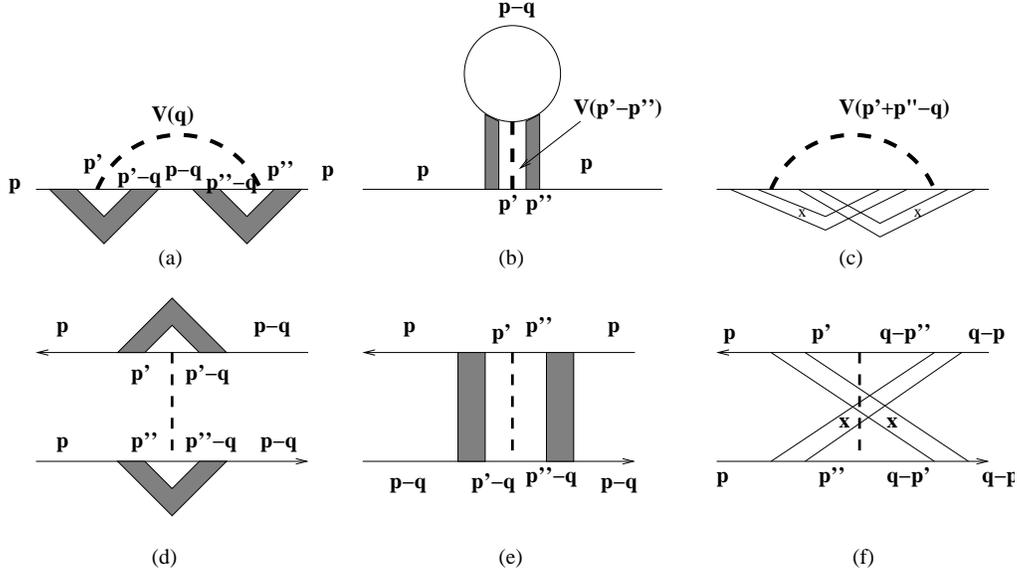}     
\caption{Diagrams for the correction to the Green's function to the lowest order in the interaction.
 (a) and (c) are exchange-type diagrams, while (b) is Hartree-type. 
 Interaction is shown as a thick dashed line. 
 The effective interaction diagrams ((d),(e),(f)) are obtained by cutting the internal
 Green's function with momentum ${\bf p}-{\bf q}$.
 The impurity "dressing" of the basic interaction
 diagrams selects three different regions of transferred momentum. In diagram (a), the momentum
 $q\sim 0$, that flows in the ladder, is also present in the interaction. 
 (See (d)). In diagram (b), instead, the different position of the impurity ladder makes
 the momentum flowing in the interaction line, $p'-p''$, unrelated to $q$ flowing in the ladder. (See (e)). 
 Finally, the crossed ladders in diagram (c) select, in the small-$q$ limit, 
 the interaction in the Cooper channel.  (See (f)).}
\label{densityofstatesdiagram}
\end{figure}
We begin by considering the diagrams for the Green's function to lowest order in the interaction.
These are shown in fig.\ref{densityofstatesdiagram}. The basic diagrams are the usual
exchange and Hartree contributions. Here, these two basic diagrams are "dressed" by the
presence of disorder. As before, this is done 
by averaging over the impurity configurations. 
To illustrate how the mechanism now works,
let us consider first the exchange diagram (a) in fig.\ref{densityofstatesdiagram}.
Its expression, before the impurity averaging,  reads
\begin{eqnarray}
\label{densityofstates1}
\delta {\cal G}({\bf r},{\bf r}';\epsilon_n)&=&-T\sum_{\omega_m}~\int {\rm d}{\bf r}_1~
{\rm d}{\bf r}_2~{\cal G}({\bf r},{\bf r}_1;\epsilon_n)
{\cal G}({\bf r}_1,{\bf r}_2;\epsilon_n -\omega_m)\nonumber\\
&\times&{\cal G}({\bf r}_2,{\bf r}';\epsilon_n){\cal V}({\bf r}_1,{\bf r}_2;\omega_m)
\end{eqnarray}
where we use the Matsubara formalism. $\epsilon_n =(2n+1)\pi T$ and $\omega_m =2m\pi T$
are fermionic and bosonic frequencies. The above expression is written in real space
and is valid for a given impurity configuration. 
The electron-electron interaction is assumed to depend on frequency,
since this will allow us to include retardation effects usually introduced by
screening effects. As a matter of convention, we use calligraphic letters
to indicate quantities depending on the Matsubara frequency. 
As a general strategy, we first  perform
the analytical continuation to real frequencies and then average over the impurity
configurations by exploiting the technique developed in the previous section.
Upon analytic continuation, ${\rm i}\epsilon_n \rightarrow \epsilon +{\rm i}0^+$\cite{abrikosov1975},
from eq.(\ref{densityofstates1}) the correction to the retarded Green's function becomes
\begin{eqnarray}
\label{densityofstates2}
\delta { G}^R({\bf r},{\bf r}';\epsilon )&=-&\int\frac{{\rm d}\omega}{2\pi {\rm i}}
~\int {\rm d}{\bf r}_1~ ~\int {\rm d}{\bf r}_2~
{ G}^R({\bf r},{\bf r}_1;\epsilon ) \left[b(\omega )
{ G}^R({\bf r}_1,{\bf r}_2;\epsilon -\omega)\right.\nonumber\\
&&\left.
({ V}^R({\bf r}_1,{\bf r}_2;\omega)-{ V}^A({\bf r}_1,{\bf r}_2;\omega))\right.\nonumber\\
&+&\left.f(\omega -\epsilon ){ V}^R({\bf r}_1,{\bf r}_2;\omega)
({ G}^R({\bf r}_1,{\bf r}_2;\epsilon -\omega)-{ G}^A({\bf r}_1,{\bf r}_2;\epsilon -\omega)  )
\right]{ G}^R({\bf r}_2,{\bf r}';\epsilon )
\end{eqnarray}
where $b$ and $f$ are the Bose and Fermi function, respectively.
$V^{R,A}$ are the analytical continuations of the dynamical interaction ${\cal V}$.
 The average over the impurity
configurations has two effects. First, each Green's function is replaced by  
expression  (\ref{selfconsistentborn}) obtained within the self-consistent Born
approximation. Secondly, one has to insert direct and crossed ladders wherever possible. 
This gives the leading approximation.
We have learned that the insertion of ladders is only 
possible when Green's functions
have poles on opposite sides of the real axis, {\sl i.e.}, when the ladder connects  a retarded
and  an advanced  Green's function. As a result we need to keep only the last term
of eq.(\ref{densityofstates2}). The correction to the density of the states then reads
\begin{eqnarray}
\label{densityofstates3}
\delta N (\epsilon )&=&
-\frac{1}{\pi}{\cal I}m \sum_{\bf q,p}
\int^{\infty}_{-\infty}\frac{{\rm d}\omega}{2\pi {\rm i}}~
f(\omega -\epsilon ){ V}^R({\bf q},\omega)
L^2 ( {\bf q}, \omega ) 
{ G}^R({\bf p},\epsilon )
{ G}^A({\bf p}-{\bf q},\epsilon -\omega)
{ G}^R({\bf p},\epsilon )\nonumber\\
&\times&\sum_{\bf p'}{ G}^R({\bf p'},\epsilon )
{ G}^A({\bf p'}-{\bf q},\epsilon -\omega)
\sum_{\bf p''}{ G}^R({\bf p''},\epsilon )
{ G}^A({\bf p''}-{\bf q},\epsilon -\omega)
\end{eqnarray}
where the average over the impurities has restored the translational invariance and
we have gone to the momentum representation. The ${\bf q}$-integral is dominated
by the diffusive pole of the impurity ladder. This implies that in the remaining
momentum integrals one can perform, as usual,  a small-${\bf q}$ expansion.
As it was also remarked in the non-interacting microscopic theory, we will perform
the {\sl fast} momenta (flowing in the Green's functions)
 integrals first, since they  contribute only to the coefficient of the
singular terms arising from the integration over the {\sl slow} momenta flowing in the ladders.
We  then set ${\bf q}=0$ in all the ${\bf p}$-, ${\bf p'}$-, and ${\bf p''}$-integrals, which
can be carried out in the standard way with the residue method and, using the results
of Appendix \ref{integrals},  one gets
\begin{equation}
\label{densityofstates4}
\delta N (\epsilon )=
\frac{1}{\pi}{\cal I}m \sum_{\bf q}\int^{\infty}_{-\infty}{\rm d}\omega~
f(\omega -\epsilon )\frac{N_0 V^R({\bf q},\omega)}
{(D{\bf q}^2-{\rm i}\omega )^2}. 
\end{equation}
At zero temperature, the Fermi function
becomes a step function giving the condition $\omega <\epsilon$. 
In the case of  short-range interaction $V^R$ remains  finite in the small frequency and momentum
limit.  Due to the presence of a double diffusive pole, the  integration in the region
$D{\bf q}^2\tau, |\omega|\tau <1$ over frequency and momenta
gives a  logarithmic divergence 
\begin{equation}
\label{densityofstates5}
\frac{\delta N (\epsilon )}{N_0}= N_0 V^R(0,0) t\ln |\epsilon \tau |\equiv V_1
t\ln |\epsilon \tau | .
\end{equation}
The Hartree diagram (b) may be evaluated in a similar way. Its expression, after impurity
averaging, reads
\begin{eqnarray}
\label{hartreedos}\delta N (\epsilon )&=&
\frac{1}{\pi}{\cal I}m \sum_{\bf q,p}
2 V_2\int^{\infty}_{-\infty}\frac{{\rm d}\omega}{2\pi {\rm i}}~
f(\omega -\epsilon )
L^2 ( {\bf q}, \omega ) \nonumber\\
&\times&{ G}^R({\bf p},\epsilon )
{ G}^A({\bf p}-{\bf q},\epsilon -\omega)
{ G}^R({\bf p},\epsilon )
\end{eqnarray}
where the relative minus sign and the factor of $2$ are  due to the extra fermionic loop
in the Hartree diagram. The interaction parameter   $V_2$ takes into account the scattering
at large angle across the Fermi surface, as shown in diagrams (b) and (e) 
of fig.\ref{densityofstatesdiagram}. 
\begin{eqnarray}
\label{hartreeinteraction}
V_2&=&\frac{1}{N_0}\sum_{\bf p',p''}
{ G}^R({\bf p'},\epsilon )
{ G}^A({\bf p'}-{\bf q},\epsilon -\omega)
{ V}^R({\bf p'}-{\bf p''},0)\nonumber\\
&\times&{ G}^R({\bf p''},\epsilon )
{ G}^A({\bf p''}-{\bf q},\epsilon -\omega).
\end{eqnarray}
We then see that the presence of the diffusive pole of the ladder,
by making the
small $q$ region more relevant,  effectively selects the electron-electron scattering
at small, $V_1$  (exchange contribution), and  large momentum transfer,
$V_2$ (Hartree contribution). A similar analysis can be carried out for
the exchange contribution with crossed ladders (diagram (c)) and
one has a third parameter $V_3$,
\begin{eqnarray}
\label{cooperinteraction}
V_3&=&\frac{1}{N_0}\sum_{\bf p',p''}
{ G}^R({\bf p'},\epsilon )
{ G}^A({\bf q}-{\bf p'},\epsilon -\omega)
{ V}^R({\bf p'}+{\bf p''},0)\nonumber\\
&\times&{ G}^R({\bf p''},\epsilon )
{ G}^A({\bf q}-{\bf p''},\epsilon -\omega).
\end{eqnarray}
Actually,
there is also the Hartree contribution with crossed ladders, but it contributes
with minus twice the same scattering amplitude. The total correction to the
density of states reads then
\begin{equation}
\label{densityofstatestotal}
\frac{\delta N (\epsilon )}{N_0}= (V_1 -2V_2-V_3) t\ln |\epsilon \tau |.
\end{equation}
We notice that $V_3$ corresponds to the 
interaction in the Cooper scattering channel(see fig.(\ref{densityofstatesdiagram})).
 Since its presence does not
change  the results qualitatively, to simplify the  exposition,
 we let it drop from our subsequent
 discussion. On  a  formal level, one may assume the presence of a small magnetic
 field which, as we have seen, by introducing a mass,  kills
 the singularity in the Cooper channel.
 The same selection of relevant momenta appear also in the perturbative calculation
 of the electrical conductivity and thermodyamic potential, as we are going to show.
 Before leaving the density of states, we notice that   
the above results are modified in the presence of long-range Coulomb interaction, which leads
to log-square singularity in two dimensions. Details are provided in  Appendix
\ref{Appendixlongrange}.

\subsubsection{Electrical conductivity}
\label{subsectionsigma}
The impurity-averaged diagrams responsible for the corrections to the
conductivity are obtained in Appendix \ref{Appendixconductivity}. 
The procedure is similar to that followed in the case of the density of states, but
there are many more diagrams. For this reason, the detailed derivation of how to
perform the impurity average and the integration over the fast momenta is left to
the Appendix \ref{Appendixconductivity}. Here we give directly the final result
for the exchange diagram containing the interaction amplitude $V_1$
before the last integration over the slow momentum and frequency. We have
\begin{eqnarray}
\label{conductivity9}
\delta \sigma &=&-\frac{2\sigma_0}{\pi~d}\sum_{\bf q}D {\bf q}^2\int_{-\infty}^{\infty}~
{\rm d}\omega
\frac{\partial }{\partial \omega}\left(\omega \coth \left(\frac{\omega}{2T}\right) \right)
{\cal I}m~\left[\frac{ V^R({\bf q},\omega)}{(D{\bf q}^2-{\rm i}\omega)^3}\right]
\end{eqnarray}
where the dimensionality factor $d$ comes from the angular integration over ${\bf q}$.
In two dimensions, in the case of short-range interaction, 
eq.(\ref{conductivity9}) yields\cite{altshuler1980,altshuler1980b}
\begin{equation}
\label{conductivity10}
\delta \sigma =-\frac{e^2}{\pi^2\hbar}
(V_1 -2V_2)\frac{1}{2}\ln \left(\frac{1}{T\tau}\right),
\end{equation}
where, as for the density of states, the factor $-2V_2$ takes into account the contribution
coming from the Hartree diagram.
In the case of Coulomb long-range forces, , at small $q$,
the screened interaction introduces
 an extra singularity (see Appendix \ref{Appendixlongrange})
$V^R (q,\omega)\approx -2\pi {\rm i}e^2 \omega /(D q^2)$.
In contrast to the case of the single-particle density of states, 
the $1/q^2$ singularity arising from the interaction
is compensated from the additional $q^2$ factor in the integrand. 
By using eq.(\ref{densityofstates9}) for $V^R({\bf q},\omega )$
one obtains for the exchange contribution
\begin{equation}
\label{conductivity11}
\delta \sigma =-\frac{e^2}{2\pi^2\hbar}
\ln \left(\frac{1}{T\tau}\right),
\end{equation} 
which has the same form as the weak localization correction.
The same will hold for both short- and long-range case in the renormalized perturbation
theory as we shall see in the next subsection.

\subsubsection{Thermodynamic potential}
\label{subsectionthermo}
\begin{figure}
\includegraphics{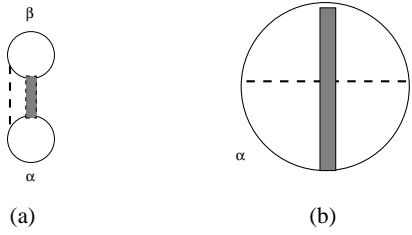}
\caption{Diagrams for the thermodynamic potential. (a) Hartree. (b) Exchange.
The ladder arising from the impurity average is already shown. Notice that in the Hartree diagram
the spin on the two lines of the ladder may differ from zero, whereas it is always
zero for the exchange one.  }
\label{thermo}
\end{figure}

Contrary to the non-interacting case, the interplay between disorder and interaction introduces
singular corrections to the specific heat and spin susceptibility.
We present here the derivation  of the correction to the
thermodynamic potential. 
The effective diagrams are shown in fig. \ref{thermo}.
We have to carry out the integration over the fast momenta
flowing in the Green's functions. After doing that, we obtain 
\begin{eqnarray}
\label{th6}
\Delta \Omega &=&- (V_1-2V_2 )T\sum_{\omega_m }\sum_{\bf q} 
\frac{ |\omega_m|}
{D{\bf q}^2+|\omega_m|}.
\end{eqnarray}
In Appendix \ref{Appendixthermo}, we provide the details of the derivation.
The Matsubara frequency sum is limited by $|\omega_m|\tau <1$. To relax the constraint
in the sum, we introduce a cutoff function
$$
\left(\frac{\tau^{-1}}{|\omega_m|+\tau^{-1}}\right)^2
$$
and the sum runs between minus and plus infinity. We may then perform analytical continuation
to get
\begin{eqnarray}
\label{th7}
\Delta \Omega& =&-\frac{1}{2}(V_1 -2V_2)\sum_{\bf q} \int_{-\infty}^{\infty}\frac{{\rm d}\omega}{\pi}
b(\omega ) {\cal I}m \left[ \left(\frac{\tau^{-1}}{-{\rm i}\omega+\tau^{-1}}\right)^2
\frac{-2{\rm i}\omega  }
{D{\bf q}^2-{\rm i}\omega }\right].
\end{eqnarray}
In two dimensions, the sum over ${\bf q}$ may be done
and eq.(\ref{th7}) becomes
\begin{eqnarray}
\label{th8}
\Delta \Omega& =&-\frac{ (V_1-2V_2)}{4\pi^2 D}\int_{-\tau^{-1}}^{\tau^{-1}}{{\rm d}\omega}~
\omega ~b(\omega)~\ln (|\omega |\tau )\nonumber\\
&=&- t(V_1-2V_2)\frac{\pi^2 N_0T^2}{3}\ln (T\tau ),
\end{eqnarray}
where we have dropped non singular terms in temperature.
The thermodynamic potential then acquires  a  logarithmic correction that implies for
the specific heat\cite{altshuler1983}
\begin{equation}
\label{th9}
\delta C_V =  C_{V,0}t(V_1-2V_2)\ln (T\tau ),
\end{equation}
where  $C_{V,0}=(2\pi^2 N_0T)/3$ is the non-interacting value. 

In order to evaluate the spin susceptibility, we must include the Zeeman coupling.
As shown in the eq.(\ref{ladderzeeman}) 
of Appendix \ref{Appendixzeeman}, only the ladder corresponding to total
spin $\pm 1$ of the incoming particle-hole pair are affected by the magnetic field
via the Zeeman energy $\omega_s =g \mu_B B$. In this respect, we notice 
that only the Hartree diagram of
fig.\ref{thermo} contributes, since in this case the total spin of the particle-hole
pair is given by the combination $\alpha -\beta$, with both indices running over $\pm 1/2$.
In the exchange diagram the total spin of the particle-hole
pair entering the ladder is $\alpha -\alpha$ and is
always zero. 
The combination of the two diagrams may be arranged as the sum of a singlet and triplet
component with respect to the total spin of the particle-hole ladder. Hence, the magnetic
field only affects the triplet component with value $M=\pm 1$.
For the purpose of isolating the magnetic-field dependent contribution
to the thermodynamic potential, it is convenient to write the difference, with and without the
magnetic field, of the triplet as 
\begin{eqnarray}
\Delta \Omega_B &=&\frac{1}{2} T\sum_{\omega_m }\sum_{M\pm 1}
\sum_{\bf q}V_2 |\omega_m|\left[\frac{1}{D{\bf q}^2+|\omega_m|-{\rm i}M\omega_s sgn(\omega ) }
-\frac{1}{D{\bf q}^2+|\omega_m| }\right]\nonumber\\
&=&\frac{1}{2}T\sum_{\omega_m }
\sum_{\bf q}V_2 |\omega_m|\left[\frac{2(D{\bf q}^2+|\omega_m|)}{(D{\bf q}^2+|\omega_m|)^2+\omega^2_s }
-\frac{2}{D{\bf q}^2+|\omega_m| }\right]\nonumber\\
&=&-\frac{1}{2}
\pi t N_0 V_2 T\sum_{\omega_m }|\omega_m| \ln \left(\frac{\omega_m^2+\omega_s^2}{\omega^2_m} \right)
\nonumber\\
&\approx&-\pi t N_0 V_2 \omega_s^2 T\sum_{\omega_m > 0 }\frac{1}{\omega_m}\nonumber\\
&=&\frac{t N_0 V_2}{2}\omega_s^2\ln (T\tau ).
\label{thzeeman}
\end{eqnarray}
Now a few words concerning the steps
leading to the final expression of eq.(\ref{thzeeman}). Since
the factor $|\omega_m|$ in the sum excludes the term with $\omega_m =0$ and 
the expression in the sum is even in $\omega_m$,  
we have rewritten the sum as twice the sum over the strictly 
positive frequencies. Then we observe that the smallest frequency is  $2\pi T$, which allows us
to make a small magnetic field expansion $\omega_s < T$.
From eq.(\ref{thzeeman}) one finally gets,
by differentianting twice with respect to the magnetic field, the correction to the
spin susceptibility
\begin{equation}
\label{spinperturbative}
\chi =-\chi_0 2 t V_2 \ln (T\tau ),
\end{equation}
where we have introduced   the non-interacting value $\chi_0 = 2 N_0 (g\mu_B /2)^2$. 
No correction is instead found for the compressibility.

The two parameters $V_1$ and $V_2$ that appear in the perturbative
expression of the density of states (\ref{densityofstatestotal}),
conductivity (\ref{conductivity10}), specific heat (\ref{th9}), and spin susceptibility
(\ref{spinperturbative}) are the natural candidates for the additional 
running couplings to be used with the dimensionless resistance $t$ to obtain the
renormalized perturbation theory in $2+\epsilon$ dimensions.


\subsection{The renormalized perturbation theory and effective Fermi-liquid description}
\label{subsectionFLT}
In this subsection we show how the perturbative results derived in the previous
one may be generalized to all orders in the interaction and how the effective
scattering amplitudes are related to the Landau parameters. We then develop
the renormalized perturbation theory for the various response functions
by making use of the Ward identities that implement the conservation laws
for charge, spin, and energy. Finally, we go back to the perturbative results which
will be generalized to all orders in the interaction and derive the group equations.
\subsubsection{Effective scattering amplitudes and Landau parameters}
According to the discussion of the previous section,
the relevant interaction terms in the Hamiltonian are reduced to
\begin{equation}
\label{effectivehamiltonian}
H_{I}=\sum_{{\bf p},{\bf p'}}\sum_{{\bf q}}'
\left[ V_1  a^{\dagger}_{\alpha{\bf p}}a^{\dagger}_{\beta{\bf p'}+{\bf q}}
a_{\beta{\bf p'}}a_{\alpha{\bf p}+{\bf q}} +V_2 
a^{\dagger}_{\alpha{\bf p}}a^{\dagger}_{\beta{\bf p'}+{\bf q}}
a_{\beta{\bf p}+{\bf q}}a_{\alpha{\bf p'}}\right],
\end{equation}
where the primed ${\bf q}$-summation will be confined to small ${\bf q}$ values
as implied when performing the disordere averaging.

\begin{figure}
\includegraphics{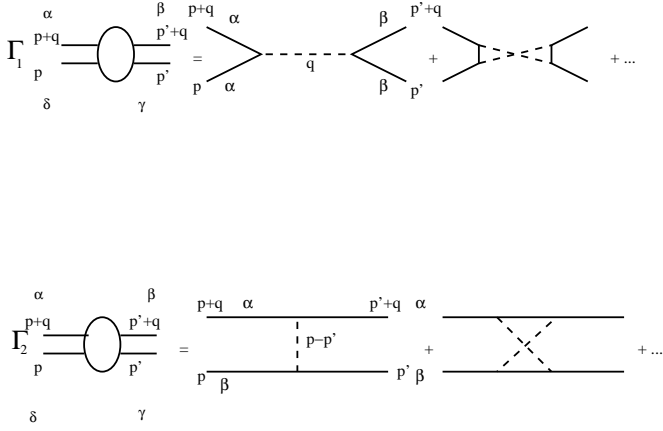}
\caption{Lowest order diagrams for the small ($\Gamma_1$)
and large  ($\Gamma_2$ ) scattering angle. Notice the different momenta flowing through
the interaction propagator. Also the two diagrams have a different spin structure.}
\label{amplitudes}
\end{figure}

Up to now our discussion has been limited to the first order in the interaction.
However, we do not want to confine our theory to small interaction and the only true
expansion parameter must be the dimensionless resistance $t=1/(2\pi g)$.
The good metal condition,
$g\gg 1$ ( $E_F \tau /\hbar \gg 1 $),
implies that the disorder only affects electron states within a small distance
$\hbar /\tau$ away from the Fermi surface. Under these circumstances, one may
go beyond the first-order interaction correction by replacing $V_1$ and $V_2$
with the Fermi-liquid scattering amplitudes $\Gamma_1$ and $\Gamma_2$,
whose lowest order diagrams are depicted in
fig.\ref{amplitudes}.

We note that in the absence of spin-flip mechanisms, the total spin of two
colliding particles  is a conserved quantity.  It is then convenient to introduce
the {\sl singlet} $\Gamma_s$ and {\sl triplet} $\Gamma_t$ scattering amplitudes. After the
selection of the relevant momentum transfer terms (cf.
eqs.(\ref{densityofstates5})-(\ref{hartreeinteraction})) the corresponding spin structures
can be decomposed as
\begin{equation}
\label{amplitudesdecomposition}
\Gamma_1 \delta_{\alpha\delta}\delta_{\beta\gamma}-\Gamma_2\delta_{\alpha\beta}\delta_{\gamma\delta}
=(\Gamma_1 -\frac{1}{2}\Gamma_2)\delta_{\alpha\delta}\delta_{\beta\gamma}-
\frac{1}{2}\Gamma_2\sigma_{\alpha\delta}\cdot \sigma_{\beta\gamma}
\end{equation}
where we have used the identity
\begin{equation}
\label{spinidentity}
\delta_{\alpha\beta}\delta_{\gamma\delta }=
\frac{1}{2}\delta_{\alpha\delta}\delta_{\beta\gamma}+
\frac{1}{2}\left[\sigma^x_{\alpha\delta}\sigma^x_{\beta\gamma}
+\sigma^y_{\alpha\delta}\sigma^y_{\beta\gamma}+
\sigma^z_{\alpha\delta}\sigma^z_{\beta\gamma} \right].
\end{equation}
We then define 
\begin{equation}
\label{densityofstates6}
\Gamma_s =\Gamma_1-\frac{1}{2}\Gamma_2,~~~\Gamma_t =\frac{1}{2}\Gamma_2.
\end{equation}
The scattering amplitudes $\Gamma_s$ and $\Gamma_t$ are 
related to the Landau Fermi-liquid parameters $F^0_s$ and $F^0_a$by\cite{abrikosov1975}
\begin{equation}
\label{densityofstates7}
\Gamma_s =\frac{1}{2N_0}\frac{F_s^0}{1+F_s^0},~~~~
\Gamma_t =-\frac{1}{2N_0}\frac{F_a^0}{1+F_a^0}.
\end{equation}
From now on, when necessary, $N_0$ is assumed to include the Landau 
effective-mass correction. In terms of the Landau parameters,  compressibility,
 spin susceptibility and specific heat are given by
\begin{eqnarray}
\label{landaufermiliquid}
\frac{\partial n}{\partial\mu}&=&\frac{2 N_0}{1+F_s^0}=2N_0 (1-2N_0\Gamma_s )\equiv 2N_0Z_s^0 \nonumber \\
\chi &=&\frac{\chi_0}{1+F_a^0}=\chi_0 (1+2N_0\Gamma_t )\equiv \chi_0 Z_t^0\nonumber\\
C_V&=&C_{V,0}\equiv C_{V,0}Z^0,
\end{eqnarray}
where $Z^0$ will  be different from one in the presence of disorder
and is here introduced for symmetry in the equations. 

\subsubsection{Renormalized response functions}
\begin{figure}
\includegraphics{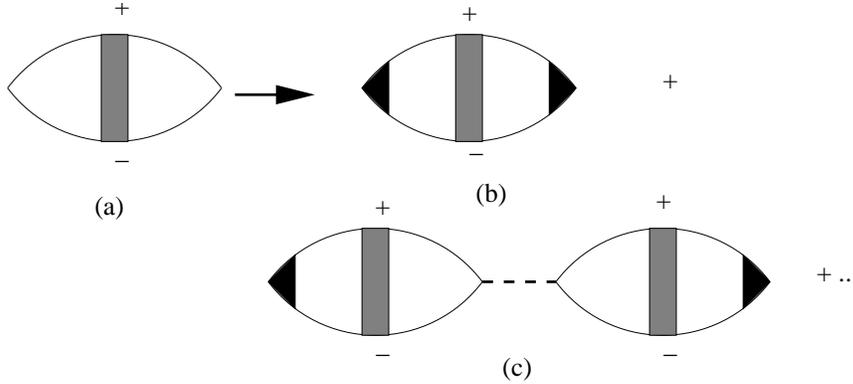}
\caption{Skeleton structure of the perturbative evaluation of the response
function in the presence of interaction. The plus and minus signs indicate the
sign of the energy flowing in the line. We remind that the ladder insertion is only possible
between a pair of retarded and advanced   Green's functions. 
The first diagram (a) is the one used in the non-interacting theory.
Diagram (b) represents how the interaction "dresses" (a). It 
 leads to i) a "dressing" of the vertex, here indicated as a black triangle, 
ii) renormalization of the ladder.  Interaction appears explicitly in diagram (c),
which has to be considered together with all the other diagrams, indicated by the dots, obtained
by the  infinite resummation of the interaction.
Depending on whether we consider the charge or spin response function one has the vertex
$\Lambda_{s,t}$ and interaction $\Gamma_{s,t}$. 
For the energy response function, there is also a vertex $\Lambda_E$, but
there is no infinite  resummation of the interaction for the
reasons explained in the text.}
\label{responsefunctionfig}
\end{figure}

We begin our discussion of the interplay between interaction and disorder in the
renormalization of the response functions, 
 by considering the density-density
response function. As in the case of the non-interacting theory of
eqs.(\ref{staticlimit}),(\ref{dynamiclimit}), we split
the response function in  static  and dynamic contributions.
The effect of the interaction may be
understood in terms of a skeleton perturbation
theory, as shown in fig.\ref{responsefunctionfig}.
The first diagram (a), due to the
ladder insertion, gives the dynamic contribution in the non-interacting case
(cf. eq.(\ref{dynamiclimit})).
Its "dressing" due to the interaction is represented by the
diagram (b), whose contribution can be written in the form 
\begin{equation}
 K^{00}_{(b)} = 
-\frac{{\rm i}\omega  2e^2 N_0\zeta^2\Lambda_s^2} {Dq^2-
{\rm i}\omega Z}\equiv 
K^{00}_{+-}\Lambda_s,
\label{densityinteractionguess}
\end{equation}
which generalizes eq.(\ref{dynamiclimit}).
We now discuss it in detail. We begin with the ladder, which is the
most important ingredient of the theory.  The ladder in fact  is the
effective propagator of the diffusive mode, responsible for the
singularities appearing in perturbation theory. It  requires a wave function
renormalization $\zeta$, a frequency (effective external field) renormalization $Z$,
and a renormalization of the diffusion constant\footnote{We leave the same symbol as before for
 the interaction-renormalized diffusion constant to keep the notation
simple. Everywhere in this subsection this is understood whenever we are
dealing with the renormalized ladder.}
\begin{equation}
\label{rp1ladder}
L(q,\omega)=\frac{1}{2\pi N_0\tau^2}\frac{1}{Dq^2-{\rm i}\omega}\rightarrow
\frac{1}{2\pi N_0\tau^2}\frac{\zeta^2}{D q^2-{\rm i}Z\omega}.
\end{equation}
In  Appendix \ref{subsectionladder}
we show that the interaction corrections to the ladder do not destroy its
diffusive pole behavior and indeed lead to the renormalized form
of eq.(\ref{rp1ladder}).  Furthermore,   the logarithmically singular  corrections to the
single-particle density of states, specific heat, and conductivity which  appear 
in two dimensions  can be absorbed in the above three renormalizations
$\zeta$, $Z$, and $D$, respectively.
Here, by exploiting the constraints given by the general conservation laws
embodied in the Ward identities, we will be able to directly express 
the ladder renormalization parameters in terms of physical quantities.

The vertex $\Lambda_s$ is  one in the non-interacting case.
In the  interacting case, it represents the vertex which, 
when multiplied by $K^{00}_{+-}$, 
gives the total dynamic part of $K^{00}$, which includes also terms ending with
two advanced ($++$) or two retarded ($--$) Green's functions. The vertex
$\Lambda_s$ is irreducible for cutting a ladder propagator.

\begin{figure}
\includegraphics{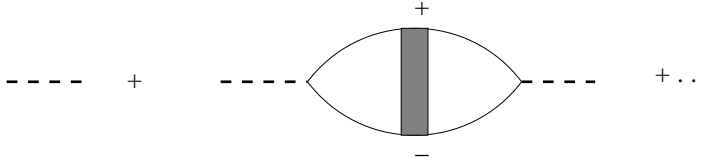}
\caption{Dynamical resummation of the interaction.}
\label{dynamicalamplitudefig}
\end{figure}

Besides "dressing" the non-interacting diagram, interaction leads to new
diagrams as diagram (c)
of fig.\ref{responsefunctionfig}, which  gives
\begin{equation}
K^{00}_{(c)}=e^2
\frac{\omega  2 N_0\zeta^2\Lambda_s} {Dq^2-
{\rm i}\omega Z}\Gamma_s 
\frac{\omega  2 N_0\zeta^2\Lambda_s} {Dq^2-
{\rm i}\omega Z}
\label{firstcorrectiontoresponsefunction}.
\end{equation}
By keeping in mind that the order in the expansion parameter $t$ is determined
by the number of integrations over the momenta flowing in the ladder propagator,
we can replace, without changing the order in $t$, 
the scattering amplitude $\Gamma_s$ by its screened form
\begin{equation}
\label{densityofstates8}
\Gamma_{s}(q,\omega )=\Gamma_{s}-\Gamma_{s}{\rm i}\frac{2\omega}{ 2\pi}(2\pi N_0 \tau )^2 L(q,\omega )
\Gamma_{s}(q,\omega )
=\Gamma_{s}
\frac{Dq^2-{\rm i}Z\omega}{Dq^2-{\rm i}Z_{s}\omega },
\end{equation}
obtained by an RPA-like infinite resummation (shown in fig.\ref{dynamicalamplitudefig})
 and using eq.(\ref{rp1ladder}). 
  In eq.(\ref{densityofstates8})
 $Z_{s}=Z-2\zeta^2N_0\Gamma_s$
  will turn to be the expression dressed by the disorder of Fermi-liquid renormalization of the
compressibility $Z_s^0$.
If we insert the  dynamical amplitude (\ref{densityofstates8})  
in the expression (\ref{firstcorrectiontoresponsefunction}) for the first interaction
correction and combine it with the contribution (\ref{densityinteractionguess})
of the first diagram, we arrive at the final form of the response function
which resums all the infinite series of diagrams indicated by dots in 
 fig. \ref{responsefunctionfig}: 
\begin{equation}
 K^{00}_{+-} = 
-\frac{{\rm i}\omega  2 e^2N_0\zeta^2\Lambda_s} {Dq^2-
{\rm i}\omega Z_s},~~~~~~~~
 K^{00} = -e^2\frac{\partial n}{\partial\mu}
+ K^{00}_{+-}\Lambda_s.
\label{rp14}
\end{equation}
We are now ready to make use  of the  Ward identities (\ref{dynamic}),(\ref{densityofstates}),
where now $K_{+-}^{00}$ is given by eq.(\ref{rp14}),
and we get
\begin{eqnarray}
\frac{N}{N_0} &=&\frac{\zeta^2\Lambda_s}{Z_s},\label{rp15}\\
\frac{\partial n}{\partial \mu}&=&2N_0\frac{\zeta^2\Lambda_s^2}{Z_s}.
\label{rp16}
\end{eqnarray}
We can now follow two alternative routes leading to the same result.
We can first remember from Appendix \ref{subsectionladder} that $N/N_0$ coincides
with $\zeta$. It then follows 
from eq.(\ref{rp15})  that $\zeta\Lambda_s=Z_s$, which when used in eq.(\ref{rp16})
gives 
\begin{equation}
\label{rp17}
\frac{\partial n}{\partial \mu}=2 N_0 Z_s.
\end{equation}
Alternatively, it can be shown that
the compressibility has no logarithmic corrections, i.e., is given by
the Fermi-liquid value in the absence of disorder with $Z_s=Z_s^0=\zeta\Lambda_s$.
 This means that,
although both $\Lambda_s$ and $\zeta$ have logarithmic corrections, the combination
$\zeta\Lambda_s$ does not.
The final expression for the density response function has the conserving form
 \begin{equation}
 \label{finaldensity}
 K^{00}(q,\omega)=-e^2\frac{\partial n}{\partial \mu}\frac{D_c q^2}{D_c q^2-{\rm i}\omega}
 \end{equation}
where  the charge diffusion constant $D_c=D/Z_s$ has been introduced.

A similar analysis can be done for the spin susceptibility and the specific heat.
First we note that the dynamical scattering
amplitude in the triplet channel reads
\begin{equation}
\label{densityofstates8bis}
\Gamma_{t}(q,\omega )=\Gamma_{t}+\Gamma_{t}{\rm i}\frac{2\omega}{ 2\pi} (2\pi N_0 \tau )^2 L(q,\omega )
\Gamma_{t}(q,\omega )
=\Gamma_{t}
\frac{Dq^2-{\rm i}Z\omega}{Dq^2-{\rm i}Z_{t}\omega },
\end{equation}
where $Z_t=Z+2N_0\zeta^2\Gamma_t$ is the expression dressed by the disorder of the  Fermi-liquid
renormalization of the spin susceptibility $Z_t^0=1+2N_0\Gamma_t$, which as for the compressibility is given by the static limit ($\omega =0, q\rightarrow 0$) of the spin-spin response function $\chi ({\bf q}, \omega )$.
To analyze the spin-spin response function, one introduces, in analogy with the density-density
response function, the i-th component of the spin density
\begin{equation}
S^i (x)=\frac{g\mu_B}{2}\psi^{\dagger}_{\alpha}(x)\sigma^i_{\alpha\beta}\psi_{\beta} (x),
\label{spindensity}
\end{equation}
and the associated spin-current density ${\bf S}^i$. 
By proceeding as in the charge density case of eq.(\ref{reduciblevertex}), 
we introduce a vertex function
\begin{equation}
\label{reduciblespinvertex}
\Lambda^{i,\mu}_{\alpha\beta}(x,x',x'')=
<T_t S^{i,\mu}(x)\psi_{\alpha} (x')\psi^{\dagger}_{\beta} (x'')>,
\end{equation}
where the first upper index indicates which component of the spin density we are dealing with 
while the
second one, $\mu$, distinguishes between density and current component.
In the presence of a magnetic field, one may derive for $\Lambda^{i,\mu}$ the following
Ward identity
\begin{equation}
q_{\mu}\Lambda^{i,\mu}_{\alpha\beta}(p,q)=\frac{g\mu_B}{2}\left[
\sigma^i_{\alpha\beta} ~G_{\beta}(p-q/2)-
G_{\alpha}(p+q/2)~\sigma^i_{\alpha\beta}\right]+{\rm i}\epsilon_{ij3}\omega_s
\Lambda^{j,0}_{\alpha\beta}(p,q),
\label{spinwardidentity}
\end{equation}
where $\epsilon_{ijk}$ is the full antisymmetric tensor and we have assumed the
magnetic field in the $z$-direction. We now consider the consequences
of eq.(\ref{spinwardidentity}) in the vanishing momentum
limit. Instead of working with the $x$ and  $y$
components, it is convenient to switch to the circularly polarized ones defined
as
\begin{eqnarray}
\Lambda^{\uparrow\downarrow,0}&=&\frac{1}{2}\left(\Lambda^{1,0}+\Lambda^{2,0}\right),\label{mplus}\\
\Lambda^{\downarrow\uparrow,0}&=&\frac{1}{2}\left(\Lambda^{1,0}-\Lambda^{2,0}\right),\label{mminus}
\end{eqnarray}
corresponding to the spin-density in the $M=\mp 1$ triplet channels, respectively. By including
the $M=0$ channel of the triplet corresponding to $\Lambda^{3,0}$, one gets the
Ward identity (\ref{spinwardidentity}) in the form
\begin{equation}
( \omega + M \omega_s )\Lambda^{M,0}_{\alpha\beta}(\epsilon ,\omega)=\frac{g\mu_B}{2}\left[
\sigma^M_{\alpha\beta} ~G_{\beta}(\epsilon -\omega /2)-
G_{\alpha}(\epsilon +\omega/2)~\sigma^M_{\alpha\beta}\right],
\label{spinwardeffective}
\end{equation}
where $\sigma^M$ are defined as in eq.(\ref{mplus}),(\ref{mminus}) and we have dropped the
explicit dependence on momentum ${\bf p}$ both in the vertex and in the Green's functions.
\footnote{Notice that the Green's function gets a spin label in the presence of
Zeeman coupling.} The dynamical resummation of the skeleton structure,
analog to eq.(\ref{rp14}) for $K^{00}$, with $\Gamma_s$ replaced by $\Gamma_t$ and
$\partial n/\partial \mu$ or $2 N_0$ by $\chi$, gives
\begin{equation}
\label{wi7bis}
 \chi_{+-}^M (q,\omega )= 
\frac{{\rm i}\omega  \chi_0\zeta^2\Lambda_t} {Dq^2-
{\rm i}\omega Z_t-{\rm i}MZ_H\omega_s},~~~~~~~~
 \chi^M (q,\omega ) = \chi
+ \chi_{+-}^M (q,\omega )\Lambda_t,
\end{equation}
where we have introduced a renormalization factor for the Zeeman energy, $Z_H$.
Equation (\ref{spinwardeffective}) together with the analog of eq.(\ref{dynamic})
for the spin-spin response function in the $M$-th channel leads to
\begin{eqnarray}
\chi^M(0,\omega )&=& \chi\frac{M\omega_s}{\omega+M\omega_s}, \label{spindynamic}\\
\chi^M_{+-}(0,\omega )&=&-\chi_0 \frac{N}{N_0}
\frac{\omega}{\omega+M\omega_s} . \label{spindensityofstates}
\end{eqnarray}
Indeed, in the limit of zero magnetic field eq.(\ref{spindynamic}) gives the total spin 
conservation by the vanishing of the response function at finite $\omega$ as $q$ goes to
zero, while eq.(\ref{spindensityofstates}) reproduces the single-particle density of
states in term of the dynamical part of the spin response function. Finally,
by making use of the  Ward identities (\ref{spindynamic}),(\ref{spindensityofstates})
in (\ref{wi7bis}),
one obtains
\begin{equation}
\label{rp18}
\chi^M (q,\omega )=\chi \frac{D_sq^2-{\rm i}M\omega_s}{D_sq^2-{\rm i}(M\omega_s +\omega)}, 
Z_t =\zeta\Lambda_t, \chi=\chi_0 Z_t ,  Z_H=Z_t,   D_s =D/Z_t.
\end{equation}
Notice that the result $Z_H=Z_t$\cite{raimondi1990}  implements, to all orders
in the expansion in the parameter $t$, the Fermi-liquid zero-order result. 

The energy-energy response function $\chi_E$ can also be  decomposed according
to the eq.(\ref{wi7bis}), where $\chi$ is replaced by $\chi_E$ and 
$Z_t$ and $2N_0\Lambda^2_t$ by $Z$ and $C_{V,0}T\Lambda^2_E$\cite{castellani1987b,castellani1988}.
\footnote{In contrast to the density and spin response functions, in the energy response function
the  renormalization parameter does not require additional terms due to the interaction
besides $Z$, i.e., the Fermi-liquid renormalization,
 the analog of diagram (c) of fig. (\ref{responsefunctionfig}), is missing.
  This occurs  since the   interaction separates the integration
at the two vertices  in the response function
diagrams.   Due to the presence of the energy in the
thermal vertex, each integration over the energy  contributes at least a term with the second power of
the frequency. Hence the terms due to the dynamical resummation of the interaction
give rise to negligible contributions going with the fourth   power in the frequency. 
} The analogous Ward identity  gives
\begin{equation}
\label{rp18bis}
Z =\zeta\Lambda_E, \\\ C_V=C_{V,0} Z, \ \ \ D_E=D/Z
\end{equation}
and the frequency renormalization $Z$ of the ladder 
is identified with the specific-heat renormalization.
In the presence of Coulomb long-range forces, to avoid double counting,
one has
to subtract the statically screened long-range Coulomb $\Gamma_0$ from
the full singlet scattering amplitude entering the ladder resummation for
the density-density response function.
Hence, 
$\Gamma_s \rightarrow \Gamma_s -\Gamma_0$ in $Z_s$,  
where
\begin{eqnarray} 
\Gamma_0 (q,  \omega =0)&=&\frac{V_C (q)\Lambda_s^2}{1+V_C(q){{{\partial n }/\partial \mu}}}
 \rightarrow_{q\rightarrow 0}
\frac{\Lambda_s^2}{{{{\partial n} /\partial \mu}}}.
\end{eqnarray}
As a consequence of eq.(\ref{rp16}), from $Z_s=Z-2N_0\zeta^2(\Gamma_s-\Gamma_0)$ we derive the constraint
\begin{equation}
\label{rp19}
Z=2N_0\zeta^2 \Gamma_s.
\end{equation}
In this way we have completed the general formulation of the effective renormalized
Fermi-liquid theory.

\subsubsection{Derivation of the group equations}
We now come back to the perturbative expressions of the electrical conductivity (\ref{conductivity10}),
specific heat (\ref{th9}) and spin susceptibility (\ref{spinperturbative}).
These expressions were derived to
first order in the interaction. However, we have seen that we may
relax this condition in two ways. First, we may use the Fermi liquid
scattering amplitudes. Second, we can  make an infinite dynamical resummation
(eqs. (\ref{densityofstates8}),(\ref{densityofstates8bis})). 
For the electrical conductivity, by inserting the dynamical scattering amplitudes
in eq.(\ref{conductivity9}) one has

\begin{eqnarray}
\frac{\delta \sigma}{\sigma}&=&
- \frac{2}{2\pi {\rm i}}\int^{\infty}_{-\infty}{\rm d}\omega 
\frac{\partial}{\omega}\left( \omega \coth \left( \frac{\Omega}{2T}\right)  \right)
\sum_{\bf q}
\frac{ D  q^2(\Gamma_s (q,\omega )-3\Gamma_t(q,\omega ))}
{(D q^2-{\rm i}Z\omega )^3 }\nonumber\\
&=&t \left[1 +\frac{Z_s}{2N_0\zeta^2\Gamma_s}
\ln \frac{Z_s}{Z}\right.\nonumber\\
&+&\left. 3\left(1 -\frac{Z_t}{2N_0\zeta^2\Gamma_t}\ln \frac{Z_t}{Z}
 \right)\right]\ln \left(T\tau \right),
\label{conductivity12}
\end{eqnarray}
which coincides with $\delta D /D$ found from the ladder renormalization
in Appendix \ref{subsectionladder}.
In the case of Coulomb interaction, when $2N_0\zeta^2\Gamma_s =Z$,  the singlet part
of eq.(\ref{conductivity12}) reproduces eq.(\ref{conductivity11}).
Furthermore for small $\Gamma_s$ and $\Gamma_t$ ($Z=1$) one recovers the
first order short range interaction result of eq.(\ref{conductivity10}).
However, in the short-range case, by allowing the group equations to flow,  $Z_s=Z_s^0$
is invariant, while, as we shall see, $Z$ and $2N_0\zeta^2\Gamma_s$  diverge
and the singlet strength becomes again universal in Eq(\ref{conductivity12}) and equal to one.

The full expression for the
correction to the thermodynamic potential is obtained by introducing in eqs.(\ref{th6}),(\ref{thzeeman})
the singlet and triplet dynamical scattering amplitudes and the renormalized ladder,
 including the magnetic-field renormalization 
$Z_H=Z_t$.
For the magnetic-field independent part one then gets
\begin{eqnarray}
\Delta\Omega=&-&T\sum_{q\omega_m}\int_{0}^{1}{\rm d}\lambda
\left[\frac{N_0\zeta^2\Gamma_s|\omega_m|}{Dq^2+(Z-\lambda 2N_0\zeta^2\Gamma_s)|\omega_m|}
\right.\nonumber\\
&-&
\left.
\frac{3{N_0 \zeta^2\Gamma}_t |\omega_m|}{Dq^2+
(Z+\lambda 2{N_0\zeta^2 \Gamma}_t)|\omega_m|}\right],
\label{th10}
\end{eqnarray}
while the field-dependent contribution  reads
\begin{eqnarray}
\Delta\Omega_B&=&T\sum_{q\omega_m}\int_{0}^{1}{\rm d}\lambda
\sum_{M\pm 1}\left[
\frac{{N_0 \zeta^2\Gamma}_t |\omega|}{Dq^2+
(Z+\lambda 2{N_0\zeta^2 \Gamma}_t)|\omega_m|-
{\rm i}M (Z+ 2{N_0\zeta^2 \Gamma}_t )\omega_s {\rm sgn} (\omega )}\right. \nonumber\\
&-&\left.
\frac{{N_0 \zeta^2\Gamma}_t |\omega|}{Dq^2+
(Z+\lambda 2{N_0\zeta^2 \Gamma}_t)|\omega_m|}
\right].
\label{thzeemanrenormalized}
\end{eqnarray}
In the above equations, due to the presence of the dynamical resummation of the
interaction we used the standard trick\cite{abrikosov1975} of multiplying the interaction
by a parameter $0<\lambda <1$. However, this must not be introduced in the amplitude
present in the magnetic field insertion since it will generate spurious diagrams.
As a result, the corrections to the specific heat and to the spin 
susceptibility are
\begin{eqnarray}
\delta C_V&=&C_{V,0}t(N_0\zeta^2\Gamma_s-3{N_0\zeta^2 \Gamma}_t)\ln (T\tau ),\label{th13bis}\\
\delta \chi&=&-\chi_0 4tN_0 \zeta^2\Gamma_t\frac{Z_t}{Z}
\ln (T\tau ).\label{th13}
\end{eqnarray} 
The renormalization of the specific heat may be interpreted as the
renormalization of the quasi-particle density of states $N_{QP}=ZN_0$. Accordingly, the
renormalization of the spin susceptibility must contain both the renormalization
of the quasi-particle mass and of the Landau parameter $F_{a}^0$. To show this we write
\begin{equation}
\label{rf14}
\chi =\chi_0 Z \frac{Z_t}{Z}= \chi_0 Z \left( 1+\frac{2N_0\zeta^2\Gamma_t}{Z}\right)\equiv 
\chi_0 Z \left( 1+\gamma_t\right)
\end{equation}
with $\gamma_t =2N_0\zeta^2\Gamma_t /Z$ being the renormalised Landau static amplitude. 
We note that $\zeta^2$ is always associated either with $\Gamma_s$ or $\Gamma_t$ and drops out from
the following group equations. It  however renormalizes the single-particle density of states, which, in the interacting case, becomes scale dependent even though in a complicated way.
Let us define the flow variable $s=-\ln T\tau$ so that $s\rightarrow\infty$ corresponds to
the infrared limit. Then we have
\begin{eqnarray}
\frac{{\rm d} Z}{{\rm d} s}&=& -\frac{t}{2}Z(1-3\gamma_t),\label{rf15}\\
\frac{{\rm d} Z_t}{{\rm d} s}&=&2tZ \gamma_t (1+\gamma_t)\label{rf16}. 
\end{eqnarray}
According to eq.(\ref{rf14}) one has
\begin{equation}
\label{rf17}
\frac{{\rm d} Z_t}{{\rm d} s}=Z\frac{{\rm d} \gamma_t}{{\rm d} s}+
(1+\gamma_t)\frac{{\rm d} Z}{{\rm d} s},
\end{equation}
from which, by using eqs.(\ref{rf15}),({\ref{rf16}), one obtains 
\begin{equation}
\label{rf18}
\frac{{\rm d} \gamma_t}{{\rm d} s}=
\frac{t}{2}(1+\gamma_t )^2,
\end{equation} 
in complete agreement with the explicit diagrammatic evaluation of the disorder induced corrections
to the scattering amplitudes\cite{finkelstein1983,castellani1984}.
By writing  the correction to the conductivity in  eq.(\ref{conductivity12}) in terms of $\gamma_t$,
the dependence on $Z$ drops out. 
One gets
\begin{equation}
\label{rf19}
\frac{{\rm d} t}{{\rm d} s}=
t^2 \left[1 
+3\left(1 -\frac{1+\gamma_t}{\gamma_t}\ln (1+\gamma_t)
 \right)\right].
\end{equation}
By resuming the weak-localization contribution, one obtains a term identical to the singlet
contribution (the first one in the square brakets). The two terms although identical,
have therefore a complete different origin and this shows up in the presence of a
magnetic field which kills the weak-localization contribution and does not affect the
singlet one.
Equations (\ref{rf18}),(\ref{rf19}) together with eq.(\ref{rf15}) are the renormalization
group equations at one-loop order for the problem of interacting disordered systems at $d=2$.
Their analysis is the task for the next section.

\section{The Renormalization Group equations}
\label{sectionRG}
By resuming our maritime metaphor, we finally have land in sight. 
In the present section, we approach the end of our trip by examining the
consequences of the physical picture we have developed in the previous
sections and briefly compare\cite{dicastro1988} the results with the experiments.
To this end we discuss here in some detail 
the solution of the renormalization group (RG) equations for the
inverse conductance $t$, triplet scattering amplitude $\gamma_t$ and the
parameter $Z$. We begin our discussion with the general case when there is no
magnetic coupling in the system. In this case the RG equations read
\begin{eqnarray}
\frac{{\rm d} t}{{\rm d} s}&=&-\epsilon \frac{t}{2}+
t^2 \left[1 
+3\left(1 -\frac{1+\gamma_t}{\gamma_t}\ln (1+\gamma_t)
 \right)\right],\label{rg1}\\
 \frac{{\rm d} \gamma_t}{{\rm d} s}&=&
\frac{t}{2}(1+\gamma_t )^2,\label{rg2}\\
\frac{{\rm d} Z}{{\rm d} s}&=& -\frac{t}{2}Z(1-3\gamma_t),\label{rg3}
\end{eqnarray}
where in eq.(\ref{rg1}) we have added the contribution due to the bare dimension due to 
Ohm's law, $\epsilon =d-2$. The first observation is that eqs.(\ref{rg1}),(\ref{rg2}) do not
depend on $Z$. After solving for $t$ and $\gamma_t$ one may successively solve eq.(\ref{rg3})
for $Z$. 

Let us consider first the case $d=2$, i.e., $\epsilon =0$.  Equation (\ref{rg2}) for $\gamma_t$
implies a continuous growth. By integrating it between $s_0$ and $s$, one has
\begin{equation}
\label{rg4}
\frac{1}{1+\gamma_t (s)}=\frac{1}{1+\gamma_t (s_0)}-\frac{1}{2}
\int_{s_0}^{s}{\rm d}s't(s'),
\end{equation}
from which one sees that $\gamma_t$ diverges at a finite value, $s_c$, of the flow parameter:
\begin{equation}
\label{rg5}
1=\frac{1}{2}(1+\gamma_t (s_0) )
\int_{s_0}^{s_c}{\rm d}s't(s').
\end{equation}
The eq.(\ref{rg1}) for $t$ says that after an initial increase for not too large $\gamma_t (s_0)$,
the growth of $\gamma_t$ makes the triplet contribution, which is antilocalizing, the
dominating one. As a result, $t$ goes through a maximum.
In fig.\ref{rgflow} we show the RG flow in terms of the variable $t/(1+t)$ and $\gamma_t /(1+\gamma_t)$.
 For all the RG trajectories
$\gamma_t =\infty$ at some finite value $s_c$, which depends on the initial values.
Due to this, one cannot seriously trust the above equations 
quantitatively. Nevertheless, the physical indication of some type of 
ferromagnetic instability is rather clear due to the diverging spin
susceptibility associated with $\gamma_t$. 
The appearance of a finite length scale may indicate a formation of local magnetic moments
on the same scale.
Furthermore,  the
dominating antilocalizing effect of the triplet while $t$ remains finite
strongly supports the possibility of a 
metallic phase at low 
temperature\cite{castellani1984b,finkelstein1984,castellani1986,castellani1998,punnose2002}, 
in contrast with the non-interacting theory
based on WL only.
Indeed, this metallic phase in $d=2$ has recently been observed
(see refs. in \cite{abrahams2001}). In any case, both experimentally and theoretically, it is
not clear whether a possible ferromagnetic phase occurs before the transition to the
insulating phase. Almost all the experimental information is based on transport measurements
and the spin susceptibility is obtained indirectly. Very recently, a new method for
measuring directly the spin susceptibility in a two dimensional electron gas
has been invented and the first result suggests that, although there is a spin susceptibility
enhancement, no ferromagnetic instability is observed\cite{prus2003}. 
\begin{figure}[h]
\resizebox{.5\textwidth}{!}
{\includegraphics{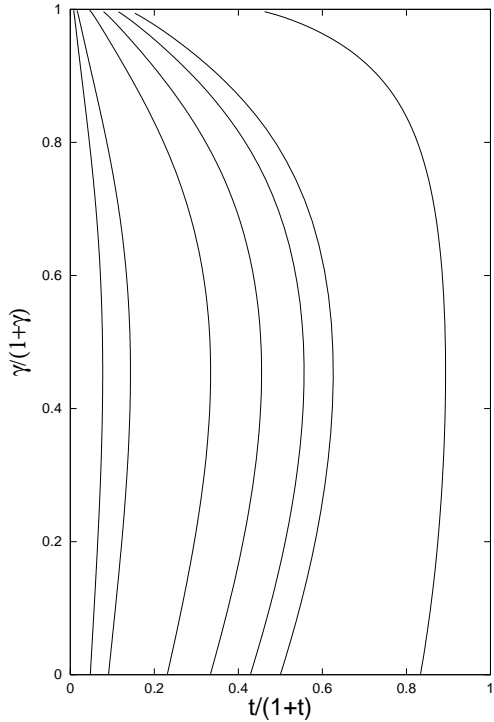}}
\caption{The RG flow, for $d=2$,
 in terms of the variable $t/(1+t)$ ($x$-axis) and $\gamma_t /(1+\gamma_t)$
($y$-axis).
In the figure the flow lines start on the $x$-axis with $\gamma_t =0$. For all of them
$\gamma_t =\infty$ at some finite value $s_c$, which depends on the initial values.}
\label{rgflow}
\end{figure}

In addition, due to the divergence of $\gamma_t$ also $Z$ goes to the strong coupling regime,
leading to an enhancement of the specific heat, which is however hardly
observable in two dimensions.
Close to the value $s_c$, one has from eq.(\ref{rg4}) that
\begin{equation}
\label{rg5bis}
\gamma_t \approx (s-s_c)^{-1},
\end{equation}
which together with the eq.(\ref{rg3}) for $Z$ gives 
\begin{equation}
\label{rg5ter}
Z\approx (s-s_c)^{-3},\\\ Z_t \approx (s-s_c)^{-4} .
\end{equation}
To convert the above behavior as function of temperature, one may reason in the
following way.  In general, the divergence of the length scale corresponds to a vanishing
temperature, as required by the diffusion law condition $L^2=D/T$. In the present case,
however, the parameter $Z$ renormalizes the temperature so that one has the
renormalized condition $L^2=D/(ZT)$. At finite length scale, the vanishing of $T$ is compensated
by the divergence of $Z$ in such a way that $T\approx (s-s_c)^{3}$. This implies for
the specific heat and spin susceptibility
\begin{equation}
\frac{C}{T} \sim T^{-1}, \\\\ \chi \sim T^{-4/3}.
\end{equation}
We may finally notice that the inclusion of the cooperon contribution would modify the above
group equations, without qualitative changes in the overall behavior (see ref.\cite{castellani1998} for details
).

In $d=3$ one has a richer scenario depending on the initial value of the running variables.
In the limit of large $\gamma_t$ and small $t$  the product $t \gamma_t$ obeys the
equation
\begin{equation}
\label{rg6}
\frac{{\rm d}(t  \gamma_t )}{{\rm d} s}=\gamma_t\frac{{\rm d} t}{{\rm d} s}+
t\frac{{\rm d} \gamma_t}{{\rm d} s}=\frac{t\gamma_t}{2}\left(t\gamma_t -\epsilon  \right),
\end{equation}
which has a fixed point for $t_c\gamma_{t,c} =\epsilon$. This condition  gives the
asymptotic expression for  a critical line in the $t-\gamma_t$ plane.
Close to this critical line, for large values of $\gamma_t$, one has the
approximate solution
\begin{eqnarray}
t(s)&=&t(s_0) e^{-\epsilon (s-s_0)/2},\label{rg7} \\
\gamma_t^{-1}(s)&=&\gamma_t^{-1}(s_0)-\frac{t(s_0)}{\epsilon}+\frac{t(s_0)}{\epsilon} 
e^{-\epsilon (s-s_0)/2}\label{rg8}.
\end{eqnarray}
One immediately sees from the above equations  that for low disorder, $t(s_0)< \epsilon /\gamma_t(s_0)$,
$t$ scales to zero and $\gamma_t$ scales to a finite value. In the high-disorder regime,
 $t(s_0)> \epsilon /\gamma_t(s_0)$, $\gamma_t$ diverges at a finite value of $s$ as in the two-dimensional
 case and $t$ stays finite.
 In fig.\ref{rgflow3} we report the flow obtained by
numerically integrating the RG equations.
\begin{figure}[h]
\resizebox{.5\textwidth}{!}
{\includegraphics{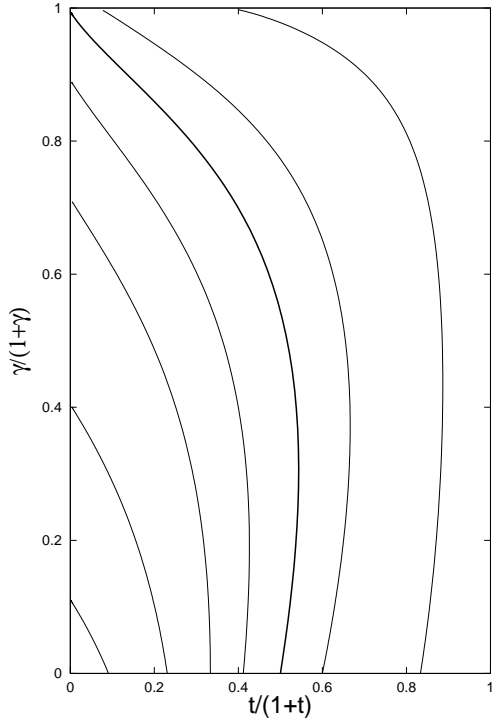}}
\caption{The RG flow, in $d=3$, in terms of the variable $t/(1+t)$ and $\gamma_t /(1+\gamma_t)$.
In the figure the flow lines start on the $x$-axis with $\gamma_t =0$. 
On the $x$-axis, there is value $t/(1+t)\sim 0.5$ below which the RG 
flows to a state with zero $t$ and finite
value of $\gamma_t$. The critical line originating from this value of $t$ is shown by
a thicker line. At large value of $\gamma_t$ this critical line is well described by
the approximate formula given in the text. 
For initial larger values of $t$, the RG flow is qualitatively similar to the two-dimensional
case.}
\label{rgflow3}
\end{figure}
We note, however, that  the strong-coupling runaway flow 
requires to go beyond the one-loop approximation we have presented here
leaving  open the issue whether this proposed scenario is realized or not.
An  approximate treatment of the two-loop correction is possible, but its 
discussion is well outside the scope of this paper. We refer the reader to 
ref. \cite{belitz1994}. 

 Along the critical line or nearby  in the low-disorder regime,
by converting to a length scale via $s=-\ln T\tau =-\ln (\tau D/L^2)\sim 2\ln (L/l)$, one has that,
 while $t$ vanishes, the conductivity stays finite
\begin{equation}
\sigma (L)=t^{-1}_c (L)L^{-\epsilon}.
\label{sigmafinite}
\end{equation}
 Equation (\ref{sigmafinite}) may be interpreted
in terms of a modification of the scaling law (\ref{scalinglaw}) due to the
scaling behavior of $t_c\sim L^{-x_t}$, i.e.,
\begin{equation}
\label{modifiedscalinglaw}
\mu =\nu (\epsilon -x_t).
\end{equation}
On the critical line $x_t=\epsilon$ and $\mu =0$. 
On the other hand $\gamma_t$ compensates the vanishing of $t$ and diverges like
\begin{equation}
\label{rg8bis}
\gamma_t \sim L^{\epsilon}.
\end{equation}
The equation for $Z$, in addition, gives now
\begin{equation}
\label{rg9}
\frac{{\rm d}Z}{{\rm d}s}\sim \frac{3}{2}\epsilon Z, \ \Rightarrow Z\sim L^{3\epsilon}.
\end{equation}
In contrast to the finite value of $\sigma$, according to the eqs. (\ref{rp18}),(\ref{rp18bis} ) the spin
and heat diffusion constants vanish, due to the divergenece of $Z_t$ and $Z$.
Higher order correction terms could modify this result leading to a vanishing conductivity also
in agreement with the experiments. Alternatively we can hypothize that the critical line does not
characterize the metal-insulator transition. The strong spin fluctuations which are the relevant physical
effect associated with $D_s\rightarrow 0$ ($\chi \rightarrow \infty$) lead instead to an instability
line before the localization takes place. In this case the system before reaching the instability,
should make a crossover to one of the universality classes with magnetic couplings, which will
be discussed later.
By assigning scaling exponents in terms of inverse length scale 
$x_Z =-3\epsilon$ and $x_{Z_2} =-4\epsilon$ to $Z$ and $Z_2$
and considering that the combination $ZT\sim L^{-(x_T+x_Z)}$ must scale as $Dq^2\sim L^{-2}$
since $D$ remains finite,
one gets
\begin{equation}
\label{rg10}
x_T =2-x_Z=2+3\epsilon,
\end{equation}
which yields for the specific heat and spin susceptibility
\begin{equation}
\label{rg11}
\frac{C}{T}\sim T^{-3\epsilon /(2+3\epsilon )}, \ \chi \sim  T^{-4\epsilon /(2+3\epsilon )}.
\end{equation}
At $\epsilon =1$, one has the temperature power laws $-3/5$ and $-4/5$.
Hence, as for the two-dimensional case, a clear prediction of the theory
is a low temperature enhancement of the specific heat and spin susceptibility,
the latter being generally stronger. A stronger enhancement of the spin susceptibility
has been indeed observed in $Si : P$\cite{paalanen1988}.
Theoretically, as mentioned before,
the divergence of the spin susceptibility at low temperature,
as predicted by the renormalization group flow, has led to the suggestion that the
system, because of the slowing-down of spin diffusion, tends to form regions of
localized magnetic moments, which would eventually drive the system into the
universality class of magnetic impurities\cite{finkelstein1984,castellani1984b}.
In a  number of
experimental papers\cite{alloul1987,lakner1989,lakner1994,schlager1997},
where the enhancements of specific heat and spin susceptibility have been
compared systematically,  some sort of an effective two-component system 
made of  localized and itinerant electrons
has been proposed to interpret  the data.


In the presence of any mechanism that inhibits the spin fluctuation enhancement,
the localizing term in the eq.(\ref{rg1}) for $t$ dominates and one has a metal-insulator transition
with $t_c\sim {\cal O}(\epsilon )$. For instance,
in the presence of a magnetic field only the ladders in the triplet channel with projection
 $M=\pm 1$ are suppressed. From eq.(\ref{th10}) for the correction to the thermodynamic
 potential, one sees immediately that the spin susceptibility is no longer singular, i.e.,
 $Z_t$ is invariant upon renormalization. By using eq.(\ref{rf17}) and eliminating the
 contribution of the triplet component with $M=\pm 1$ in  eq.(\ref{rg3}) for $Z$ one
 gets
 \begin{eqnarray}
 \frac{{\rm d}Z}{{\rm d}s}&=&-\frac{t}{2}Z\left( 1-\gamma_{t}\right),\label{bfieldZ}\\
 \frac{{\rm d}\gamma_t}{{\rm d}s}&=&\frac{t}{2}(1-\gamma_t^2)\label{bfieldgammat}. 
 \end{eqnarray}
 The above equations have a
 fixed point $\gamma_{t}^* =1$ with a constant $Z$. 
  It is now direct to obtain the equation for the parameter $t$.
After suppressing the $M=\pm 1$ triplet channel contributions in eq.(\ref{rg1}) and by
using the fixed-point condition  $\gamma_{t}^* =1$, one obtains 
\begin{equation}
\frac{{\rm d}t}{{\rm d}s}=-\epsilon \frac{t}{2}+(2-2\ln 2)t^2\label{bfieldt},
\end{equation}
which has a fixed point $t^* =\epsilon /(2(2-2\ln 2))$ and  gives $\mu =1$
as in the non-magnetic impurity case for the non-interacting system.

In the case of magnetic impurities or spin-orbit
scattering, we have seen in subsection \ref{spins} that only the ladder in the singlet channel
remains diffusive (cf. eqs.(\ref{spinsingletd}) and (\ref{orbitsingletd})). 

Since all triplet
channels are massive, $\gamma_t$ drops out in the equations for $Z$ and $t$, which now read
\begin{eqnarray}
\frac{{\rm d}t}{{\rm d}s}&=&-\epsilon \frac{t}{2}+t^2,\label{rgtspin}\\
\frac{{\rm d}Z}{{\rm d}s}&=&-\frac{t}{2}Z\label{rgZspin}.
\end{eqnarray}
The equation for $t$ gives the fixed point $t^*=\epsilon /2$ and conductivity scaling
exponent $\mu =1$.  By using the fixed point value for $t$ in the equation for $Z$, one
obtains
\begin{equation}
Z\approx e^{-(\epsilon /4 )\ln s}=T^{\epsilon / 4}\label{ZvsTspin}.
\end{equation}
Notice that the identical behavior for the cases of magnetic impurities and spin-orbit scattering
only holds when neglecting the contribution of the pure localization effects. When the latter is also
taken into account nothing happens for the magnetic impurities case, since all cooperon ladders
are massive (cf. eqs.(\ref{spinsingletc}) and (\ref{spintripletc})). For the spin-orbit case,
on the other hand, the ladder in the cooperon singlet channel is still diffusive and contributes
by minus one half to the standard localization term, as we have discussed at the end of subsection
\ref{spins}. Hence the combination of the antilocalizing contribution from pure interference
 with the localizing term due to interaction in the singlet channel does not change the qualitative
 behavior of $t$  and gives $\mu =1$, even though the fixed point value and the approach to it
 will differ giving $Z\sim T^{\epsilon /2}$. This is relevant in the 
 experiments\cite{okuma1988,nishida1984,katsumoto1987,katsumoto1988}
 (already discussed in points 2),3), and 4) of  subsect. \ref{experiments}), where
 a value of $\mu =1$ is observed both in the absence and presence of magnetic field.
 This is exactly what is predicted by the present theory of combined disorder and interaction
 effects, where a  magnetic field simply controls the contribution of the antilocalizing pure
 interference effect in the Cooper channel and changes the approach to the fixed point.
 Such a change is indeed observed in \cite{katsumoto1987}.
 From a theoretical point of view, we finally comment that in order  to perform
 a quantitative analysis in the presence of both diffuson and cooperon diffusive channels
 requires however the inclusion of the interaction
 amplitude $V_3$ in the  Cooper channel. For details we refer the reader to
 \cite{castellani1984c,finkelstein1990}.
  We hence see that all the
universality classes share the same conductivity exponent $\mu =1$, but differ as far as
the behavior of $Z$ (and hence of the specific heat) is concerned. To the best of our knowledge
there are no experiments available to check this last prediction.

We are at the end of our journey through the fascinating world of disordered electron
systems. It is time to draw a few conclusions. We have seen that the non-interacting theory
is not sufficient to interpret the existing experiments. 
In two dimensions the most relevant result is the prediction of the metallic phase,
which is observed to be suppressed by the magnetic field. However, a full account
of the experimental situation is far from being reached.
In three dimensions,
the predictions of the
theory of disordered interacting electron systems   agree with the experiments
whenever there is a magnetic coupling in the system and most of the
puzzles met while discussing the non-interacting case are resolved.
In the general case, with no
magnetic coupling present, although the strong enhancement for the specific heat and
spin susceptibility predicted by the theory appear to be confirmed by the experiments,
a deeper understanding is clearly needed and further theoretical and experimental work
is required with particular emphasis on the magnetic instability problem.

\acknowledgments
\noindent
R.~Raimondi acknowledges partial financial support from MIUR under grant COFIN 2002022534.
C.~ Di Castro acknowledges partial financial support from MIUR under grant COFIN 2001023848.

\appendix

\section{Useful integrals}
\label{integrals}
In the evaluation of diagrams, as we have seen, we leave the ${\bf q}$-integration
at the end, i.e, the integration over the momenta flowing in the ladder.
Due to the presence of the diffusive pole, which makes the small-${\bf q}$ region dominant,
the remaining integrals over the fast momenta can be expanded in powers of ${\bf q}$ and
$\omega$. To this end it is useful to expand the Green's function as
\begin{equation}
G({\bf p}+{\bf q},\epsilon+\omega )=G\left[1-(\omega -{\bf v}\cdot {\bf q})G+
 (\omega -{\bf v}\cdot {\bf q})^2G^2+...\right],
\end{equation}
where on the right-hand side $G=G({\bf p}, \epsilon )$.
Then, the integration over the fast momenta involves  integrals containing 
products of retarded and advanced Green's functions with the same argument. 
Finally, by using the residue theorem and the formula for the residue of poles we get
the useful formula
\begin{equation}
(m,n)=\sum_{\bf p}(G^R)^m (G^A)^n =
(-1)^m{\rm i}^{n+m}2\pi N_0 \frac{n (n+1)...(n+m-1)}{(m-1)!}\tau^{n+m-1}.
\end{equation}
The most frequent cases are
\begin{eqnarray}
(1,1)&=&2 \pi N_0 \tau \\
(2,2)&=&2 \ 2\pi N_0 \tau^3\\
(1,2)&=&{\rm i} \ 2\pi N_0 \tau^2\\
(2,1)&=&-{\rm i} \ 2\pi N_0 \tau^2\\
(1,3)&=&- \ 2\pi N_0 \tau^3\\
(3,1)&=&- \ 2\pi N_0 \tau^3\\
(1,4)&=&-{\rm i} \ 2\pi N_0 \tau^4\\
(4,1)&=&{\rm i} \ 2\pi N_0 \tau^4\\
(2,4)&=&-4 \ 2\pi N_0 \tau^5\\
(4,2)&=&-4 \ 2\pi N_0 \tau^5\\
(3,3)&=&6 \ 2\pi N_0 \tau^5.
\end{eqnarray}
Integrals containing scalar products are evaluated as
\begin{equation}
\sum_{\bf p}({\bf p}\cdot {\bf q})^2(G^R)^m (G^A)^n =
\frac{p_F^2 q^2}{d}\sum_{\bf p}(G^R)^m (G^A)^n .
\end{equation}

\section{Magnetic impurities}
\label{magneticimpurities}
To see this,
let us consider the following term in the Hamiltonian
\begin{equation}
\label{magneticdisorder}
H_{disorder}=\int {\rm d}~{\bf r}~
\psi^{\dagger}_{\alpha}( {\bf r})
\left[ u({\bf r})\delta_{\alpha \beta}+u_s({\bf r}){\bf S}\cdot \sigma_{\alpha\beta}\right]
\psi_{\beta} ({\bf r}),
\end{equation}
where ${\bf S}$ is the spin of a paramagnetic  impurity located at ${\bf r}$ having a scattering amplitude
$u_s({\bf r})$, whereas $u({\bf r})$ is the scattering amplitude due to non magnetic impurities
and already taken into account within the self-consistent Born approximation. The first step
amounts to recompute the Green's function in the presence of the full term eq.(\ref{magneticdisorder}).
One gets for the Green's function
\begin{equation}
\label{greenmagneticimpurities}
G^R({\bf p},\epsilon )=\left[\epsilon -\xi_{\bf p}+\frac{\rm i}{2}
\left( \frac{1}{\tau}+\frac{1}{\tau_s}\right) \right]^{-1},
\end{equation}
where $\tau^{-1} =2\pi N_0 {\overline { u^2}}({\bf r})$, 
$\tau_s^{-1} =2\pi N_0 {\overline { u_s^2}}({\bf r}){ {S(S+1)}}$
and for the single impurity line
\begin{equation}
\label{impuritylinemagnetic}
U_{\alpha\beta\gamma\delta}=\frac{1}{2\pi N_0 \tau}
\left[ \delta_{\alpha\beta}\delta_{\gamma\delta}+
\frac{\tau}{3\tau_s}\sigma_{\alpha\beta}\cdot \sigma_{\gamma\delta}\right],
\end{equation}
where the meaning of the spin greek indices is shown in fig.\ref{impurityline}.
\begin{figure}
\includegraphics{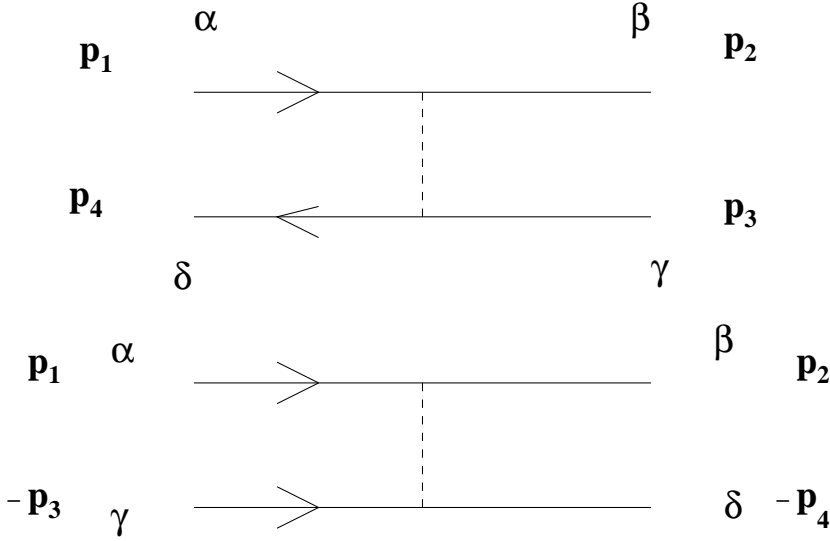}     
\caption{A single impurity line with the spin and momentum structure which are related to the diffuson and
cooperon ladder. (See fig.\ref{cooperonladder}).}
\label{impurityline}
\end{figure}
To find the expression for the diffuson and cooperon ladders in the presence
of magnetic impurities, it is convenient to exploit the conservation of the total
spin of the pair (particle-hole for the diffuson ($\alpha -\delta$) and particle-particle for the cooperon
($\alpha +\gamma$)).
To this end we use the  singlet and triplet spin projection operator, which for the  diffuson are given  by
\begin{eqnarray}
\label{spinoperatorsdiffuson}
S^{p-h}_{\alpha\beta\gamma\delta}&=&\frac{1}{2}\delta_{\alpha\delta}\delta_{\beta\gamma}
= \frac{1}{4}\left[ \delta_{\alpha\beta}\delta_{\gamma\delta}+
\sigma_{\alpha\beta}\cdot \sigma_{\gamma\delta}\right],\\
T^{p-h}_{\alpha\beta\gamma\delta}&=&\frac{1}{2}\sigma_{\alpha\delta}\cdot \sigma_{\beta\gamma}
=\frac{1}{4}\left[3 \delta_{\alpha\beta}\delta_{\gamma\delta}-
\sigma_{\alpha\beta}\cdot \sigma_{\gamma\delta}\right],
\end{eqnarray}
and for the cooperon by inverting one of the Green's function:
\begin{eqnarray}
\label{spinoperatorscooperon}
S^{p-p}_{\alpha\beta\gamma\delta}&=& \frac{1}{4}\left[ \delta_{\alpha\beta}\delta_{\gamma\delta}-
\sigma_{\alpha\beta}\cdot \sigma_{\gamma\delta}\right],\\
T^{p-p}_{\alpha\beta\gamma\delta}&=&\frac{1}{4}\left[3 \delta_{\alpha\beta}\delta_{\gamma\delta}+
\sigma_{\alpha\beta}\cdot \sigma_{\gamma\delta}\right],
\end{eqnarray}
where $S$ and $T$ stand for singlet and triplet, respectively.
If we indicate with the $L^{(0)S,T}$ and $L^{(0)}_{c}$ the 
 the single-impurity line contribution for both
singlet (S) and triplet (T) components,   
\begin{eqnarray}
L^{(0)S} = Tr(US^{p-h})&=&\frac{1}{2\pi N_0 \tau}\left(1+\frac{\tau}{\tau_s} \right),\\
L^{(0)T}= Tr(UT^{p-h})&=&\frac{1}{2\pi N_0 \tau}\left(1-\frac{\tau}{3\tau_s} \right),\\
L^{(0)S}_{c}= Tr(US^{p-p})&=&\frac{1}{2\pi N_0 \tau}\left(1-\frac{\tau}{\tau_s} \right),\\
L^{(0)T}_{c}=Tr(UT^{p-p})&=&\frac{1}{2\pi N_0 \tau}\left(1+\frac{\tau}{3\tau_s} \right),
\end{eqnarray}
one has for  $L^{S,T}$ and $L^{S,T}_{c}$ the  diffuson  and cooperon  ladders 
\begin{equation}
\label{magneticladders}
L^{S,T}=\frac{L^{(0)S,T}}{1-\Pi L^{(0)S,T}}, \ 
L^{S,T}_{c}=\frac{L^{(0)S,T}_{c}}{1-\Pi L^{(0)S,T}_{c}}.
\end{equation}
The trace sign implies summation over four spin indices. 
The quantity $\Pi$ is defined as in eq.(\ref{fundamentalintegral})
with the appropriate Green's function for this case. Its evalution leads to eq.(\ref{fundamental3})
with $1/\tau$ replaced by $1/ \tau +1/\tau_s\approx 1/\tau (1-\tau /\tau_s )$.

\section{Spin-orbit scattering}
\label{spinorbitscattering}
In the presence of spin-orbit interaction, in the scattering Hamiltonian one adds a term
\begin{equation}
H_{SO}=
\frac{\hbar}{4m^2c^2}\sum_{\bf R}
\psi_{\alpha}^{\dagger}({\bf r})\left[
\sigma_{\alpha\beta}\cdot \nabla V({\bf r}-{\bf R})\wedge {\bf p}\right]
\psi ({\bf r})
\end{equation}
where ${\bf R}$ indicates a ion site and $V({\bf r}-{\bf R})$ is the corresponding potential.
The matrix element between states of momentum ${\bf p}$ and ${\bf p}'$ is
\begin{eqnarray}
<{\bf p}|H_{SO}|{\bf p}'>&=&\frac{\hbar}{4m^2c^2}\sum_{\bf R}\sigma_{\alpha\beta}\cdot
\int {\rm d}{\bf r}~ e^{-{\rm i}{\bf p}\cdot {\bf r}}\left[\nabla V({\bf r}-{\bf R})\wedge {\bf p}\right]
e^{{\rm i}{\bf p}'\cdot {\bf r}}\\
&\equiv&{\rm i}u_{so}\sum_{\bf R}e^{{\rm i}({\bf p}-{\bf p}')\cdot {\bf R}}
\sigma_{\alpha\beta}\cdot \frac{({\bf p}\wedge {\bf p}')}{p_F^2}.
\end{eqnarray}

Since the scattering depends on the momenta of the particles involved, one has to consider
separately the contribution to the self-energy and to the single-impurity line in a ladder
resummation. For the self-energy one has
\begin{eqnarray}
\Sigma_{so\alpha\beta}({\bf p},\epsilon )&=&-{\overline{ u^2}}_{so}
\sum_{\bf p'} \frac{({\bf p}\wedge {\bf p}')}{p_F^2} \cdot \sigma_{\alpha\gamma}
 \frac{({\bf p}'\wedge {\bf p})}{p_F^2}\cdot \sigma_{\gamma\beta} G({\bf p'},\epsilon )\\
&\equiv&-{\rm i}\frac{sign (\epsilon )}{2\tau_{so}}\delta_{\alpha\beta}
\end{eqnarray}
so that the Green's function reads
\begin{equation}
G^R({\bf p},\epsilon )=\left[\epsilon -\xi_{\bf p}+\frac{\rm i}{2}
\left( \frac{1}{\tau}+\frac{1}{\tau_{so}}\right) \right]^{-1}.
\end{equation}
For the contribution of a single impurity line one has, by performing the angle average (indicated with a bar) over
${\bf p}$ and ${\bf p'}$,
\begin{equation}
U_{\alpha\beta\gamma\delta}({\bf p}_1,{\bf p}_2,{\bf p}_3,{\bf p}_4)=
-{\overline{ u^2}}_{so}\overline{\frac{({\bf p}_1\wedge {\bf p}_2)}{p_F^2}\cdot \sigma_{\alpha\beta}
\frac{({\bf p}_3\wedge {\bf p}_4)}{p_F^2}\cdot \sigma_{\gamma\delta}}
\end{equation}
with the condition that the total momentum is conserved, ${\bf p}_1+{\bf p}_3={\bf p}_2+{\bf p}_4$.
In the case of the diffuson ladder one has ${\bf p}_1\sim {\bf p}_4$, while for the
cooperon ${\bf p}_1\sim -{\bf p}_3$ (See fig.\ref{impurityline}). As a result
\begin{eqnarray}
U^{p-h}_{\alpha\beta\gamma\delta}&=&\frac{1}{2\pi N_0 \tau_{so}}
\sigma_{\alpha\beta}\cdot \sigma_{\gamma\delta}\\
U^{p-p}_{\alpha\beta\gamma\delta}&=&-\frac{1}{2\pi N_0 \tau_{so}}
\sigma_{\alpha\beta}\cdot \sigma_{\gamma\delta}.
\end{eqnarray}
As in the case of magnetic impurities one calculates the single-impurity line contribution
\begin{eqnarray}
L^{(0)S}&=&\frac{1}{2\pi N_0 \tau}\left(1+\frac{\tau}{\tau_{so}} \right)\\
L^{(0)T}&=&\frac{1}{2\pi N_0 \tau}\left(1-\frac{\tau}{d\tau_{so}} \right)\\
L^{(0)S}_{c}&=&\frac{1}{2\pi N_0 \tau}\left(1+\frac{\tau}{\tau_{so}} \right)\\
L^{(0)T}_{c}&=&\frac{1}{2\pi N_0 \tau}\left(1-\frac{\tau}{d\tau_{so}} \right)
\end{eqnarray}
In the diagram giving the weak localization correction, in general
the spin structure has the form $\delta_{\alpha\delta}\delta_{\gamma\beta}$.
By using the projection operators developed in the previous Appendix, one
can separate the singlet and triplet contribution to the weak-localization
correction in the Cooper channel
\begin{eqnarray}
Tr(S^{p-p}_{\alpha\beta\gamma\delta}\delta_{\alpha\delta}\delta_{\gamma\beta})&=&-1\\
Tr(T^{p-p}_{\alpha\beta\gamma\delta}\delta_{\alpha\delta}\delta_{\gamma\beta})&=&3.
\end{eqnarray}
We see that, while the triplet is localizing,  the singlet has an opposite effect.
In the absence of spin-orbit scattering both the singlet and triplet are massless and
 since the triplet contribution is three times larger, its effect prevails. In the presence of
 spin-orbit scattering, the triplet becomes massive and does not contribute to the logarithmic singularity
 term in $d=2$. The latter therefore changes sign and becomes antilocalizing. 
\section{The long-range interaction case}
\label{Appendixlongrange}
The effective screened interaction is given by
\begin{equation}
\label{densityofstates9}
V^R( q,\omega)=\frac{V_C(q)}{1+V_C(q) K^{00}(q,\omega)}=
\frac{2\pi e^2}{q}\frac{Dq^2-{\rm i}\omega}{Dq^2+Dq\kappa-{\rm i}\omega},
\end{equation}
where we used the two-dimensional expressions for the $V_C=2\pi e^2/q$ and the Thomas-Fermi
inverse screening length $\kappa =4\pi e^2N_0$. In the energy and momentum region
given by $\omega <Dq\kappa$ and $Dq^2<\omega$, the correction to the density of states
becomes
\begin{eqnarray}
\label{densityofstates10}
\delta N (\epsilon )&=&
\frac{1}{2\pi}{\cal I}m 
\int^{\infty}_{-\infty}{\rm d}\omega~
f(\omega -\epsilon )\sum_{\omega /D\kappa<|{\bf q}|<\sqrt{\omega /D}}
\frac{1}
{Dq^2(Dq^2-{\rm i}\omega )}\\
&=&-\frac{t}{4}\ln \left(|\epsilon |\tau \right)
\ln \left(\frac{|\epsilon |}{\tau D^2\kappa^4} \right). 
\end{eqnarray}
We see that the correction becomes log-square! This is a peculiar feature of the
single-particle density of states. In fact, all other physical quantities that we deal with
in these lecture notes acquire logarithmic corrections only even in the presence
of Coulomb interaction. 
As we shall see in Appendix \ref{subsectionladder}, the density of states can be reabsorbed
into the definition of the scattering amplitudes and drops out from the renormalization group
equations.
\section{Details on the evaluation of the interaction correction to the conductivity}
\label{Appendixconductivity}
In this Appendix, we show how to obtain the correction to the electrical conductivity
due to the combined effect of disorder and interaction.
The diagrams contributing to electrical conductivity to lowest order in the
interaction are shown in fig.\ref{conductivitydiagrams}. Diagrams (a) and (d) are 
obtained by inserting a self-energy correction into the Green's function and one
has to consider also the symmetric ones with the self-energy insertion
in the bottom electron line. Diagrams (b) and (c) are due to vertex corrections.
We begin our discussion with diagrams (a) and (b). 
 The extension to diagrams (c) and (d) is straightforward.
\begin{figure}
\includegraphics{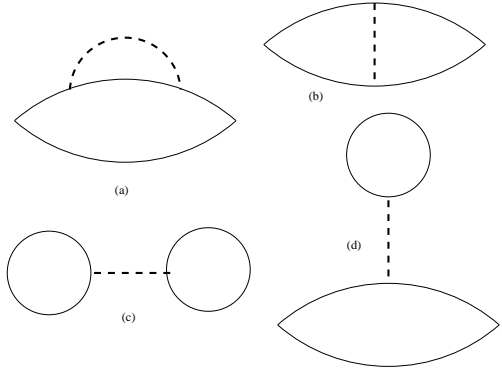}     
\caption{Diagrams for the conductivity to the lowest order in the interaction.
(a) and (b) Exchange, (c) and (d) Hartree. Interaction is shown as a thick dashed line.}
\label{conductivitydiagrams}
\end{figure}
The expression for diagram (a) reads
\begin{eqnarray}
\label{conductivity1}
R^{ij}_{(a)}({\bf r},{\bf r}';\Omega )&=& 2T\sum_{\epsilon_n }
\gamma^i ({\bf r})\gamma^j({\bf r}')
\int{\rm d}{\bf r}_1{\rm d}{\bf r}_2
 T\sum_{\omega_m}{\cal V}({\bf r}_1,{\bf r}_2; \omega_{m})
 \nonumber\\
&& {\cal G }( {\bf r}, {\bf r}_1; \epsilon_n +\Omega_{\nu})
{\cal G}({\bf r}_1,{\bf r}_2; \epsilon_n +\Omega_{\nu} -\omega_m)
{\cal G}({\bf r}_2,{\bf r}'; \epsilon_n +\Omega_{\nu}){\cal G}({\bf r'},{\bf r}; \epsilon_n).
\end{eqnarray}
To this expression one has to add the one corresponding to having the
interaction line in the bottom electron line.
$\gamma^i ({\bf r})$ is the real-space representation of the current vertices.
We do not need to write here its explicit expression since at the end, after the impurity
average, we recover translational invariance and go back to the momentum representation.
Diagram (b) is instead given by
\begin{eqnarray}
\label{conductivity2}
R^{ij}_{(b)}({\bf r},{\bf r}';\Omega )&=& 2T\sum_{\epsilon_n }
\gamma^i ({\bf r})\gamma^j({\bf r}')\int{\rm d}{\bf r}_1{\rm d}{\bf r}_2
 {\cal G }( {\bf r}, {\bf r}_1; \epsilon_n +\Omega_{\nu})
{\cal G}({\bf r}_2,{\bf r}; \epsilon_n)
\nonumber\\
&&
T\sum_{\omega_m}
{\cal G } ( {\bf r}_1, {\bf r}'; \epsilon_n +\Omega_{\nu}-\omega_m)
{\cal G}({\bf r'},{\bf r}_2; \epsilon_n  -\omega_m)
{\cal V}({\bf r}_1,{\bf r}_2; \omega_{m}).
\end{eqnarray}
In both the eqs.(\ref{conductivity1}),(\ref{conductivity2}), the Matsubara
frequency sums  are transformed to contour integrals in the complex plane
by means of standard manipulations. Then one gets for diagram (a)
(including the other diagram with the two electron lines interchanged)
\begin{eqnarray}
\label{conductivity3}
R^{ij}_{(a)}({\bf r},{\bf r}';\Omega )&=& -2
\gamma^i ({\bf r})\gamma^j({\bf r}')
\int{\rm d}{\bf r}_1{\rm d}{\bf r}_2
\int_{-\infty}^{\infty}\frac{{\rm d}\epsilon}{2\pi{\rm i}}
\int_{-\infty}^{\infty}\frac{{\rm d}\omega}{2\pi{\rm i}}\left[ \right.\nonumber\\
&&  f (\epsilon-\Omega )b(\omega) (V^R -V^A)_{\omega}
G^R_{\epsilon}G^R_{\epsilon -\omega}G^R_{\epsilon} (G^R_{\epsilon -\Omega}-
G^A_{\epsilon -\Omega})\nonumber\\
&+&f (\epsilon-\Omega )f(\omega -\epsilon) V^R_{\omega}
G^R_{\epsilon}( G^R_{\epsilon -\omega}-G^A_{\epsilon -\omega}) 
G^R_{\epsilon} (G^R_{\epsilon -\Omega}-
G^A_{\epsilon -\Omega})\nonumber\\
&+&f (\epsilon )b(\omega) (V^R -V^A)_{\omega}
G^R_{\epsilon}G^R_{\epsilon -\omega}G^R_{\epsilon} 
G^A_{\epsilon -\Omega}\nonumber\\
&+&f (\epsilon )f(\omega -\epsilon) V^R_{\omega}
G^R_{\epsilon} (G^R_{\epsilon -\omega}-G^A_{\epsilon -\omega} )G^R_{\epsilon} 
G^A_{\epsilon -\Omega}\nonumber\\
&-&f (\epsilon )b(\omega) (V^R -V^A)_{\omega}
G^A_{\epsilon}G^A_{\epsilon -\omega}G^A_{\epsilon} 
G^A_{\epsilon -\Omega}\nonumber\\
&-&f (\epsilon )f(\omega -\epsilon ) V^A_{\omega}
G^A_{\epsilon} (G^R_{\epsilon -\omega} -G^A_{\epsilon -\omega}) G^A_{\epsilon} 
G^A_{\epsilon -\Omega}\nonumber\\
&+&  f (\epsilon-\Omega )b(\omega) (V^R -V^A)_{\omega}
G^R_{\epsilon}G^R_{\epsilon -\omega -\Omega}G^R_{\epsilon -\Omega} G^R_{\epsilon -\Omega}\nonumber\\
&+&f (\epsilon-\Omega )f(\omega -\epsilon +\Omega) V^R_{\omega}
G^R_{\epsilon}  G^R_{\epsilon -\Omega} 
(G^R_{\epsilon -\omega -\Omega} -G^A_{\epsilon -\omega -\Omega }) G^R_{\epsilon -\Omega}\nonumber\\
&-&f (\epsilon -\Omega )b(\omega) (V^R -V^A)_{\omega}
G^R_{\epsilon} G^A_{\epsilon -\Omega}G^A_{\epsilon -\omega -\Omega} 
G^A_{\epsilon -\Omega}\nonumber\\
&-&f (\epsilon -\Omega )f(\omega -\epsilon +\Omega) V^A_{\omega}
G^R_{\epsilon} G^A_{\epsilon -\Omega} (G^R_{\epsilon -\omega -\Omega} -G^A_{\epsilon -\omega} ) 
G^A_{\epsilon -\Omega}\nonumber\\
&+&f (\epsilon )b(\omega) (V^R -V^A)_{\omega}
( G^R_{\epsilon }- G^A_{\epsilon}) G^A_{\epsilon -\Omega }G^A_{\epsilon -\omega -\Omega} 
G^A_{\epsilon -\Omega}\nonumber\\
&+&\left.f (\epsilon )f(\omega -\epsilon  +\Omega ) V^A_{\omega}
 ( G^R_{\epsilon }- G^A_{\epsilon}) G^A_{\epsilon -\Omega}
 ( G^R_{\epsilon -\omega -\Omega} -G^A_{\epsilon -\omega -\Omega }) 
G^A_{\epsilon -\Omega}\right].
\end{eqnarray}
Diagram (b) gives
\begin{eqnarray}
\label{conductivity4}
R^{ij}_{(b)}({\bf r},{\bf r}';\Omega )&=& -2
\gamma^i ({\bf r})\gamma^j({\bf r}')
\int{\rm d}{\bf r}_1{\rm d}{\bf r}_2
\int_{-\infty}^{\infty}\frac{{\rm d}\epsilon}{2\pi{\rm i}}
\int_{-\infty}^{\infty}\frac{{\rm d}\omega}{2\pi{\rm i}}\left[ \right.\nonumber\\
&&f(\epsilon )b (\omega ) (V^R-V^A)_{\omega } G^R_{\epsilon +\Omega}G^R_{\epsilon}
G^R_{\epsilon +\Omega-\omega}G^R_{\epsilon -\omega}\nonumber\\
&&f(\epsilon )f (\omega -\epsilon) V^R)_{\omega } G^R_{\epsilon +\Omega}G^R_{\epsilon}
G^R_{\epsilon +\Omega-\omega}
(G^R_{\epsilon -\omega}-G^A_{\epsilon -\omega})\nonumber\\
&&f(\epsilon )f (\omega -\epsilon -\Omega) V^R_{\omega } G^R_{\epsilon +\Omega}G^R_{\epsilon}
(G^R_{\epsilon +\Omega-\omega}-G^A_{\epsilon +\Omega-\omega})G^A_{\epsilon -\omega}\nonumber\\
&-&f(\epsilon )b (\omega ) (V^R-V^A)_{\omega } G^R_{\epsilon +\Omega}G^A_{\epsilon}
G^R_{\epsilon +\Omega-\omega}G^A_{\epsilon -\omega}\nonumber\\
&-&f(\epsilon )f (\omega -\epsilon) V^A_{\omega } G^R_{\epsilon +\Omega}G^A_{\epsilon}
G^R_{\epsilon +\Omega-\omega}
(G^R_{\epsilon -\omega}-G^A_{\epsilon -\omega})\nonumber\\
&-&f(\epsilon )f (\omega -\epsilon -\Omega) V^R_{\omega } G^R_{\epsilon +\Omega}G^A_{\epsilon}
(G^R_{\epsilon +\Omega-\omega}-G^A_{\epsilon +\Omega-\omega})G^A_{\epsilon -\omega}\nonumber\\
&+&f(\epsilon +\Omega)b (\omega ) (V^R-V^A)_{\omega } G^R_{\epsilon +\Omega}G^A_{\epsilon}
G^R_{\epsilon +\Omega-\omega}G^A_{\epsilon -\omega}\nonumber\\
&+&f(\epsilon +\Omega)f (\omega -\epsilon) V^A_{\omega } G^R_{\epsilon +\Omega}G^A_{\epsilon}
G^R_{\epsilon +\Omega-\omega}(G^R_{\epsilon -\omega}-G^A_{\epsilon -\omega})\nonumber\\
&+&f(\epsilon +\Omega)f (\omega -\epsilon -\Omega) V^R_{\omega } G^R_{\epsilon +\Omega}
G^A_{\epsilon}(G^R_{\epsilon +\Omega-\omega}-G^A_{\epsilon +\Omega-\omega})
G^A_{\epsilon -\omega}\nonumber\\
&-&f(\epsilon +\Omega)b (\omega ) (V^R-V^A)_{\omega } G^A_{\epsilon +\Omega}G^A_{\epsilon}
G^A_{\epsilon +\Omega-\omega}G^A_{\epsilon -\omega}\nonumber\\
&-&f(\epsilon +\Omega)f (\omega -\epsilon) V^A_{\omega } G^A_{\epsilon +\Omega}G^A_{\epsilon}
G^R_{\epsilon +\Omega-\omega}(G^R_{\epsilon -\omega}-G^A_{\epsilon -\omega})\nonumber\\
&-&\left. f(\epsilon +\Omega)f (\omega -\epsilon -\Omega) V^A_{\omega } G^A_{\epsilon +\Omega}
G^A_{\epsilon}(G^R_{\epsilon +\Omega-\omega}-G^A_{\epsilon +\Omega-\omega})G^A_{\epsilon -\omega}\right].
\end{eqnarray}

Since we are discussing the frequency structure, in eqs.(\ref{conductivity3}) 
and (\ref{conductivity4}) we have dropped the explicit dependence
on space coordinates. 
 The Green's functions are presented in the
same order as in eqs.(\ref{conductivity1}) and (\ref{conductivity2}) where the space
dependence is shown.  In eqs.(\ref{conductivity3}) and (\ref{conductivity4}) we may now
perform the impurity average. 
First, averaging impurity pairs belonging to the same Green's function line implies
the replacement of the Green's function with its self-consistent Born
approximation expression. Secondly, we have to perform the average of impurity pairs
belonging to different Green's function lines. 
This can be performed by arranging the Green's functions on the sides
of a square and inserting ladders wherever possible.
At the leading order in the expansion parameter, we neglect, as a rule, all the diagrams
in which a crossing of impurity lines occurs.
Depending on the sequence of retarded and advanced Green's functions around the sides
of the square, we may insert two or three ladders. For instance, terms with four retarded Green's functions
give zero since all poles lie on the same side of the real axis. 
In eq.(\ref{conductivity3}), the terms that allow two or three ladder insertions are
the second, fourth, sixth, eigthth, tenth, and twelfth: 
\begin{eqnarray}
\label{conductivity5}
R^{ij}_{(a)}({\bf 0},\Omega )&=& -2\sum_{\bf q}
\int_{-\infty}^{\infty}\frac{{\rm d}\epsilon}{2\pi{\rm i}}
\int_{-\infty}^{\infty}\frac{{\rm d}\omega}{2\pi{\rm i}}\left[ \right.\nonumber\\
&+&f (\epsilon-\Omega )f(\omega -\epsilon) V^R_{\omega}({\bf q})\overline{
\gamma^i \gamma^j G^R_{\epsilon}( G^R_{\epsilon -\omega}-G^A_{\epsilon -\omega}) 
G^R_{\epsilon} (G^R_{\epsilon -\Omega}-
G^A_{\epsilon -\Omega})}\nonumber\\
&+&f (\epsilon )f(\omega -\epsilon) V^R_{\omega}({\bf q})
\overline{\gamma^i \gamma^j
G^R_{\epsilon} (G^R_{\epsilon -\omega}-G^A_{\epsilon -\omega} )G^R_{\epsilon} 
G^A_{\epsilon -\Omega}}\nonumber\\
&-&f (\epsilon )f(\omega -\epsilon ) V^A_{\omega}({\bf q})\overline{ \gamma^i \gamma^j
G^A_{\epsilon} (G^R_{\epsilon -\omega} -G^A_{\epsilon -\omega}) G^A_{\epsilon} 
G^A_{\epsilon -\Omega}}\nonumber\\
&+&f (\epsilon-\Omega )f(\omega -\epsilon +\Omega) V^R_{\omega}({\bf q}) \overline{
\gamma^i \gamma^j 
G^R_{\epsilon}  G^R_{\epsilon -\Omega} 
(G^R_{\epsilon -\omega -\Omega} -G^A_{\epsilon -\omega -\Omega }) G^R_{\epsilon -\Omega}}\nonumber\\
&-&f (\epsilon -\Omega )f(\omega -\epsilon +\Omega) V^A_{\omega}
\overline{
\gamma^i \gamma^j
G^R_{\epsilon} G^A_{\epsilon -\Omega} (G^R_{\epsilon -\omega -\Omega} -G^A_{\epsilon -\omega} ) 
G^A_{\epsilon -\Omega}}\nonumber\\
&+&\left.f (\epsilon )f(\omega -\epsilon  +\Omega ) V^A_{\omega}({\bf q})\overline{
\gamma^i \gamma^j 
( G^R_{\epsilon }- G^A_{\epsilon}) G^A_{\epsilon -\Omega}
 ( G^R_{\epsilon -\omega -\Omega} -G^A_{\epsilon -\omega -\Omega }) 
G^A_{\epsilon -\Omega}}\right].
\end{eqnarray}
In the same way, from eq.(\ref{conductivity4}) we pick up  the third and the eleventh term 
\begin{eqnarray}
\label{conductivity6}
R^{ij}_{(b)}({\bf 0},\Omega )&=& -2\sum_{\bf q}
\int_{-\infty}^{\infty}\frac{{\rm d}\epsilon}{2\pi{\rm i}}
\int_{-\infty}^{\infty}\frac{{\rm d}\omega}{2\pi{\rm i}}\left[ \right.\nonumber\\
&+&f(\epsilon )f (\omega -\epsilon -\Omega) V^R_{\omega }({\bf q})
\overline{\gamma^i \gamma^j G^R_{\epsilon +\Omega}G^R_{\epsilon}
(G^R_{\epsilon +\Omega-\omega}-G^A_{\epsilon +\Omega-\omega})G^A_{\epsilon -\omega}
}\nonumber\\
&-&\left. f(\epsilon +\Omega)f (\omega -\epsilon) V^A_{\omega }({\bf q}) 
\overline{\gamma^i \gamma^j G^A_{\epsilon +\Omega}G^A_{\epsilon}
G^R_{\epsilon +\Omega-\omega}(G^R_{\epsilon -\omega}-G^A_{\epsilon -\omega})
}\right].
\end{eqnarray}
The bar over the products of Green's function indicates the impurity average. 
Notice that in eqs.(\ref{conductivity5}) and (\ref{conductivity6}) the current vertices
appear under the impurity average bar. After the average and the restoration of traslational
invariance, there appear summations over momenta. The sum over the {\sl slow}
momenta that enter the interaction and the ladders are performed at the end, while
the sum over the {\sl fast} momenta entering the Green's functions are performed
with the help of residue theorem within the approximations explained in detail 
in  Appendix \ref{integrals}. To this end, all frequencies in the Green's function 
can be set to zero in the leading order in the diffusive regime expansion
$\omega \tau <1$. As a result, the impurity average of the product of Green's functions 
does not depend on the energy. Since, the ladders depend only on the {\sl slow}
frequencies $\omega$ and $\Omega$, we can perform the $\epsilon$-integration at once
by using the useful identity
\begin{equation}
\label{usefulidentity}
\int_{-\infty}^{\infty}{\rm d}~\epsilon ~f(\epsilon )~f(\omega -\epsilon )=
\frac{\omega}{2}\left(\coth \left( \frac{\omega}{2T}\right)-1\right)
\equiv F (\omega ).
\end{equation}
We may finally consider explicitly the impurity average.
To illustrate the procedure, let us consider the first term in eq.(\ref{conductivity5}).
It contains four products of four Green's functions each. The first product
$G^R_{\epsilon} G^R_{\epsilon -\omega} 
G^R_{\epsilon} G^R_{\epsilon -\Omega}$ gives zero upon averaging.
The term $G^R_{\epsilon} G^R_{\epsilon -\omega} 
G^R_{\epsilon}  G^A_{\epsilon -\Omega}$ vanishes because of the vector nature of the current vertices.
There remain the terms
$G^R_{\epsilon}G^A_{\epsilon -\omega} 
G^R_{\epsilon} G^R_{\epsilon -\Omega}$
and $G^R_{\epsilon}G^A_{\epsilon -\omega} 
G^R_{\epsilon}  G^A_{\epsilon -\Omega}$. Upon averaging, the first term
gives rise to an effective diagram with  three ladders, corresponding to 
(a)  of fig.\ref{effectiveconductivitydiagrams}.
The second term, on the other hand, yields  effective diagrams with two ladders only,
corresponding to (c) and (d) in fig.\ref{effectiveconductivitydiagrams}. 
\begin{figure}
\includegraphics{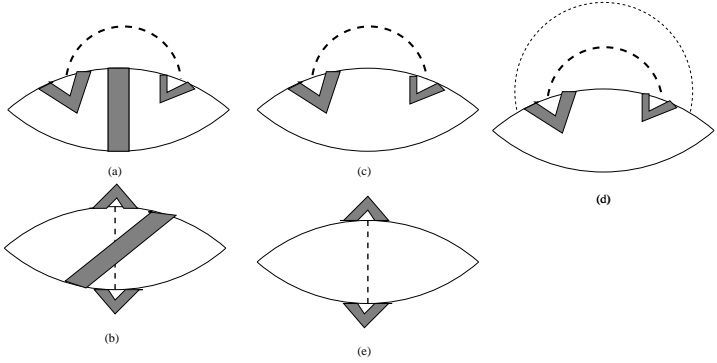}
\caption{Effective conductivity diagrams.}
\label{effectiveconductivitydiagrams}
\end{figure}
By following this line of reasoning, one obtains all the diagrams of 
fig.\ref{effectiveconductivitydiagrams}, including those obtained by interchanging the
top and bottom Green's function lines. One key point to notice is that the diagrams
with two ladders cancel each other, i.e., the sum of (c),(d), and (e).
This cancellation is shown in detail in ref.\cite{altshuler1980b}. 
To see it, let us consider the terms in eqs.(\ref{conductivity3}-\ref{conductivity4})
which contain the retarded interaction $V^R_{\omega}$ (a similar analysis can be done
for the terms containing $V^A_{\omega}$). 
By using eq.(\ref{usefulidentity}), 
we get from eqs.(\ref{conductivity5}-\ref{conductivity6})
\begin{eqnarray}
&&-2\sum_{\bf q}
\int_{-\infty}^{\infty}\frac{{\rm d}\omega}{2\pi{\rm i}}V^R_{\omega}
\left( F(\omega -\Omega )\left[\overline{\gamma^i \gamma^j G^R_{\epsilon}G^A_{\epsilon -\omega} 
G^R_{\epsilon} G^A_{\epsilon -\Omega}} -\overline{\gamma^i \gamma^j G^R_{\epsilon}G^A_{\epsilon -\omega} 
G^R_{\epsilon} G^R_{\epsilon -\Omega}}\right] \right. \nonumber\\
&&\left. -F(\omega)\overline{\gamma^i \gamma^j G^R_{\epsilon}G^A_{\epsilon -\omega} 
G^R_{\epsilon} G^A_{\epsilon -\Omega}}-
F(\omega)\overline{\gamma^i \gamma^j G^R_{\epsilon}G^A_{\epsilon -\omega} 
G^R_{\epsilon} G^R_{\epsilon -\Omega}}\right.\nonumber\\
&&\left.-F(\omega -\Omega )\overline{\gamma^i \gamma^j G^R_{\epsilon +\Omega}G^R_{\epsilon } 
G^A_{\epsilon +\Omega -\omega} G^A_{\epsilon -\omega}}\right)\nonumber,
\end{eqnarray}
which is readily seen to vanish by considering  the following averages:
\begin{eqnarray}
\overline{ \gamma^i \gamma^j G^R_{\epsilon}G^A_{\epsilon -\omega} 
G^R_{\epsilon} G^R_{\epsilon -\Omega}}&=&-\delta_{ij}\frac{v_F^2}{d} 
2\pi N_0 \tau^3 \left(\frac{1}{2\pi N_0\tau^2}\frac{1}{Dq^2-{\rm
i}\omega}\right)^2\nonumber \\
\overline{\gamma^i \gamma^j G^R_{\epsilon}G^A_{\epsilon -\omega} 
G^R_{\epsilon} G^A_{\epsilon -\Omega}}&=&\delta_{ij}\frac{v_F^2}{d}
 2\pi N_0 \tau^3 \left(\frac{1}{2\pi N_0\tau^2}\frac{1}{Dq^2-{\rm
i}\omega}\right)^2
\nonumber \\
\overline{\gamma^i \gamma^j G^R_{\epsilon +\Omega}G^R_{\epsilon } 
G^A_{\epsilon +\Omega -\omega} G^A_{\epsilon -\omega}}&=&
\delta_{ij}\frac{v_F^2}{d}4\pi N_0 \tau^3 \left(\frac{1}{2\pi N_0\tau^2}\frac{1}{Dq^2-{\rm
i}\omega}\right)^2. \nonumber 
\end{eqnarray}
As a final comment, we note that each individual diagram with two ladders, in the presence
of long-range forces, will suffer from the strong singularity as in the case of the
density of states, as discussed in Appendix \ref{Appendixlongrange}. However,
it has been shown that the singularity due to the long-range forces can be incorporated
into a gauge factor, which drops out in the evaluation of gauge-invariant quantities.
This is the origin of the cancellation of the diagrams with two ladders\cite{kopietz1998}.

Let us now turn our attention to the  diagrams with three ladders.
The  impurity average needed for   the 
diagram(a) of  fig.\ref{effectiveconductivitydiagrams} is
\begin{eqnarray}
\label{conductivity7}
\overline{
\gamma^i \gamma^j
G^R_{\epsilon}G^A_{\epsilon -\omega} 
G^R_{\epsilon} G^R_{\epsilon -\Omega}}&=& \sum_{\bf p}\gamma^i
G^R_{\epsilon}({\bf p})G^A_{\epsilon -\omega} ({\bf p})
 G^R_{\epsilon -\Omega}({\bf p -q})\nonumber\\
 && \sum_{\bf p'}\gamma^j
G^R_{\epsilon} ({\bf p'})G^A_{\epsilon -\omega} ({\bf p' })
 G^R_{\epsilon -\Omega}({\bf p' -q})
  \nonumber(\Gamma^0({\bf q},\omega))^2L({\bf q},\omega-\Omega)\nonumber\\
&=&  \frac{(-4\pi e N_0 D \tau^2 q^i)(-4\pi e N_0 D \tau^2 q^j)}{2\pi N_0 \tau^4}
  \frac{1}{(D{\bf q}^2-{\rm i}\omega)^2(D{\bf q}^2-{\rm i}(\omega -\Omega))}\nonumber\\
 &=&4\pi\sigma_0  \frac{D q^iq^j}{(D{\bf q}^2-{\rm i}\omega)^2(D{\bf q}^2-{\rm i}(\omega -\Omega))}
\end{eqnarray}
where the factors in round brackets, in the second line, arise from the integration of 
the three Green's functions with a current vertex. Notice that in diagram (b) 
of fig.\ref{effectiveconductivitydiagrams},
the two integrations over products of three Green's functions produce an  opposite sign.
This gives an overall minus sign for diagram (b) with respect to (a).
In the last line $\sigma_0 =2e^2 N_0 D$ is the Drude conductivity.
By collecting in eqs.(\ref{conductivity5}) (first, third, fourth, and sixth line)
 and (\ref{conductivity6}) all the terms  
giving rise to diagrams with three ladders we get
\begin{eqnarray}
\label{conductivity8}
R^{ij}({\bf 0},\Omega )&=&-4\sigma_0\sum_{\bf q}\int_{-\infty}^{\infty}~\frac{{\rm d}\omega}{2\pi}
\left[ \right.\nonumber\\
&&F(\omega -\Omega )V^R_{\omega}({\bf q})
\frac{D q^iq^j}{(D{\bf q}^2-{\rm i}\omega)^2(D{\bf q}^2-{\rm i}(\omega -\Omega))}\nonumber\\
&+&F(\omega  )~~V^A_{\omega}({\bf q})
\frac{D q^iq^j}{(D{\bf q}^2+{\rm i}\omega)^2(D{\bf q}^2+{\rm i}(\omega -\Omega))}\nonumber\\
&+&F(\omega  )~~V^R_{\omega}({\bf q})
\frac{D q^iq^j}{(D{\bf q}^2-{\rm i}\omega)^2(D{\bf q}^2-{\rm i}(\omega +\Omega))}\nonumber\\
&+&F(\omega +\Omega )V^A_{\omega}({\bf q})
\frac{D q^iq^j}{(D{\bf q}^2+{\rm i}\omega)^2(D{\bf q}^2+{\rm i}(\omega +\Omega))}\nonumber\\
&-&F(\omega -\Omega )V^R_{\omega}({\bf q})
\frac{D q^iq^j}{(D{\bf q}^2-{\rm i}\omega)^2(D{\bf q}^2-{\rm i}(\omega +\Omega))}\nonumber\\
&-&F(\omega -\Omega )V^R_{\omega}({\bf q})
\frac{D q^iq^j}{(D{\bf q}^2-{\rm i}\omega)^2(D{\bf q}^2-{\rm i}(\omega -\Omega))}\nonumber\\
&-&F(\omega +\Omega )V^A_{\omega}({\bf q})
\frac{D q^iq^j}{(D{\bf q}^2+{\rm i}\omega)^2(D{\bf q}^2+{\rm i}(\omega +\Omega))}\nonumber\\
&-&F(\omega +\Omega )V^A_{\omega}({\bf q})
\frac{D q^iq^j}{(D{\bf q}^2+{\rm i}\omega)^2(D{\bf q}^2+{\rm i}(\omega -\Omega))}\left. \right]
\end{eqnarray}
Notice that the first term cancels with the sixth and the fourth with the seventh.
By diving by $-{\rm i}\Omega$ and sending $\Omega$ to zero
we get eq.(\ref{conductivity9}) quoted in the text. 

\section{Details of the evaluation of the thermodynamic potential}
\label{Appendixthermo}
To begin with, let us consider, at fixed impurity configuration,
the first-order exchange interaction correction to the
thermodynamic potential (see the second diagram in fig. \ref{thermo}, without 
the inserted ladder)
\begin{equation}
\label{th1}
\Delta \Omega =\int_0^1\frac{{\rm d}\lambda}{\lambda}\frac{1}{2} T^2\sum_{\omega_m ,\epsilon_n}
\int {\rm d}{\bf r} {\rm d}{\bf r}'
 {\cal V}({\bf r}-{\bf r}', \omega_m, \lambda){\cal G}({\bf r},{\bf r}';\epsilon_n)
{\cal G}({\bf r}',{\bf r};\epsilon_n +\omega_m),
\end{equation}
where we have used the standard trick\cite{abrikosov1975} to multiply the interaction by a parameter
$0< \lambda <1$
$$
{\cal V}({\bf r}-{\bf r}', \omega_m, \lambda) =\lambda
{\cal V}({\bf r}-{\bf r}', \omega_m).
$$
The sum over the Fermionic Matsubara frequency gives
\begin{eqnarray}
\label{th2}
&&T\sum_{\epsilon_n}{\cal G}({\bf r},{\bf r}';\epsilon_n)
{\cal G}({\bf r}',{\bf r};\epsilon_n +\omega_m)|_{{\rm i}\omega_m =\omega +{\rm i}0^+}=\nonumber\\
&&\int_{-\infty}^{\infty}\frac{{\rm d}\epsilon }{2\pi {\rm i}}f(\epsilon )
\left[ (G^R({\bf r},{\bf r}';\epsilon)-G^A ({\bf r},{\bf r}';\epsilon) )
G^R ({\bf r}',{\bf r};\epsilon +\omega)\right.\nonumber\\
&&+\left. G^A ({\bf r},{\bf r}';\epsilon -\omega ) 
(G^R({\bf r}',{\bf r};\epsilon  )-G^A ({\bf r}',{\bf r};\epsilon ) )\right].
\end{eqnarray}
By performing the impurity average we need to keep only terms with both retarded and
advanced Green's functions. This selects in eq.(\ref{th2}) the term
\begin{eqnarray}
\label{th3}
&&\int_{-\infty}^{\infty}\frac{{\rm d}\epsilon }{2\pi {\rm i}}
f(\epsilon ) 
\left[ -\overline{G^A ({\bf r},{\bf r}';\epsilon) 
G^R ({\bf r}',{\bf r};\epsilon +\omega)}+ \overline{G^A ({\bf r},{\bf r}';\epsilon -\omega ) 
G^R({\bf r}',{\bf r};\epsilon  )}\right]\nonumber\\
&=&\int_{-\infty}^{\infty}\frac{{\rm d}\epsilon }{2\pi {\rm i}}
\left[f(\epsilon +\omega ) - f(\epsilon ) \right]
\overline{G^A ({\bf r},{\bf r}';\epsilon  ) 
G^R({\bf r}',{\bf r};\epsilon  +\omega)}\nonumber\\
&=&\frac{{\rm i}\omega}{2\pi}(2\pi N_0\tau)^2L({\bf q},\omega),
\end{eqnarray}
where, making use of the fact that the impurity average of the product of Green's functions
does not depend on the energy, we have performed the integration over the frequency.
As a result, eq.(\ref{th1}) becomes
\begin{equation}
\label{th4}
\Delta \Omega =-\int_0^1\frac{{\rm d}\lambda}{\lambda} T\sum_{\omega_m }
\sum_{\bf q}\frac{N_0{\cal V}({\bf q},\omega_m,\lambda )|\omega_m|}{D{\bf q}^2+|\omega_m|},
\end{equation}
from which, by taking $N_0{\cal V}({\bf q},\omega_m,\lambda )=\lambda (V_1-2V_2)$,
one obtains the eq.(\ref{th6}) of the main text.
Whence the dynamical amplitude resummation is inserted into the diagrams of 
fig.\ref{thermo} one obtains the final expressions (\ref{th10}),(\ref{thzeemanrenormalized})
for the thermodynamic potential.

\section{Ladder in the presence of Zeeman coupling}
\label{Appendixzeeman}
In the presence of a magnetic field, electron energies are changed by the Zeeman energy,
so that the Green's function reads
\begin{equation}
\label{greenmagneticfield}
G_{\alpha}^R({\bf p},\epsilon )=\left[\epsilon -\xi_{\bf p}+
\alpha\omega_s
+\frac{\rm i}{2\tau}
  \right]^{-1},
\end{equation}
where $\omega_s =g \mu_B B$ with $g$ the Land\`e factor and $\mu_B$ the Bohr magneton.
The spin projection takes values $\alpha =\pm 1/2$. One sees that the Zeeman energy
$\alpha \omega_s$ enters the Green's function as an energy in shift. This allows to
get immediately the ladder in the presence of magnetic Zeeman coupling, since the 
energy difference $\omega$ is shifted by the difference of the Zeeman energies of the
particle-hole pair. For instance, by making reference to the spin structure of
fig.\ref{impurityline} and taking into account the spin conservation along a Green's function
line, one has
\begin{equation}
\label{ladderzeeman}
L_{\alpha \delta}(q,\omega )=\frac{1}{2\pi N_0 \tau^2}\frac{1}{Dq^2-{\rm i}\omega 
-{\rm i}(\alpha -\delta )\omega_s }.
\end{equation}
This show that only the triplet components with total spin projection $\alpha -\delta \equiv M =\pm 1$
are affected by the magnetic field.
\section{The ladder renormalization}
\label{subsectionladder}
In this Appendix we show that the logarithmic corrections found for the
physical quantities can be absorbed into a renormalization of the parameters
characterizing the ladder propagator. This identification is the formal
basis of the renormalizability of the effective field theory, whose physical
meaning is discussed in the text via the Ward identities. 
Let us assume that the ladder, in the presence of interaction, gets renormalized
as
\begin{equation}
\label{rp1}
L(q,\omega)=\frac{1}{2\pi N_0\tau^2}\frac{1}{Dq^2-{\rm i}\omega}\rightarrow
\frac{1}{2\pi N_0\tau^2}\frac{\zeta^2}{D_Rq^2-{\rm i}Z\omega}
\end{equation}
where $\zeta$, $D_R$, and $Z$ represent the effective wave function renormalization,
the renormalized diffusion coefficient, and the renormalization of the frequency.
By expanding 
\begin{equation}
\label{rp2}
\frac{\zeta^2}{D_Rq^2-{\rm i}Z\omega}-\frac{1}{Dq^2-{\rm i}\omega}
=
\frac{(2\delta\zeta - \delta D /D )Dq^2-{\rm i}(2\delta\zeta - \delta Z )\omega}{(Dq^2-{\rm i}\omega )^2} 
\equiv
\frac{\Sigma_L ( q,\omega )}{(Dq^2-{\rm i}\omega )^2} 
\end{equation}
and the last equation defines the ladder self-energy.
\begin{figure}
\includegraphics{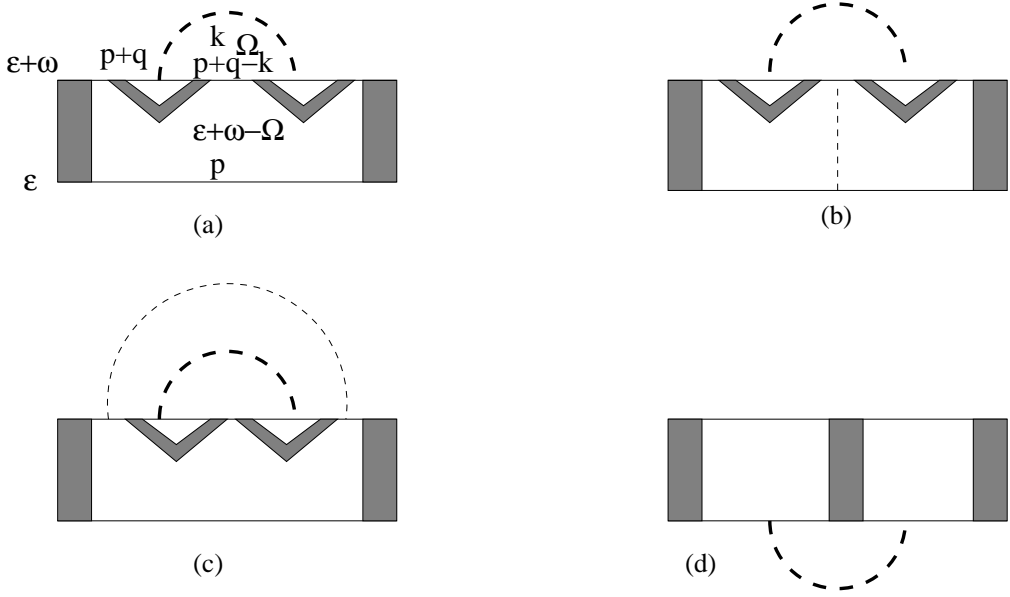}
\caption{Diagrams for the ladder self-energy. A similar set of diagrams is generated by the interchange
of the interaction line (tick-dashed line) between top and bottom Green's function lines.}
\label{ladderselfenergy}
\end{figure}
The diagrams for the self-energy are shown in fig.\ref{ladderselfenergy}.
The first step is the integration over the {\sl fast} momenta running within
the Green's functions. This integration amounts to the evaluation
of  several integrals of products
of the type $(G^R)^m(G^A)^n$, whose result is given in  Appendix \ref{integrals}.
 For diagrams (a-c), in the small $k$, $\Omega$, $q$ and $\omega$ limit, we obtain 
\begin{equation}
\label{rp3}
I_{abc}=(2\pi N_0 \tau )^2 
2\pi N_0 \tau^4\left[D (q^2+k^2)-{\rm i}(\omega +\Omega)-2{\bf q}\cdot {\bf k} \right],
\end{equation}
where the first factor $(2\pi N_0 \tau )^2 $ represents the two integrations over the two
Green's functions at the interaction vertices. The rest gives the integration over the remaining
Green's functions.
In a similar way, integration over Green's functions gives for diagram (d) 
\begin{equation}
\label{rp3bis}
I_d =-(2\pi N_0 \tau^2 )^3.
\end{equation}
The  diagrams (a),(b),(c) shown in fig.\ref{ladderselfenergy} finally yield 
\begin{equation}
\label{rp4}
\Sigma_{L,abc}=-2T\sum_{\epsilon_n+\omega_{\nu} <\Omega_m}\sum_{\bf k}
\frac{I_{abc} ({\bf k},{\bf q},\Omega_m , \omega_{\nu} ){\cal V}({\bf k},\Omega_m)}
{(2\pi N_0\tau^2)^3(Dk^2+|\Omega_m |)^2}
\end{equation}
and diagram (d)
\begin{equation}
\label{rp5}
\Sigma_{L,d}=2T\sum_{\epsilon_n <\Omega_m}\sum_{\bf k}
\frac{I_d {\cal V}({\bf k},\Omega_m)}
{(2\pi N_0 \tau^2)^3(D({\bf k}+{\bf q})^2+|\Omega_m +\omega_{\nu} |) }.
\end{equation}
In eqs.(\ref{rp4}-\ref{rp5}) we have used Matsubara frequencies.
The relative minus sign comes from the integration over the {\sl fast} momenta.
The factor of $2$ is due to the fact that there is another set of diagrams generated
by interchanging the interaction line between the two electron lines.
One may check that the sum of eqs.(\ref{rp4}-\ref{rp5}) vanishes in the limit
$q=0$ and $\omega =0$. For small, but finite external momentum and frequency,
we rewrite eq.(\ref{rp4}) in the form
\begin{eqnarray}
\Sigma_{L,abc}&=&-2T\sum_{\epsilon_n+\omega_{\nu} <\Omega_m}\sum_{\bf k}
\frac{{\cal V}({\bf k},\Omega_m)}{Dk^2+|\Omega |}\nonumber \\
&-&2T\sum_{\epsilon_n+\omega_{\nu} <\Omega_m}\sum_{\bf k}
\frac{(Dq^2+|\omega_{\nu} | ){\cal V}({\bf k},\Omega_m ) }{(Dk^2+|\Omega_m | )^2}
\label{rp6}
\end{eqnarray}
and eq.(\ref{rp5}) as
\begin{eqnarray}
\Sigma_{L,d}&=&2T\sum_{\epsilon_n <\Omega_m < \epsilon_n+\omega_{\nu}}\sum_{\bf k}
\frac{{\cal V}({\bf k},\Omega_m)}{D({\bf k}+{\bf q})^2+|\Omega_m +\omega_{\nu} | }\nonumber\\
&+&2T\sum_{\epsilon_n+\omega_{\nu} <\Omega_m }\sum_{\bf k}\left[
\frac{{\cal V}({\bf k},\Omega_m)}{D{\bf k}^2+|\Omega_m  |}+
\frac{{\cal V}({\bf k},\Omega_m)}{D({\bf k}+{\bf q})^2+|\Omega_m +\omega_{\nu} |}
-\frac{{\cal V}({\bf k},\Omega_m)}{D{\bf k}^2+|\Omega_m  | }\right]\label{rp7},
\end{eqnarray}
where  less divergent terms  have been neglected.
The first term in the square brakets of eq.(\ref{rp7}) cancels with the first term in eq.(\ref{rp6}).
Let us analyze the first term of eq.(\ref{rp7}). By transforming the Matsubara sum into an integral
in the complex plane and analytically continuing $\epsilon \rightarrow -{\rm i}\epsilon$
and $\epsilon +\omega\rightarrow -{\rm i}(\epsilon +\omega )$, it reads
\begin{eqnarray}
\Sigma_{L,d}^{1}&=&2T\sum_{\epsilon_n <\Omega_m < \epsilon_n+\omega_{\nu}}\sum_{\bf k}
\frac{{\cal V}({\bf k},\Omega_m)}{D({\bf k}+{\bf q})^2+| \Omega_m +\omega_{\nu}| }\nonumber\\
&\approx&-\frac{2}{2\pi {\rm i}}\int_{-\infty}^{\infty}{\rm d}\Omega \left[ f(\Omega -\epsilon -\omega)-
f(\Omega -\epsilon )\right]
\sum_{\bf k}
\frac{V^R (0,0)}{D{\bf k}^2-{\rm i}(\Omega +\omega ) }\nonumber\\
&=&-\frac{2\omega}{2\pi{\rm i}}\sum_{\bf k}
\frac{V^R (0,0)}{D{\bf k}^2-{\rm i}(\epsilon +\omega ) }\nonumber\\
&=&{\rm i}\omega \frac{V_1-2V_2}{4\pi^2D}
\ln \left(\frac{1}{\epsilon\tau} \right)\nonumber\\
&=&-{\rm i}\omega \ {\rm I}_3,\label{rp8}
\end{eqnarray}
having used $f(\Omega -\epsilon -\omega)-
f(\Omega -\epsilon )\approx \omega \ \partial_{\Omega}f(\Omega -\epsilon )$.
In the last line we have also included the contribution of the Hartree diagrams.
The difference between the second and third term in the square brakets of eq.(\ref{rp7})
may be expanded in powers of $q$ and $\omega_{\nu}$. The lowest order term reads
\begin{eqnarray}
\Sigma_{L,d}^2&=& 2T\sum_{\epsilon_n+\omega_{\nu} <\Omega_m }\sum_{\bf k}{\cal V}({\bf k},\Omega_m)\left[
\frac{2D^2q^2 {\bf k}^2}{(D{\bf k}^2+|\Omega_m  |)^3 }-
\frac{(Dq^2+|\omega_{\nu}|)}{(D{\bf k}^2+|\Omega_m  |)^2 }\right]\nonumber\\
&=&- \frac{2}{2\pi {\rm i}}\int^{\infty}_{-\infty}{\rm d}\Omega f(\Omega -\epsilon -\omega)
\sum_{\bf k}
\left[\frac{Dq^2 D {\bf k}^2V^R (0,0)}{(D{\bf k}^2-{\rm i}\Omega )^3 }-
\frac{(Dq^2-{\rm i}\omega ) V^R (0,0)}{(D{\bf k}^2-{\rm i}\Omega )^2 }\right]\nonumber\\
&=&Dq^2\frac{V_1-2V_2}{4\pi^2 D}
\ln \left(\frac{1}{\epsilon\tau} \right)-(Dq^2-{\rm i}\omega )\frac{V_1-2V_2}{4\pi^2 D}
\ln \left(\frac{1}{\epsilon\tau} \right) \nonumber\\
&=& Dq^2 \ {\rm I}_2-(Dq^2-{\rm i}\omega ){\rm I}_1.
\label{rp9}
\end{eqnarray}
Finally, the second term in eq.(\ref{rp6}) reads
\begin{eqnarray}
\Sigma_{L,abc}&=&
-(Dq^2+|\omega_{\nu} | )
2T\sum_{\epsilon_n+\omega_{\nu} <\Omega_m}\sum_{\bf k}
\frac{{\cal V}({\bf k},\Omega_m ) }{(Dk^2+|\Omega_m | )^2}\nonumber\\
&=&-(Dq^2 -{\rm i}\omega)\frac{2}{2\pi{\rm i}}
\int^{\infty}_{-\infty}{\rm d}\Omega f(\Omega -\epsilon -\omega)
\sum_{\bf k}
\frac{V^R  (0,0)}{(Dk^2-{\rm i}\Omega  )^2}\nonumber\\
&=&-(Dq^2 -{\rm i}\omega)
\frac{V_1-2V_2}{4\pi^2 D}
\ln \left(\frac{1}{\epsilon\tau} \right)\nonumber\\
&=&-(Dq^2-{\rm i}\omega ) \ {\rm I}_1.
\label{rp10}
\end{eqnarray}
One sees that the by inserting the self-energy results (\ref{rp8}-\ref{rp10}) into eq.(\ref{rp2})
one gets
\begin{eqnarray}
\zeta &=&1-{\rm I}_1\label{rp11},\\
\frac{D_R}{D}&=&1-{\rm I}_2\label{rp12},\\
Z&=&1-{\rm I}_3\label{rp13}.
\end{eqnarray}
In the main text the renormalized diffusion coefficient $D_R$ will be renamed $D$.
We conclude this Appendix by remarking that the
above renormalizations coincide with the 
perturbative corrections of the single-particle density of states,
eq.(\ref{densityofstatestotal}), conductivity, eq.(\ref{conductivity10}) and 
specific heat, eq.(\ref{th9}), satisfying at this order the Ward identities
identifications.


\begin{thebibliography}{000}
\bibitem{bergmann1984}\BY{Bergmann~G.}
      \IN{Phys.~Rep.}{ 107}{1984}{1}.
      
\bibitem{lee1985} \BY{Lee~P.~A. \atque Ramakrishnan~T.~V.}
      \IN{Rev.~Mod.~Phys.}{57}{1985}{287}.
      
\bibitem{altshuler1985}\BY{Altshuler~B.~L. \atque Aronov~A.~G.}
      in \TITLE{ Electron-Electron Interactions in Disordered Systems},
      edited by \NAME{Pollak~M. \atque Efros~A.~L.} (North-Holland,
      Amsterdam) 1985, p.~1. 
      
\bibitem{castellani1985}\BY{Castellani~C. \atque Di~Castro~C.}
in \TITLE{Lecture Notes in Physics~ 216} edited by \NAME{Garrido~L.}
(Springer-Verlag) 1985.

\bibitem{castellani1985b}\BY{Castellani~C., Di~Castro~C. \atque Strinati~G.}
in \TITLE{Lecture Notes in Physics~ 268} edited by \NAME{Garrido~L.}
(Springer-Verlag) 1985.

\bibitem{finkelstein1990}\BY{Finkelstein~A.~M.~} 
      \IN{Soviet Scientific Reviews}{14}{1990}{ }.
                 
\bibitem{kramer1993} \BY{Kramer~B. and MacKinnon~A.} 
\IN{Rep.~Prog.~Phys.}{ 56}{1993}{1469} . 

\bibitem{belitz1994}\BY{Belitz~D. \atque Kirkpatrick~T.~R.}
      \IN{Rev.~Mod.~Phys.}{66}{1994}{261}.   
\bibitem{ioffe1960}\BY{Ioffe~A.~F. \atque  Regel~A.~R.}
\IN{Prog.~Semicond.}{4}{1960}{237}.

\bibitem{mott1972}\BY{Mott~N.~F.}\IN{Philos. Magazine}{26}{1972}{1015}.
\bibitem{dolan1979} \BY{Dolan~  G.~J. \atque Osheroff~ D.~D.}
 \IN{ Phys.~ Rev. ~Lett.}{ 43}{1979}{ 721}.
 
\bibitem{rosenbaum1983}  \BY{Rosenbaum~ T. ~F., 
 Milligan~R.~F.,  Paalanen~M.~A., Thomas~G.~A.,  Bhatt~R.~N., \atque 
 Lin~W.} \IN{ Phys. Rev. B }{27}{1983}{7509}. 
 
 \bibitem{rosenbaum1980}\BY{Rosenbaum~T.~F.   Andres~K.,  Thomas~G.~A., 
  \atque  Bhatt~R.~N.}
  \IN{Phys.~Rev.~Lett.}{ 45}{1980}{ 1723}.
  

 
  
 \bibitem{stupp1993}\BY{Stupp~H.,  Hornung~M., Lakner~M., 
 Madel~O., \atque v.L\"ohneysen~H.}
 \IN{Phys.~Rev.~Lett.}{ 71}{1993}{ 2634}.
 
 \bibitem{rosenbaum1994}\BY{Rosenbaum~ T. ~F., Thomas~G.~A., 
 \atque  Paalanen~M.~A., }
 \IN{Phys.~Rev.~Lett.}{ 72}{1994}{ 2121}.
 
 \bibitem{stupp1994}\BY{Stupp~H.,  Hornung~M., Lakner~M., 
 Madel~O., \atque v.L\"ohneysen~H.}
 \IN{Phys.~Rev.~Lett.}{ 72}{1994}{ 2122}.
 
 \bibitem{castner1994}\BY{Castner~T.~G.}
 \IN{Phys.~Rev.~Lett.}{ 73}{1994}{ 3600}.
 
 \bibitem{stupp1994b}\BY{Stupp~H.,  Hornung~M., Lakner~M., 
 Madel~O., \atque v.L\"ohneysen~H.}
 \IN{Phys.~Rev.~Lett.}{ 73}{1994}{ 3601}.
 
 \bibitem{thomas1982}\BY{Thomas~G.~A., Ootuka~Y., Katsumoto~S.,
 Kobayashi~S., \atque Sasaki~S.}
\IN{Phys.~Rev.~B}{ 25}{1982}{ 4288}.
 
 
 \bibitem{hertel1983}   \BY{Hertel ~G.,  Bishop ~D.~J., Spencer~ E.~G.,
  Rowell~ J.~M., \atque Dynes~ R.~C.}
 \IN{ Phys.~ Rev. ~Lett. }{ 50}{1983}{ 743}.
 
 
 \bibitem{yamaguchi1983}\BY{Yamaguchi~M., Nishida~N.,  Furubayashi~T., 
 Morigaki~K.,
 Ishimoto~H., \atque Ono~K.}
 \IN{Physica~ B (Amsterdam)}{118}{1983}{694}.
  
\bibitem{rhode1987}\BY{Rhode~M. \atque Micklitz~H.}
\IN{Phys.~Rev.~B}{ 36}{1987}{ 7572}. 

\bibitem{mcmillan1981}\BY{McMillan~W.~L. \atque Mockel~J.}
 \IN{ Phys.~ Rev. ~Lett. }{ 46}{1981}{ 556}.
 
 
 \bibitem{kobayashi1979}\BY{Kobayashi~ S., Ikeata~ S., Kobayashi~ S.,
 \atque Sasaki~ W.}
 \IN{Solid State Communications}{ 32}{1979}{ 1174}.
 \bibitem{thomas1981} \BY{Thomas~ G. ~A., Ootuka~ Y., Kobayashi~ S., \atque
  Sasaki~ W.}
 \IN{Phys. ~Rev.~ B}{ 24}{1981}{ 4886}. 
 
 \bibitem{paalanen1988} \BY{Paalanen~M.~A.,  Graebner~J.~E., 
  Bhatt~R.~N.,  \atque Sachdev~S.}(1988) 
  \IN{Phys.~Rev.~Lett.}{ 61}{1988}{ 597}.
  \bibitem{lakner1989}\BY{Lakner~M. \atque v.L\"ohneysen~H.}
 \IN{Phys.~Rev.~Lett.}{63}{1989}{648}.
 
 
 \bibitem{ikehata1985}  \BY{Ikeata~ S. \atque Kobayashi~ S.}
  \IN{Solid State Communications}{ 56}{1985}{ 607}. 
  
 \bibitem{paalanen1986}  \BY{Paalanen~M.~A., Sachdev~S., Bhatt~R.~N., 
 \atque Ruckenstein~A.~E.}
  \IN{Phys.~Rev.~Lett.}{57}{1986}{ 2061}.  
  
  \bibitem{alloul1987}\BY{Alloul~H. \atque Dellouve~P.}
  \IN{Phys.~Rev.~Lett.}{ 59}{1987}{ 578}.
  \bibitem{hirsh1992} \BY{ Hirsch~M.~ J.  Holcomb~D. ~F.,   Bhatt~R.~ N.,
  \atque Paalanen~M.A. }
  \IN{ Phys.~Rev.~Lett.}{ 68}{1992}{ 1418}. 
  
  \bibitem{schlager1997} \BY{Schlager~H.~G. \atque v. L\"ohneysen~H.}
   \IN{Europhys.~Lett.}{40}{1997}{661}.
 
\bibitem{kravchenko1995} \BY{Kravchenko~S.~V., Mason~W.~E., Bower~G.~E.,
Furneaux~J.~E., Pudalov~V.~M.~, \atque D'Iorio~M.}
\IN{Phys.~Rev.~B}{51}{1995}{7038}. 
 
 \bibitem{abrahams2001} \BY{Abrahams~E., Kravchenko~S.~V., \atque Sarachik~M.~P.}
\IN{ Rev. Mod. Phys.}{ 73}{2001}{ 251}.

 \bibitem{kravchenko2004} \BY{Kravchenko~S.~V., \atque Sarachik~M.~P.}
\IN{ Rep. Prog. Phys.}{ 67}{2004}{ 1}.
 
\bibitem{anderson1958}\BY{Anderson ~P.~W.}
\IN{Physical Review }{109}{1958}{1492}.

\bibitem{mott1967} \BY{Mott~N.~F.} \IN{Advances in Physics (Philosophical
Magazine Supplement}{16}{1967}{49}.



\bibitem{edwards1972}\BY{Edwards~J.~T. \atque Thouless~ D.~.J}
\IN{J. Phys. C}{5}{1972}{807}.

\bibitem{licciardello1975}\BY{Licciardello~D.~C. \atque Thouless~ D.~.J}
\IN{Phys. Rev. Lett.}{35}{1975}{1475}.
 

      
\bibitem{abrahams1979}\BY{Abrahams~E., Anderson~P.~W., LicciardelloD.~C.,
      \atque Ramakrishnan~T.~V.~}\IN{ Phys.~Rev.~Lett.}{42}{1979}{673}.
      
\bibitem{wegner1976}\BY{Wegner~F.}\IN{Z.~Phys.~B}{25}{1976}{327}. 



\bibitem{langer1966}\BY{Langer~J.~S. \atque Neal~ T.}
\IN{Phys. Rev. Lett.}{16}{1966}{984} .     

      

      
             

\bibitem{castellani1983}\BY{Castellani ~C., Di~Castro~C., Forgacs~G., \atque Tabet~E.}
\IN{J. Phys. C: Solid State Phys.}{16}{1983}{159}.

\bibitem{abrikosov1975} \BY{A. A. Abrikosov, L. P. Gorkov, and I. E. 
Dzyaloshinski} \TITLE{\it Methods of Quantum Field Theory 
in Statistical Physics},
Dover Publications Inc, New York (1975), Sec 1.3.

\bibitem{zala2001}\BY{Zala~G.,  Narozhny~B. N., \atque  Aleiner~I. L. }
\IN{Physical Review B}{64}{2001}{214204}.

\bibitem{gorkov1979}\BY{Gorkov~L.~P., Larkin~A.~I., \atque Khmelnitskii~ D.~E.}
      \IN{Pis'ma Zh.\ Eksp.\ Teor.\ Fiz.}{ 30}{1979}{248} 
      [\IN{JETP Lett.}{ 30}{1979}{ 228}].
      
\bibitem{okuma1988}\BY{ Okuma~ S., Komori~ F., \atque Kobayashi~ S.}
in \TITLE{Anderson Localization. Proceedings of the International Symposium}
 edited by \NAME{Ando, T.; Fukuyama, H.}
(Springer-Verlag) 1988, p.78.
 
\bibitem{nishida1984}\BY{ Nishida~ N., Furubayashi~ T. \atque Yamaguchi~ M.} 
\IN{ Solid State Physics}{19}{1984}{410}.



\bibitem{katsumoto1987}\BY{Katsumoto~ S., Komori~ F. Sano~N., 
\atque Kobayashi~S.}
\IN{Journal of the Physical Society of Japan}{56}{1987}{2259}.

\bibitem{katsumoto1988}\BY{Katsumoto~S.}
in \TITLE{Anderson Localization. Proceedings of the International Symposium}
 edited by \NAME{Ando, T.; Fukuyama, H.}
(Springer-Verlag) 1988, p.45.

\bibitem{paalanen1982} \BY{Paalanen~M.~A., Rosenbaum~ T. ~F., 
   Thomas~G.~A.,  Bhatt~R.~N., \atque 
 Lin~W.} \IN{ Phys.~ Rev.~ Lett. }{48}{1982}{1284}.

\bibitem{shafarman1989}\BY{Shafarman~W.~N., Koon~D.~W., \atque Castner~T.~G.}
\IN{Phys.~ Rev.~ B}{40}{1989}{1216}.
\bibitem{dai1991}\BY{  Dai~P.,  Zhang~Y., \atque Sarachik~ M.P.}
\IN{Phys.~ Rev.~ Lett.}{ 67}{1991}{ 1914}.
\bibitem{dai1993}\BY{  Dai~P.,  Zhang~Y., Bogdanovich~S., \atque Sarachik~ M.P.}
 \IN{Phys.~ Rev.~ B}{48}{1993}{4941}.   
 
  \bibitem{attaccalite2002}\BY{Attaccalite~C., Moroni~S.,  Gori-Giorgi~P., 
  \atque  Bachelet~G.B.  }
 \IN{Phys.~ Rev.~ Lett.}{88}{2002}{ 256601}.

 
 \bibitem{simonian1997}\BY{Simonian~D.  Kravchenko~S.~V., \atque  Sarachik~M.  }
 \IN{Phys.~ Rev.~ Lett.}{79}{1997}{2304}.
 
 
   
\bibitem{altshuler1979} \BY{Altshuler~B.~L. \atque Aronov~A.~G.} 
\IN{Pis'ma Zh.~Eksp.~Teor.~Fiz.}{30}{1979}{514}
        [\IN{JETP Lett.}{30}{1979}{482}].
	
\bibitem{altshuler1980}\BY{Altshuler~B.~L., Aronov~A.~G., \atque Lee~P.~A.~}
      \IN{Phys.~Rev.~Lett.}{44}{1980}{1288}.
      
\bibitem{altshuler1980b}\BY{Altshuler~B.~L., Khmelnitskii~ D.~E., 
Larkin~A.~I., \atque Lee~P.~A.~}
      \IN{Phys.~Rev.~B}{22}{1980}{5142}.
      

	
\bibitem{altshuler1983b} \BY{Altshuler~B.~L. \atque Aronov~A.~G.} 
\IN{Solid State Communications}{46}{1983}{429}.	


\bibitem{finkelstein1983}\BY{Finkelstein~A.~M.~} 
      \IN{Zh.~Eksp.~Teor.~Fiz.}{84}{1983}{ 168}
      [\IN{Sov.~Phys.~JETP }{57}{1983}{97}].
      
\bibitem{castellani1984}\BY{Castellani~C., Di~Castro~C., Lee~P.~A., \atque Ma~M.} 
      \IN{Phys.~Rev.~B}{ 30}{1984}{527}.  
      
\bibitem{castellani1986b}\BY{Castellani~C. \atque Di~Castro~C.} 
      \IN{Phys.~Rev.~B}{ 34}{1986}{5935}.
      
\bibitem{castellani1987}\BY{Castellani~C., Kotliar~G., \atque Lee~P.~A.} 
      \IN{Phys.~Rev.~Lett.}{ 59}{1987}{323}.      

\bibitem{castellani1983b}\BY{Castellani ~C., Di~Castro~C., Forgacs~G., \atque Tabet~E.}
\IN{Nucler Phys. B}{225}{1983}{441}.

\bibitem{altshuler1983} \BY{Altshuler~B.~L.,  Aronov~A.~G., \atque Zyuzin~ A.~Yu.~} 
\IN{ Zh.~Eksp.~Teor.~Fiz.}{84}{1983}{1537}
        [\IN{Sov. Phys. JETP }{57}{1983}{889}].
      
\bibitem{castellani1986}\BY{Castellani~C., Di~Castro~C., Lee~P.~A.,  Ma~M., Sorella~S., \atque Tabet~E.} 
      \IN{Phys.~Rev.~B}{ 33}{1986}{6169}.
            
 \bibitem{raimondi1990}\BY{Raimondi~R.,  Castellani~C., \atque Di~Castro~C.}
\IN{Physical Review B}{42}{1990}{4724}.                
 
 \bibitem{castellani1987b}\BY{Castellani~C.,  Di~Castro~C., Kotliar~G., Lee~P.A.,
 \atque Strinati~ G.}
\IN{Physical Review Letters}{59}{1987}{477}. 

\bibitem{castellani1988}\BY{Castellani~C.,  Di~Castro~C., Kotliar~G., Lee~P.A.,
 \atque Strinati~ G.}
\IN{Physical Review B}{37}{1988}{9046}. 

\bibitem{dicastro1988}    \BY{ Di~Castro~C.}  
in \TITLE{Anderson Localization. Proceedings of the International Symposium}
 edited by \NAME{Ando, T.; Fukuyama, H.}
(Springer-Verlag) 1988, p.96.     

 \bibitem{finkelstein1984}\BY{Finkelstein~A.~M.~} \IN{Z. Phys. B}{56}{1984}{189}.  

\bibitem{castellani1984b}\BY{Castellani~C., Di~Castro~C., Lee~P.~A.,  Ma~M., Sorella~S., \atque Tabet~E.} 
      \IN{Phys.~Rev.~B}{ 30}{1984}{1596}.  
 
 
 
\bibitem{castellani1998} \BY{Castellani~C., Di~Castro~C., \atque Lee~P.~A.}
\IN{ Phys. Rev. B}{ 57}{1998}{ 9381}.

\bibitem{punnose2002} \BY{ Punnoose~A. \atque  Finkel'stein~A.~M.}
\IN{ Phys. Rev. Lett.}{88}{2002}{ 016802}.  

\bibitem{prus2003}\BY{Prus~O., Yaish~Y., Reznikov~M., Sivan~U., and
Pudalov~V.}
\IN{Phys.~Rev.~B}{67}{2003}{205407}.

\bibitem{lakner1994} \BY{Lakner~M.,  v.L\"ohneysen~H., Langenfeld~A., \atque
W\"olfle~P.}
 \IN{Phys.~Rev.~B}{50}{1994}{17064}.
 
 \bibitem{castellani1984c}\BY{Castellani ~C., Di~Castro~C., Forgacs~G., \atque
Sorella~S.}
\IN{Solid State Communications}{52}{1984}{261}.
 
\bibitem{kopietz1998}\BY{Kopietz~P.} 
      \IN{Phys.~Rev.~Lett.}{ 81}{1998}{2120}.
      

      
\end{thebibliography}
\end{document}